\pdfoutput=1

\documentclass[11pt,twoside,a4paper,cmspaper,final,collab]{cms-tdr}
\def\svnVersion{189584}\def\cmsCernNo{2013-027}\def\cmsCernDate{\today}\def\cmsMessage{Submitted to the Journal of High Energy Physics}

\begin{document}\cmsNoteHeader{HIG-12-035}

\hyphenation{had-ron-i-za-tion}
\hyphenation{cal-or-i-me-ter}
\hyphenation{de-vices}

\RCS$Revision: 182208 $
\RCS$HeadURL: svn+ssh://svn.cern.ch/reps/tdr2/papers/HIG-12-035/trunk/HIG-12-035.tex $
\RCS$Id: HIG-12-035.tex 182208 2013-04-22 12:54:25Z sboutle $
\newlength\cmsFigWidth
\ifthenelse{\boolean{cms@external}}{\setlength\cmsFigWidth{0.85\columnwidth}}{\setlength\cmsFigWidth{0.4\textwidth}}
\ifthenelse{\boolean{cms@external}}{\providecommand{\cmsLeft}{top}}{\providecommand{\cmsLeft}{left}}
\ifthenelse{\boolean{cms@external}}{\providecommand{\cmsRight}{bottom}}{\providecommand{\cmsRight}{right}}

\renewcommand{\GeVcc}{\GeV\xspace}
\renewcommand{\GeVc}{\GeV\xspace}

\cmsNoteHeader{HIG-12-035} 
\title{Search for the standard model Higgs boson produced in association with a top-quark pair in pp collisions at the LHC}

\date{\today}

\abstract{
A search for the standard model Higgs boson produced in association
with a top-quark pair is presented using data samples corresponding to
an integrated luminosity of $5.0\fbinv$ ($5.1\fbinv$) collected in
$\Pp\Pp$ collisions at the center-of-mass energy of $7\TeV$ ($8
\TeV$). Events are considered where the top-quark pair decays to
either one lepton+jets ($\ttbar \to \ell\cPgn
\cPq \cPaq^{\prime} \bbbar $) or dileptons ($\ttbar \to \ell^{+} \cPgn \ell^{-} \cPagn
\bbbar$), $\ell$ being an electron or a muon. The search is optimized
for the decay mode $\PH \to \bbbar$.
The largest background to the $\ttbar\PH$ signal is
top-quark pair production with additional jets. Artificial neural
networks are used to discriminate between signal and background
events.  Combining the results from the $7\TeV$ and $8\TeV$ samples, the
observed (expected) limit on the cross section for Higgs boson
production in association with top-quark pairs for a Higgs boson mass
of $125\GeVcc$ is 5.8 (5.2) times the standard model expectation.}

\hypersetup{%
pdfauthor={CMS Collaboration},%
pdftitle={Search for the standard model Higgs boson produced in association with a top-quark pair in pp collisions at the LHC},%
pdfsubject={CMS},%
pdfkeywords={CMS, physics, higgs, top}}

\maketitle 
\section{Introduction}
\label{sec:intro}
With the recent observation~\cite{ATLAS:2012gk,CMS:2012gu} at the
Large Hadron Collider (LHC) of a new, Higgs-like particle with a mass
of approximately $125\GeVcc$, the focus of searches for the
standard model (SM) Higgs boson has shifted to evaluating the
consistency of this new particle with SM expectations. A key component
in this effort will be to determine whether the new particle's
observed couplings to other fundamental particles match the
predictions for a SM Higgs boson. A deviation from expectations could
provide hints of physics beyond the standard model.

In the SM, the dominant production mechanism for the Higgs boson at
the LHC arises from gluon fusion, via the Higgs boson coupling to
gluons through a heavy quark loop.  However, with sufficient data,
other production mechanisms, such as Higgs boson production via vector
boson fusion or in association with a $\PW$ boson, $\cPZ$ boson, or
$\ttbar$ pair, should also be observable.  Furthermore, there are a
number of decay channels available to a SM Higgs boson with a mass of
approximately $125\GeVcc$.  Although the dominant decay mode at this
mass is to a pair of bottom quarks, decays to $\PW \PW$, $\cPZ \cPZ$,
$\Pgt \Pgt$, and $\cPgg \cPgg$ are also experimentally accessible.
The SM provides precise predictions for these production and decay
rates that depend on the coupling strength of the Higgs boson to the
other fundamental particles of the SM.

To date, the only combinations of production mechanism and decay mode
that have been established at greater than three standard deviation
($\sigma$) significance for this newly observed particle are direct
production, with the new particle decaying either to a pair of photons
or a pair of $\PW$ or $\cPZ$ bosons.  In all three of these cases, the
observed rates are in agreement with SM expectations for Higgs boson
production within the experimental uncertainties. However, establishing
the complete consistency of the couplings of this newly observed
particle with SM expectations for the Higgs boson involves measuring
the rate of production across all the various possible production and
decay channels discussed above.

The analysis described herein focuses on the search for a Higgs boson
produced in association with a pair of top quarks ($\ttbar \PH$
production) conducted at the Compact Muon Solenoid (CMS) experiment.
The analysis considers Higgs boson masses between $110$ and $140
\GeVcc$.  The search is optimized for Higgs boson decays to a
bottom-quark pair, but we do not exclude events from other Higgs boson
decay modes.  The rate at which this process occurs depends on the
largest of the fermionic couplings to the Higgs boson, namely the
couplings to the top and bottom quarks.  These two key couplings will
be particularly important in probing the new particle's consistency
with SM expectations.

The $\ttbar \PH$ vertex is the most challenging one to probe directly.
Measuring the rate of Higgs boson production through the gluon fusion
process provides an indirect measurement of the coupling between the
top quark and the Higgs boson because this production mechanism is
dominated by a top-quark loop that couples the gluons to the Higgs
boson~\cite{PhysRevLett.40.692}.  Likewise, the decay of the Higgs
boson to two photons receives a significant contribution from a
top-quark loop, although the loop involving $\PW$ bosons dominates in
this process~\cite{Actis:2008ts}.  However, extraction of the coupling
between the top quark and the Higgs boson in this way relies on the
assumption that there are no new massive fundamental particles beyond
those of the SM that contribute in the loop.  Unless the Higgs boson
is very heavy, it will not decay to top quarks.  Therefore, for the
mass range most favored for the SM Higgs~\cite{LEPewkfits}, and for
$125\GeVcc$ in particular, $\ttbar \PH$ production is the only way to
probe the $\ttbar
\PH$ vertex in a model-independent manner~\cite{Maltoni:2002jr,Belyaev:2002ua}.

In contrast, there are several processes that can be used to probe the
coupling of this new particle to bottom quarks.  Because of the large
$\bbbar$ background from multijet production, it is not experimentally
feasible to probe $\PH \to \bbbar$ in Higgs boson production via gluon
fusion.  Instead, the search is typically made using associated
production involving either a $\PW$ or a $\cPZ$ boson (V$\PH$
production).  Although $\ttbar \PH$ production has a smaller expected
cross section, this signature provides a probe that is complementary
to the V$\PH$ channel: they both provide information about the
coupling between the bottom quark and the Higgs boson, but the
dominant backgrounds are very different, $\ttbar+\text{jets}$ production
instead of $\PW+\text{jets}$ production.

An observation of $\ttbar \PH$ production, depending on the measured
properties, might be consistent with the SM Higgs boson or could
indicate something more exotic~\cite{Degrande:2012gr,Carmona:2012jk}.
Since the expected SM rates in this channel are very small, a sizeable
excess would be clear evidence for new physics. A previous search at
the Tevatron~\cite{Collaboration:2012bk}, the first such search
conducted at a hadron collider, showed no significant excess over SM
expectation.

This paper is organized as follows. Section~\ref{sec:cms} describes
the CMS apparatus. Section~\ref{sec:samples} describes the data and
simulation samples utilized in the analysis, while
Section~\ref{sec:selection} discusses the object identification, event
reconstruction and selection. The extraction of the $\ttbar
\PH$ signal is discussed in Section~\ref{sec:mva}, followed by a description
of the impact of systematic uncertainties encountered in the analysis
in Section~\ref{sec:systematics}. The results of this search are
reported in Section~\ref{sec:results} and followed by a summary in
Section~\ref{sec:conclusions}.

\section{The CMS detector}
\label{sec:cms}
The CMS detector consists of the following main components. A
superconducting solenoid occupies the central region of the CMS
detector, providing an axial magnetic field of 3.8\unit{T} parallel to
the beam direction. The silicon pixel and strip tracker, the crystal
electromagnetic calorimeter and the brass/scintillator hadron
calorimeter are located in concentric layers within the solenoid.  These
layers provide coverage out to $|\eta|= 2.5$, where pseudorapidity is defined as
$\eta = - \ln \left[\tan \left(\theta/2\right)\right]$.  A
quartz-fiber Cherenkov hadron forward calorimeter extends further
to $|\eta|< 5.2$. The CMS experiment
uses a right-handed coordinate system, with the origin at the nominal
interaction point, the $x$ axis pointing to the center of the LHC ring, the
$y$ axis pointing up (perpendicular to the LHC plane), and the $z$
axis along the counterclockwise beam direction. The polar angle $\theta$
is measured from the positive $z$ axis and the azimuthal angle $\phi$
is measured in the $x$-$y$ plane in radians. Muons are detected by gas-ionization
detectors embedded in the steel flux return yoke outside the
solenoid. The first level of the CMS trigger system, composed of
custom hardware processors, is designed to select the most interesting
events in less than 3\mus using information from the calorimeters
and muon detectors. The high-level trigger processor farm further
decreases the event rate to a few hundred Hz for data storage. More details
about the CMS detector can be found in Ref.~\cite{Chatrchyan:2008zzk}.

\section{Data and simulation samples}
\label{sec:samples}
This search is performed with samples of proton-proton collisions at
$\sqrt{s} = 7\TeV$ and $8\TeV$, collected with the CMS detector in 2011
and 2012, respectively.  These data correspond to a total integrated
luminosity of $5.0\fbinv$ at $7\TeV$ and $5.1\fbinv$ at $8\TeV$.

All background and signal processes are modeled using Monte Carlo (MC)
simulations from \MADGRAPH 5.1.1 \cite{Alwall:2011uj}, \PYTHIA 6.4.24
\cite{Sjostrand:2006za}, and \POWHEG 1.0 \cite{Alioli:2010xd} event
generators, depending on the physics process. The MC samples use
CTEQ6L1~\cite{Pumplin:2002vw} parton distribution functions (PDFs) of
the proton, except for the \POWHEG samples, which use CTEQ6M. The
$\ttbar \PH$ signal events are generated using \PYTHIA. The main
background $\ttbar$ sample is generated with \MADGRAPH, with matrix
elements corresponding to up to three additional partons which are
then matched to parton showers produced by \PYTHIA. The additional partons generated with the $\ttbar$ sample include b and c quarks in addition to light flavored quarks and gluons.  Decays of $\Pgt$
leptons are handled with \TAUOLA 2.75
\cite{Davidson:2010rw}. \MADGRAPH is also used to simulate $\ttbar
\PW$, $\ttbar \cPZ$, $\PW$ + jets, and Drell--Yan (DY) processes, with
up to 4 partons in the final state. The DY contribution includes all
$\cPZ/\cPgg^* \to \ell\ell$ processes with the dilepton invariant mass
$m_{\ell\ell} > 10\GeVcc$.  Single-top production is modeled with the
next-to-leading order (NLO) generator \POWHEG combined with
\PYTHIA. Electroweak diboson processes ($\PW \PW$, $\PW \cPZ$, and
$\cPZ \cPZ$) are simulated using \PYTHIA.

All background and signal process rates are estimated using
NLO or higher theoretical predictions. The $\ttbar \PH$ cross
section~\cite{Raitio:1978pt,Ng:1983jm,Kunszt:1984ri,Beenakker:2001rj,Beenakker:2002nc,Dawson:2002tg,Dawson:2003zu,Dittmaier:2011ti} and Higgs branching
fractions~\cite{Djouadi:1997yw,Djouadi:2006bz,Bredenstein:2006rh,Bredenstein:2006ha} used in the analysis have NLO
accuracy.
The $\ttbar$ and diboson cross sections are calculated at NLO with
\MCFM~\cite{Cacciari:2008zb,Campbell:2011bn,Campbell:2011cu}.  The
single-top-quark production rates are normalized to an approximately
next-to-next-to-leading order (NNLO)
calculation~\cite{Kidonakis:2012db,Kidonakis:2011wy,Kidonakis:2010tc,Kidonakis:2010ux}.
The $\PW$+jets and DY+jets rates are normalized to inclusive NNLO
cross sections from \textsc{fewz}~\cite{Gavin:2010az,Gavin:2012sy}.  The
$\ttbar \PW$ and $\ttbar \cPZ$ rates are normalized to the NLO
predictions from Refs.~\cite{Campbell:2012dh,Garzelli:2012bn}. These
cross sections are allowed to vary within their uncertainties in the
fit we use to calculate the limit.

Effects from additional pp interactions in the same bunch crossing (pileup) are modeled by adding
simulated minimum-bias events (generated with \PYTHIA) to the
simulated processes. The CMS detector response is simulated using the
\GEANTfour software package \cite{Agostinelli:2002hh}. The pileup
multiplicity distribution in MC is reweighted to reflect the
luminosity profile of the observed pp collisions. We apply an
additional correction factor to account for residual differences in
the jet transverse momentum ($\pt$) spectrum due to pileup; the
event-by-event correction factor is based on the difference between
simulation and data in the distribution of the scalar sum of the
transverse momenta of the jets in the event. We include a systematic
shape uncertainty in association with this correction factor. In
addition to correcting the MC due to pileup, we also apply jet energy
resolution corrections~\cite{Chatrchyan:2011ds} and lepton and trigger
efficiency scale factors to the MC events.

\section{Event reconstruction and selection}
\label{sec:selection}
\begin{figure}[t]
 \begin{center}
   \includegraphics[width=0.40\textwidth]{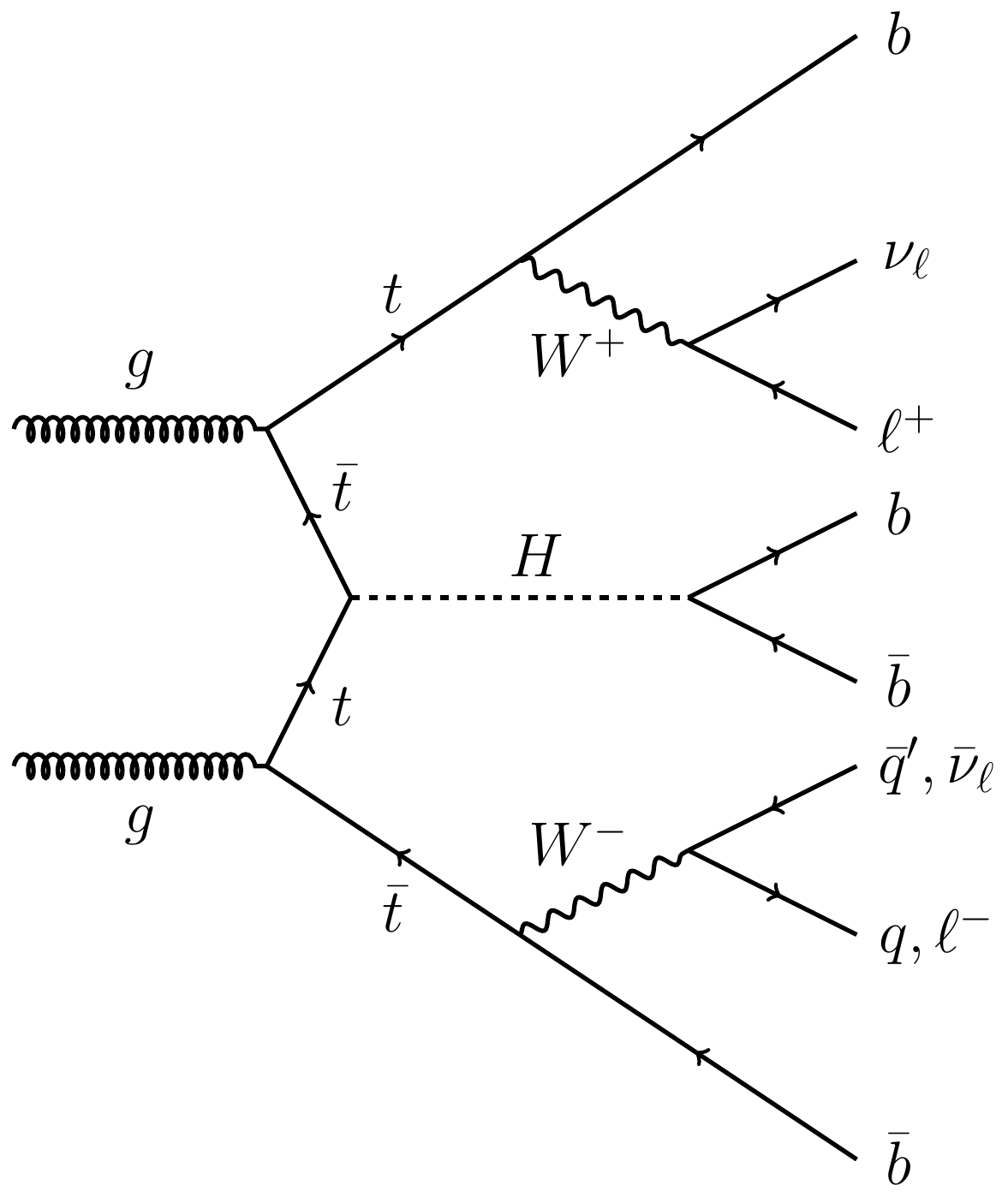}
   \caption{A leading-order Feynman diagram for $\ttbar \PH$ production, illustrating the two top-quark pair system decay channels considered here, and the $\PH \to \bbbar$ decay mode for which the analysis is optimized.}
   \label{fig:ttHFeynman}
 \end{center}
\end{figure}

This analysis selects events consistent with the production of a Higgs
boson in association with a top-quark pair (see Fig.~\ref{fig:ttHFeynman}). In the SM, the top quark
is expected to decay to a $\PW$ boson and a bottom quark nearly 100\%
of the time. Hence different $\ttbar$ decay modes can be identified
according to the subsequent decays of the $\PW$ bosons. Here we
consider two $\ttbar$ decay modes: the lepton+jets mode ($\ttbar \to
\ell \Pgn \cPq\cPaq^{\prime} \bbbar$), where one $\PW$ boson decays
leptonically, and the dilepton mode ($\ttbar \to \ell^{+} \Pgn
\ell^{-} \Pagn \bbbar$), where both $\PW$ bosons do so.  For the lepton+jets case, we select
events containing an energetic, isolated, electron or muon, and at
least four energetic jets, two or more of which should be identified
as originating from a b quark (b-tagged)~\cite{CMS:2012hd}.  For
the dilepton case, we require a pair of oppositely charged energetic
leptons (two electrons, two muons, or one electron and one muon) and
two or more jets, with at least two of the jets being b-tagged.

Object reconstruction is based on the particle flow (PF)
algorithm~\cite{CMS-PAS-PFT-09-001}, which combines the information
from all CMS subdetectors to identify and reconstruct individual
objects including muons, electrons, photons, and charged and neutral
hadrons produced in an event. To minimize the impact of pileup,
charged particles are required to originate from the primary vertex,
which is identified as the reconstructed vertex with the largest value
of $\Sigma \pt^2$, where the summation includes all tracks associated
with that vertex.  In both channels, a significant amount of missing
transverse energy ($\MET$) should be present due to the presence of
neutrinos, however no explicit requirement on the $\MET$ is used in
the event selection. The $\MET$ vector is calculated as the negative
of the vectorial sum of the transverse momenta of all particles. For
both channels, we use a common set of criteria for selecting
individual objects (electrons, muons, and jets) which is described
below.

In the lepton+jets channel, the data were recorded with triggers
requiring the presence of either a single muon or electron. The
trigger muon candidate was required to be isolated from other activity
in the event and to have $\pt > 24\GeVc$ for both the 2011 and 2012
data-taking periods. In 2011, the trigger electron candidate was
required to have transverse energy $\et > 25\GeV$ and to be produced
in association with at least three jets with $\pt > 30\GeVc$, whereas
in 2012, a single-electron trigger with minimum $\et$ threshold of $27
\GeV$ was used. In the dilepton channel, the data were recorded with
triggers requiring any combination of electrons and muons, one lepton
with $\pt > 17\GeVc$ and another with $\pt > 8\GeVc$. The offline
object selection detailed below is designed to select events in the
plateau of the trigger efficiency turn-on curve.

Muons are reconstructed using information from the tracking detectors
and the muon chambers \cite{Chatrchyan:2012xi}. Tight muons must
satisfy additional quality criteria based on the number of hits
associated with the muon candidate in the pixel, strip, and
muon detectors. For lepton+jets events, tight muons are required to
have $\pt>30\GeVc$ and $|\eta|<2.1$ to ensure the full trigger
efficiency. For dilepton events, tight muons are required to have $\pt
> 20\GeVc$ and $|\eta|<2.1$.  Loose muons in both channels are
required to have $\pt>10\GeVc$ and $|\eta|<2.4$.  The muon isolation
is assessed by calculating the scalar sum of the $\pt$ of
charged particles from the same primary vertex and neutral particles
in a cone of $\Delta R = \sqrt{\left(\Delta \eta\right)^2 + \left(\Delta \phi\right)^2} = 0.4$
around the muon direction, excluding the muon itself; the resulting
sum is corrected for the effects of neutral hadrons from pileup
interactions.  The ratio of this corrected isolation sum to the muon
$\pt$ is the relative isolation of the muon.  For tight muons, the
relative isolation is required to be less than 0.12.  For loose muons,
this ratio must be less than 0.2.

Electrons are reconstructed using both calorimeter and tracking
information \cite{CMS-PAS-EGM-10-004}. Any electron that can be paired
with an oppositely charged particle consistent with the conversion of
an energetic photon is rejected. Tight electrons in lepton+jets events
are required to have $\et>30\GeV$, while in dilepton events they must
have $\et>20\GeV$.  Loose electrons must have $\et>10\GeV$.  All
electrons are required to have $|\eta|<2.5$.  Electrons that fall into
the transition region between the barrel and endcap of the
electromagnetic calorimeter ($1.442<|\eta|<1.566$) are rejected
because the reconstruction of an electron object in this region is not
optimal.  The isolation for electrons is calculated in a similar
manner to muon isolation; however, for electrons the isolation sum is
calculated in a cone of $\Delta R = 0.3$.  In the same way as for
muons, the relative isolation is the ratio of this corrected isolation
sum to the electron $\et$.  Tight electrons must have a relative
isolation less than 0.1, while loose electrons must have a relative
isolation less than 0.2.

In both channels of this search, all events are required to contain at
least one tight lepton, either a muon or an electron. The second lepton
in the dilepton channel may be loose or tight, while in the lepton+jets
channel events with a second loose lepton are rejected to ensure the
same events do not enter both channels.

Jets are reconstructed by clustering the charged and neutral PF
particles using the anti-$\kt$ algorithm with a distance parameter of
0.5~\cite{Cacciari:2005hq,Cacciari:2008gp}. Particles identified as
isolated muons and electrons are expected to come from $\PW$ decays
and are excluded from the clustering. Non-isolated muons and electrons
are expected to come from $\cPqb$-decays and are included in the
clustering. The momentum of a jet is determined from the vector sum of
all particle momenta in the jet candidate and is scaled according to
jet energy corrections, based on simulation, jet plus photon data
events and dijet data events~\cite{Chatrchyan:2011ds}.  Charged PF
particles not associated with the primary event vertex are ignored
when reconstructing jets. The neutral component coming from pileup
events is removed by applying a residual energy correction following
the area-based procedure described in
Refs.~\cite{Cacciari:2008gn,Cacciari:2007fd}. In the lepton+jets
channel, we require at least three jets with $\pt > 40\GeVc$ and a
fourth jet with $\pt > 30\GeVc$. In the dilepton analysis, we require
at least two jets with $\pt > 30\GeVc$. All jets must have a
pseudorapidity in the range $|\eta| < 2.4$.

Jets are identified as originating from a b quark using the combined
secondary vertex (CSV) algorithm~\cite{CMS:2012hd}.  This
algorithm combines information about the impact parameter of tracks
and reconstructed secondary vertices within the jets in a multivariate
algorithm designed to separate jets containing the decay products of
bottom-flavored hadrons from jets originating from charm quarks, light
quarks, or gluons.  The CSV algorithm provides a continuous output
discriminant; high values of the CSV discriminant indicate that the
jet is more consistent with being a b jet, while low values indicate
the jet is more likely a light-quark jet.  To select b-tagged jets, a
selection is placed on the CSV discriminant distribution such that the
efficiency is 70\% (20\%) for jets originating from a b (c) quark and
the probability of tagging jets originating from light quarks or
gluons is 2\%. In addition, the CSV discriminant values for the
selected jets are used in the signal extraction as described in
Section~\ref{sec:mva}.  For MC events, the CSV discriminant values of
each jet are adjusted so that the proportion of b jets, c jets, and
light-quark jets of different $\eta$ and $\pt$ values passing each of
three CSV working points (tight,medium, and loose) is the same in data
and MC. The adjustment factor is computed
using a linear interpolation between CSV working points.

\begin{figure}[hbtp!]
 \begin{center}
   \includegraphics[width=0.60\textwidth]{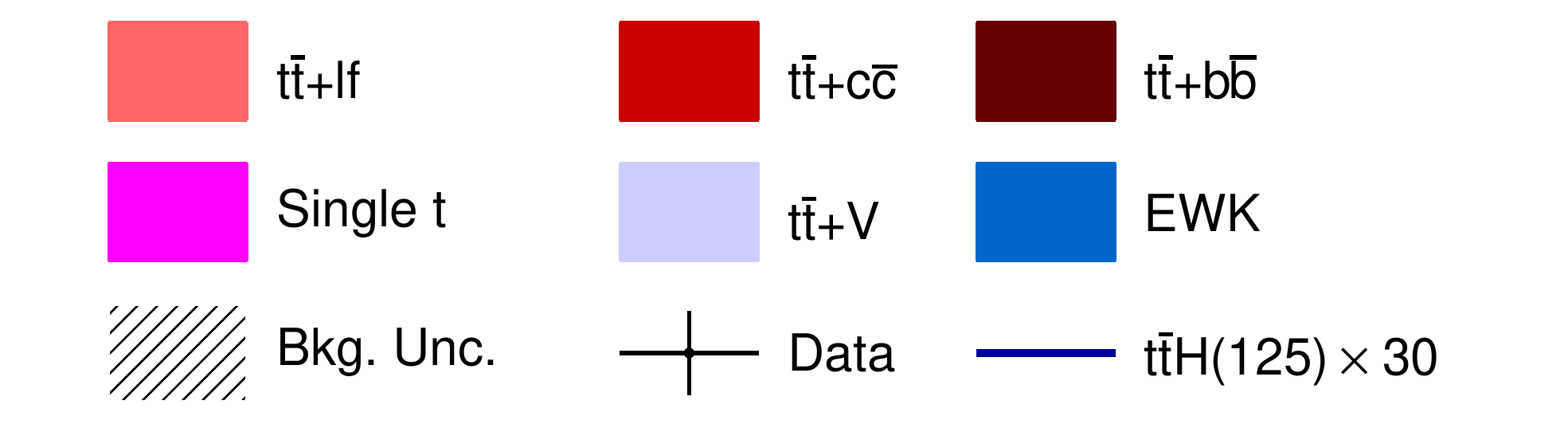} \\
   \includegraphics[width=0.35\textwidth]{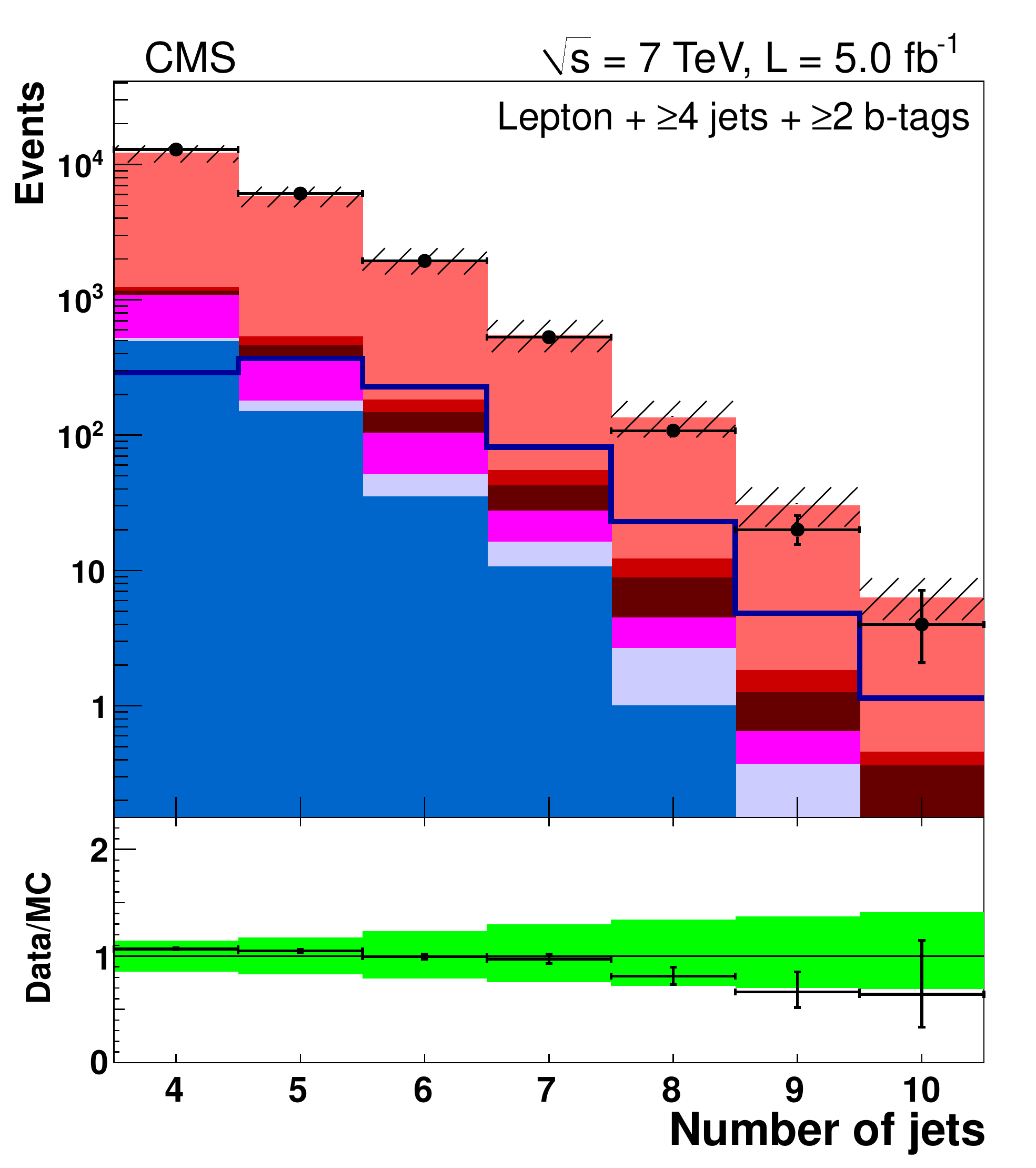}
   \includegraphics[width=0.35\textwidth]{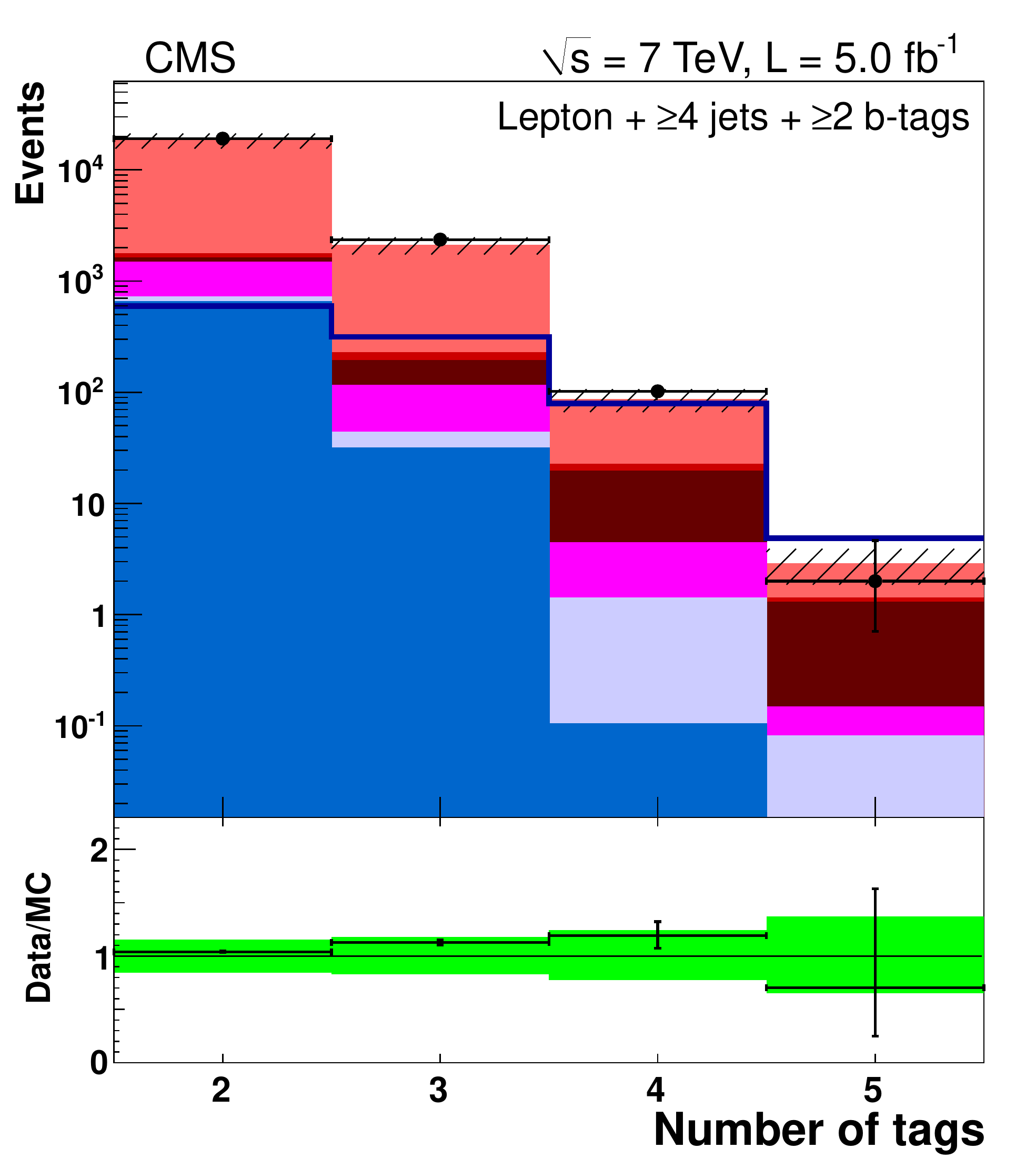}
   \includegraphics[width=0.35\textwidth]{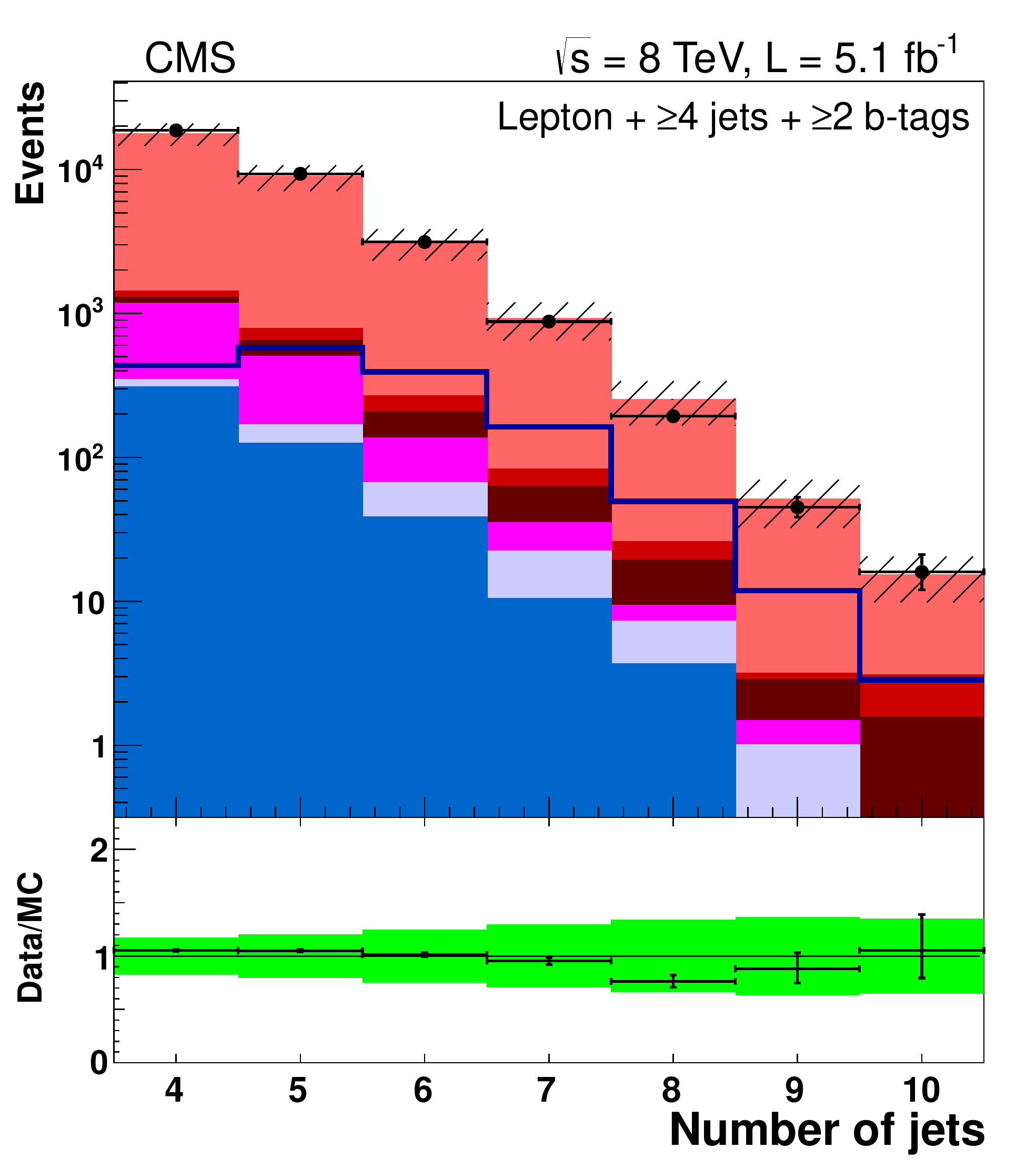}
   \includegraphics[width=0.35\textwidth]{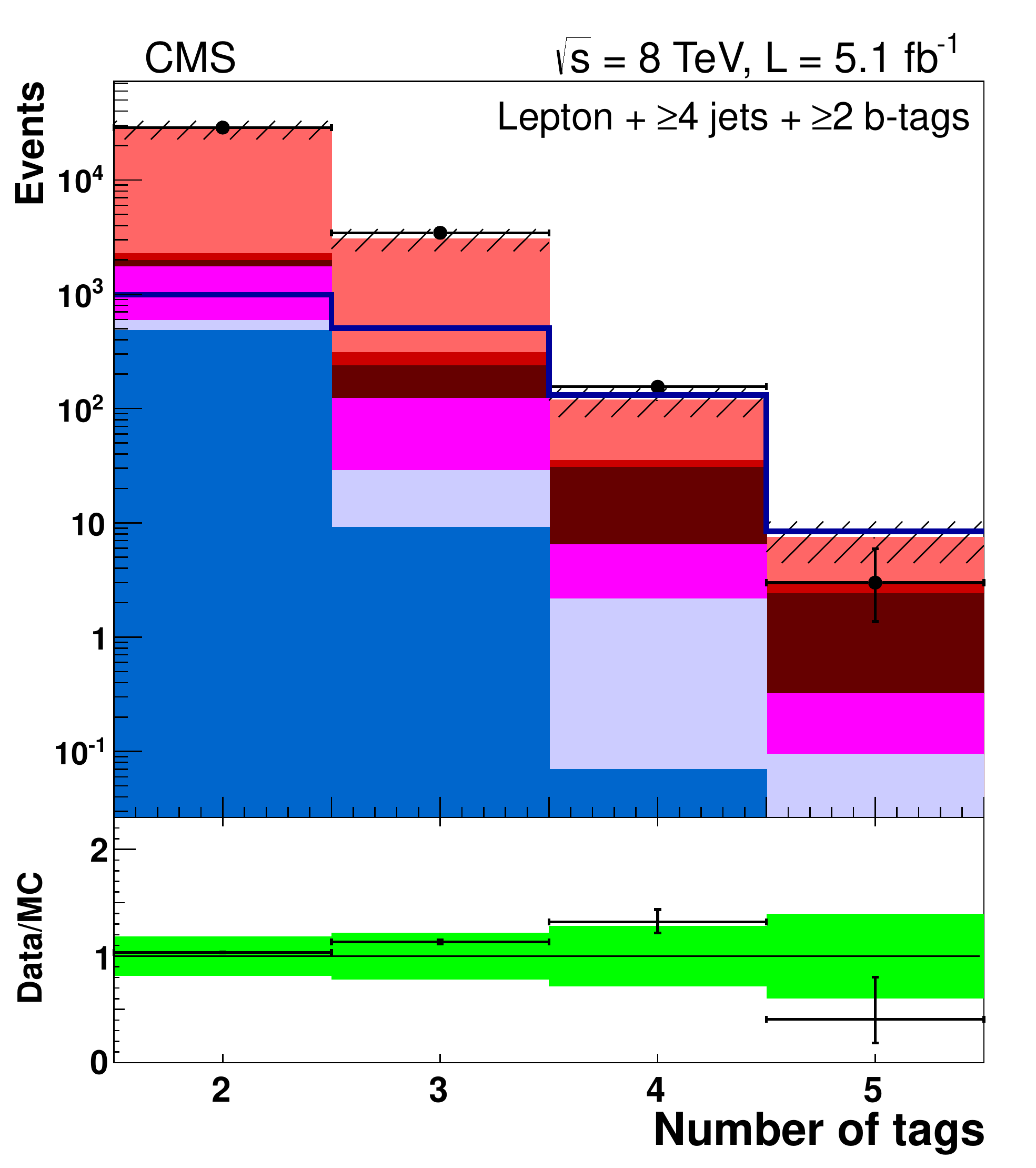}
   \caption{Number of jets (left) and number of b-tagged jets
   (right) in data and simulation for events with $\geq$4~jets +
   $\geq$2~b-tags in the lepton+jets channel at $7\TeV$ (top) and $8\TeV$
   (bottom). The background is normalized to the SM expectation;
   the uncertainty band (shown as a hatched band in the
   stack plot and a green band in the ratio plot) includes statistical
   and systematic uncertainties that affect both the rate and shape of
   the background distributions.  The $\ttbar \PH$ signal ($m_{\PH} =
   125\GeVcc$) is normalized to 30 $\times$ SM expectation.}
   \label{fig:njetsntags_LJ}
 \end{center}
\end{figure}

Figure~\ref{fig:njetsntags_LJ} shows the jet and b-tagged jet
multiplicities for events selected in the lepton+jets channel.  For
both lepton+jets and dilepton channels, signal $\ttbar \PH$ events are
generally characterized by having more jets and more tags than the
background processes.  To increase the sensitivity of this analysis,
we separate the selected events into different categories based on the
number of jets and tags.  For lepton+jets events, we use the following
seven categories: $\geq$6~jets + 2~b-tags, 4~jets + 3~b-tags, 5~jets +
3~b-tags, $\geq$6~jets + 3~b-tags, 4~jets + 4~b-tags, 5~jets +
$\geq$4~b-tags, and $\geq$6~jets + $\geq$4~b-tags.  For dilepton
events, only two categories are used: 2~jets + 2~b-tags and
$\geq$3~jets + $\geq$3~b-tags.
Tables~\ref{tab:dataMC_LJeventyield_7TeV}--\ref{tab:dataMC_DILEPeventyield_7and8TeV}
show the predicted signal, background, and observed yields in each
category for the lepton+jets and dilepton channels. Background
estimates are obtained from MC after the appropriate corrections and
scale factors have been applied, as described above. Given the event
selection criteria and the large jet and b-tag multiplicity
requirements in the lepton+jets channel, the background from QCD
multijet production is negligible. Uncertainties in signal and
background yields include both statistical and systematic
sources. Sources of systematic uncertainty are described in
Section~\ref{sec:systematics}. In
Tables~\ref{tab:dataMC_LJeventyield_7TeV}--\ref{tab:dataMC_DILEPeventyield_7and8TeV},
the $\ttbar+\text{jets}$ background is separated into the $\ttbar+\bbbar$,
$\ttbar+\ccbar$, and $\ttbar$+light flavor ($lf$) components.  The
categories with higher jet and tag multiplicities are the most sensitive
to signal. We include less sensitive categories in order to better
constrain the background.

The choice of event selection categories outlined above is optimized
for the $\PH \to \bbbar$ decay mode.  However, in the higher end of
our search range---including $m_{\PH}=125~\GeVcc$---other decay
modes, especially $\PW \PW$ and $\Pgt \Pgt$, can have significant
standard model branching fractions.  For the purposes of this search, we define any
$\ttbar \PH$ event as signal, regardless of the Higgs boson decay.
For most of the event selection categories defined above, the
contribution from the decay modes other than $\PH \to \bbbar$ is less
than 10\%.  The largest contribution from the non-$\bbbar$ decay modes
arises in the $\geq$6~jets + 2~b-tags lepton+jets category where almost
50\% of the events come from decay modes other than $\PH \to \bbbar$.
In that category $\PH \to \PW \PW$ dominates the non-$\bbbar$
contribution. With the current optimization, the impact of the
non-$\bbbar$ decay modes to the analysis sensitivity is negligible
as the contribution from $\PH \to \bbbar$ in the most sensitive
categories is $> 95\%$.

\begin{table}[htp]
  \centering
  \small
    \topcaption{Expected event yields for backgrounds (bkg), signal, and number of observed events in the lepton+jets channel in $7\TeV$ data.}
    \label{tab:dataMC_LJeventyield_7TeV}
    \begin{tabular}{|l|c|c|c|c|c|c|c|} \hline
& $\geq$6 jets & 4 jets & 5 jets & $\geq$6 jets & 4 jets & 5 jets & $\geq$6 jets \\
& 2 b-tags & 3 b-tags & 3 b-tags & 3 b-tags & 4 b-tags & $\geq$4 b-tags & $\geq$4 b-tags \\ \hline \hline
$\ttbar \PH(125)$ & 6.1 $\pm$ 0.9 & 2.7 $\pm$ 1.1 & 4.0 $\pm$ 1.6 & 3.8 $\pm$ 1.6 & 0.4 $\pm$ 0.2 & 1.1 $\pm$ 0.4 & 1.4 $\pm$ 0.6 \\
 \hline
$\ttbar+$lf & 2040 $\pm$ 520 & 940 $\pm$ 170 & 590 $\pm$ 120 & 346 $\pm$ 92 & 15.7 $\pm$ 3.3 & 22.8 $\pm$ 5.3 & 26.1 $\pm$ 7.7 \\
$\ttbar+\bbbar$ & 31 $\pm$ 17 & 26 $\pm$ 13 & 28 $\pm$ 15 & 24 $\pm$ 13 & 2.1 $\pm$ 1.1 & 5.7 $\pm$ 3.1 & 8.4 $\pm$ 4.8 \\
$\ttbar+\ccbar$ & 37.5 $\pm$ 9.5 & 10.1 $\pm$ 1.9 & 12.8 $\pm$ 2.7 & 11.8 $\pm$ 3.2 & 0.5 $\pm$ 0.1 & 1.0 $\pm$ 0.3 & 1.5 $\pm$ 0.5 \\
$\ttbar$~V & 18.4 $\pm$ 3.5 & 3.2 $\pm$ 0.6 & 4.3 $\pm$ 0.8 & 4.5 $\pm$ 0.9 & 0.2 $\pm$ 0.0 & 0.5 $\pm$ 0.1 & 0.7 $\pm$ 0.2 \\
Single $\cPqt$ & 54.8 $\pm$ 7.0 & 40.0 $\pm$ 5.1 & 21.8 $\pm$ 3.3 & 9.6 $\pm$ 1.6 & 1.2 $\pm$ 0.4 & 1.0 $\pm$ 0.3 & 0.8 $\pm$ 0.3 \\
V+jets & 41 $\pm$ 26 & 21 $\pm$ 11 & 4.9 $\pm$ 4.8 & 0.5 $\pm$ 0.6 & 0.0 $\pm$ 0.0 & 0.0 $\pm$ 0.0 & 0.1 $\pm$ 0.1 \\
Diboson & 0.6 $\pm$ 0.2 & 0.7 $\pm$ 0.2 & 0.7 $\pm$ 0.2 & 0.3 $\pm$ 0.2 & 0.0 $\pm$ 0.0 & 0.0 $\pm$ 0.0 & 0.0 $\pm$ 0.0 \\
 \hline
Total bkg & 2230 $\pm$ 540 & 1040 $\pm$ 180 & 660 $\pm$ 130 & 396 $\pm$ 99 & 19.7 $\pm$ 4.1 & 30.9 $\pm$ 7.3 & 38 $\pm$ 11 \\
 \hline
Data & 2137 & 1214 & 736 & 413 & 18 & 37 & 49 \\
\hline
\end{tabular}
\vspace{10 mm}
  \small
    \topcaption{Expected event yields for backgrounds (bkg), signal, and number of observed events in the lepton+jets channel in $8\TeV$ data.}
    \label{tab:dataMC_LJeventyield_8TeV}
    \begin{tabular}{|l|c|c|c|c|c|c|c|} \hline
& $\geq$6 jets & 4 jets & 5 jets & $\geq$6 jets & 4 jets & 5 jets & $\geq$6 jets \\
& 2 b-tags & 3 b-tags & 3 b-tags & 3 b-tags & 4 b-tags & $\geq$4 b-tags & $\geq$4 b-tags \\ \hline \hline
$\ttbar \PH(125)$ & 11.7 $\pm$ 1.9 & 3.9 $\pm$ 1.8 & 6.1 $\pm$ 2.8 & 6.9 $\pm$ 3.1 & 0.6 $\pm$ 0.3 & 1.5 $\pm$ 0.7 & 2.5 $\pm$ 1.2 \\
 \hline
$\ttbar+$lf & 3460 $\pm$ 940 & 1320 $\pm$ 280 & 870 $\pm$ 210 & 570 $\pm$ 170 & 18.0 $\pm$ 5.1 & 27.6 $\pm$ 8.6 & 41 $\pm$ 15 \\
$\ttbar+\bbbar$ & 61 $\pm$ 34 & 35 $\pm$ 19 & 43 $\pm$ 24 & 35 $\pm$ 20 & 2.5 $\pm$ 1.7 & 8.4 $\pm$ 5.3 & 15.4 $\pm$ 9.4 \\
$\ttbar+\ccbar$ & 62 $\pm$ 17 & 19.6 $\pm$ 5.1 & 25.0 $\pm$ 6.9 & 25.9 $\pm$ 7.7 & 0.6 $\pm$ 0.4 & 0.8 $\pm$ 0.9 & 3.7 $\pm$ 1.8 \\
$\ttbar$~V & 35.7 $\pm$ 7.5 & 4.5 $\pm$ 1.1 & 6.1 $\pm$ 1.4 & 8.6 $\pm$ 2.1 & 0.1 $\pm$ 0.1 & 0.7 $\pm$ 0.2 & 1.5 $\pm$ 0.4 \\
Single $\cPqt$ & 79 $\pm$ 18 & 56 $\pm$ 11 & 25.6 $\pm$ 6.2 & 10.3 $\pm$ 2.9 & 0.3 $\pm$ 0.6 & 3.1 $\pm$ 2.2 & 1.0 $\pm$ 0.6 \\
V+jets & 53 $\pm$ 40 & 5.9 $\pm$ 6.0 & 0.8 $\pm$ 0.9 & 0.0 $\pm$ 0.0 & 0.0 $\pm$ 0.0 & 0.0 $\pm$ 0.0 & 0.0 $\pm$ 0.0 \\
Diboson & 1.2 $\pm$ 0.4 & 1.8 $\pm$ 0.6 & 0.5 $\pm$ 0.2 & 0.2 $\pm$ 0.1 & 0.0 $\pm$ 0.0 & 0.0 $\pm$ 0.0 & 0.0 $\pm$ 0.0 \\
 \hline
Total bkg & 3760 $\pm$ 980 & 1440 $\pm$ 300 & 970 $\pm$ 230 & 650 $\pm$ 190 & 21.5 $\pm$ 6.1 & 41 $\pm$ 12 & 63 $\pm$ 21 \\
 \hline
Data & 3503 & 1646 & 1116 & 686 & 28 & 56 & 74 \\
\hline
\end{tabular}
\end{table}

\begin{table}[hp!]
  \centering
  \small
    \topcaption{Expected event yields for backgrounds (bkg), signal, and number of observed events in the dilepton channel in $7\TeV$ and $8\TeV$ data.}
    \label{tab:dataMC_DILEPeventyield_7and8TeV}
    \begin{tabular}{|l|c|c|c|c|} \hline
& \multicolumn{2}{c|}{$7\TeV$ Data} & \multicolumn{2}{c|}{$8\TeV$ Data} \\
& 2~jets + 2~b-tags & $\geq$3~jets + $\geq$3~b-tags & 2~jets + 2~b-tags & $\geq$3~jets + $\geq$3~b-tags \\ \hline \hline
$\ttbar \PH(125)$ & 0.5 $\pm$ 0.2 & 2.1 $\pm$ 0.9 & 0.7 $\pm$ 0.3 & 3.3 $\pm$ 1.5\\
 \hline
$\ttbar+$lf & 3280 $\pm$ 590 & 109 $\pm$ 25 & 4100 $\pm$ 780 & 135 $\pm$ 34\\
$\ttbar+\bbbar$ & 6.5 $\pm$ 3.4 & 16.1 $\pm$ 8.6 & 7.6 $\pm$ 4.2 & 25 $\pm$ 14\\
$\ttbar+\ccbar$ & 5.1 $\pm$ 1.0 & 7.5 $\pm$ 1.8 & 10.1 $\pm$ 2.8 & 14.1 $\pm$ 4.1\\
$\ttbar$~V & 2.6 $\pm$ 0.5 & 2.3 $\pm$ 0.5 & 3.5 $\pm$ 0.8 & 3.8 $\pm$ 0.9\\
Single $\cPqt$ & 99 $\pm$ 11 & 3.9 $\pm$ 0.8 & 129 $\pm$ 18 & 6.2 $\pm$ 2.4\\
V+jets & 810 $\pm$ 190 & 23.5 $\pm$ 9.7 & 830 $\pm$ 200 & 29 $\pm$ 13\\
Diboson & 25.8 $\pm$ 2.7 & 0.6 $\pm$ 0.1 & 29.2 $\pm$ 3.7 & 0.7 $\pm$ 0.2\\
 \hline
Total bkg & 4230 $\pm$ 660 & 163 $\pm$ 35 & 5110 $\pm$ 860 & 215 $\pm$ 48\\
 \hline
Data & 4303 & 185 & 5406 & 251\\
\hline
\end{tabular}
\end{table}

\section{Signal extraction}
\label{sec:mva}

Artificial neural networks (ANNs)~\cite{ANNBook} are used in all
categories of the analysis to further discriminate signal from
background and improve signal sensitivity. Separate ANNs are trained
for each jet-tag category, and the choice of input variables is
optimized for each as well.  The ANN input variables considered are
related to object kinematics, event shape, and the discriminant output
from the b-tagging algorithm. A total of 24 input variables has been
considered and are listed in column 1 of Table~\ref{tab:inputs}. The
inputs are selected from a ranked list based on initial separation
between signal and background.  The separation of the individual
variables is evaluated using a separation benchmark $\langle S^2
\rangle$~\cite{Hocker:2007ht} defined as follows:

\begin{equation} \langle S^2 \rangle = \frac{1}{2}
 \int{\frac{\left(\hat{y}_S(y) - \hat{y}_B(y)\right)^2}{\hat{y}_S(y) +
 \hat{y}_B(y)} \rd y},
\end{equation}

where $y$ is the input variable, and $\hat{y}_S$ and $\hat{y}_B$ are
the signal and background probability density functions for that input
variable in the signal and background samples, respectively. The
maximum number of input variables considered is determined by the
statistics in the simulated samples used for ANN training. The number
of variables per category is determined by reducing the number of
variables until the minimum number of variables needed to maintain
roughly the same ANN performance is reached. In the lepton+jets
categories, the use of approximately 10 input variables
yields stable performance; using fewer inputs exhibits degraded
discrimination power, and using more inputs exhibits little
improvement in performance in most categories. A similar exercise was
done for the dilepton categories. The choice of input variables for
each jet-tag category used in the $8\TeV$ analysis is summarized in
Table~\ref{tab:inputs}; the input variables for each category in the
$7\TeV$ analysis are very similar.  The input variables used in the
ANN can be broken down into several classes, as detailed below.

\begin{table}[htp]

  \centering
  \small
    \topcaption{The ANN inputs for the nine jet-tag categories in the $8\TeV$ $\ttbar
\PH$ analysis in the lepton+jets and dilepton channels. The choice of inputs
is optimized for each category. Definitions of the variables are given
in the text. The best input variable for each jet-tag category is denoted by
$\bigstar$.}
    \label{tab:inputs}
    \begin{tabular}{|l|c|c|c|c|c|c|c||c|c|} \hline
& \multicolumn{7}{c||}{Lepton+Jets}&\multicolumn{2}{c|}{Dilepton} \\ \hline
Jets& $\geq$6 & 4 & 5 & $\geq$6 & 4 & 5 & $\geq$6 & 2 & $\geq$3 \\
Tags & 2 & 3 & 3 & 3 & 4 & $\geq$4 & $\geq$4 & 2 & $\geq$3 \\ \hline \hline
Jet 1 $\pt$ & & $\checkmark$ & $\checkmark$ & & $\checkmark$ & & & $\bigstar$ & $\checkmark$ \\
Jet 2 $\pt$ & & $\checkmark$ & $\checkmark$ & & & & & &  \\
Jet 3 $\pt$ & $\checkmark$ & $\checkmark$ & $\checkmark$ & & & $\checkmark$ & & &  \\
Jet 4 $\pt$ & $\checkmark$ & $\checkmark$ & $\checkmark$ & & & $\checkmark$ & & &  \\
$N_{\text{jets}}$ & & & & & & & & & $\checkmark$ \\
$\pt(\ell,\MET,\text{jets})$ & & $\bigstar$ & $\checkmark$ & & $\checkmark$ & $\checkmark$ & & $\checkmark$ & $\checkmark$ \\
$M(\ell,\MET,\text{jets})$ & $\checkmark$ & $\checkmark$ & & $\checkmark$ & $\checkmark$ & & $\checkmark$ & &  \\
Average $M((\text{j}_m^{\text{untag}},\text{j}_n^{\text{untag}}))$ & $\checkmark$ & & & $\checkmark$ & & & & &  \\
$M((\text{j}_m^{\text{tag}},\text{j}_n^{\text{tag}})_{\text{closest}})$ & & & & & & & $\checkmark$ & &  \\
$M((\text{j}_m^{\text{tag}},\text{j}_n^{\text{tag}})_{\text{best}})$ & & & & & & & $\checkmark$ & &  \\
Average $\Delta R(\text{j}_m^{\text{tag}},\text{j}_n^{\text{tag}})$ & & & & $\checkmark$ & $\checkmark$ & $\checkmark$ & $\checkmark$ & &  \\
Minimum $\Delta R(\text{j}_m^{\text{tag}},\text{j}_n^{\text{tag}})$ & & & $\checkmark$ & & & & & $\checkmark$ & $\checkmark$  \\
$\Delta R(\ell,\text{j}_{\text{closest}})$ & & & & & & $\checkmark$ & $\checkmark$ & $\checkmark$ & $\checkmark$ \\
Sphericity & $\checkmark$ & & & $\checkmark$ & & & $\checkmark$ & &  \\
Aplanarity & $\checkmark$ & & & & $\checkmark$ & & & &  \\
$H_0$ & $\checkmark$ & & & & & & & &   \\
$H_1$ & $\checkmark$ & & & & $\checkmark$ & & & & \\
$H_2$ & & & & $\checkmark$ & & & $\checkmark$ & &   \\
$H_3$ & $\bigstar$ & & & $\checkmark$ & & & $\checkmark$ & &  \\
$\mu^{\text{CSV}}$ & $\checkmark$ & $\checkmark$ & $\bigstar$ & $\bigstar$ & $\bigstar$ & $\bigstar$ & $\bigstar$ & $\checkmark$ & $\bigstar$ \\
$(\sigma_n^{\text{CSV}})^2$ & & $\checkmark$ & $\checkmark$ & $\checkmark$ & $\checkmark$ & $\checkmark$ & & &  \\
Highest CSV value & & & & & & $\checkmark$ & & &  \\
2$^{nd}$-highest CSV value & & $\checkmark$ & $\checkmark$ & $\checkmark$ & $\checkmark$ & $\checkmark$ & $\checkmark$ & &  \\
Lowest CSV value & & $\checkmark$ & $\checkmark$ & $\checkmark$ & $\checkmark$ & $\checkmark$ & $\checkmark$ & &  \\ \hline
 \end{tabular}
\end{table}

The first class of variables are those that are basic kinematic
properties of single objects in the event or combinations of objects.
These variables include the $\pt$ of the leading four jets, and the
$\pt$ and mass of the system defined by the vector sum of the
lepton(s) momenta, the $\MET$ vector, and the momenta of the jets in
the event ($\pt(\ell,\MET,\text{jets})$ and
$M(\ell,\MET,\text{jets})$, respectively), all of which favor larger
values for $\ttbar \PH$ signal than for the backgrounds. The number of
jets is used in the $\geq$3~jets + $\geq$3~b-tags category in the
dilepton analysis since $\ttbar \PH$ signal favors larger jet
multiplicity than background.

A related class of variables involves looking at the kinematic
properties of pairs of jets.  The $\PH \to \bbbar$ decay produces jets
that have a large invariant mass even if the jets fail the $\cPqb$-tag
selection.  Other untagged jets in the event tend to come from
hadronic $\PW$ decay and initial- or final-state radiation, and tend
to have a small invariant mass compared to the jets from the Higgs
boson decay.  For this reason, some signal discrimination is provided
by examining the invariant mass of pairs of untagged jets in
lepton+jets categories with six or more jets but fewer than four
b-tagged jets.

Likewise, the 6-jet category with four or more tags uses two variables
that rely specifically on the $\PH \to \bbbar$ hypothesis: the
invariant mass of the tagged-jet pair with the smallest opening angle
($M((\text{j}_m^{\text{tag}},\text{j}_n^{\text{tag}})_{\text{closest}})$),
and the ``best Higgs mass''
($M((\text{j}_m^{\text{tag}},\text{j}_n^{\text{tag}})_{\text{best}})$),
the invariant mass constructed from the two tagged jets least likely
to be a part of the $\ttbar$ system as determined by a minimum
$\chi^2$ search among all the jet, lepton, and $\MET$ combinations in
the event, using the $\PW$ and top masses as kinematic
constraints. The
$M((\text{j}_m^{\text{tag}},\text{j}_n^{\text{tag}})_{\text{closest}})$
distribution for both signal and background has a peak near the same
value; however, the distribution is wider in the case of signal,
offering some discriminating power.  In signal events, the ``best
Higgs mass'' is highly correlated with the Higgs boson mass.  Although
the peak is broadened by events where the wrong jets are associated
with the Higgs boson decay, this variable still provides some power in
discriminating signal from background. The $\geq$6~jets + $\geq$4~b-tags
uses 11 variables instead of the typical 10 because it was shown that
the addition of the ``best Higgs mass'' variable, uniquely designed for this
jet-tag category, offers a non-negligible increase in
expected ANN performance.

Another class of variables exploits differences in the ``shape'' of
events between signal and background. In general, production of an
extra massive object, in addition to top quarks tends to make $\ttbar
\PH$ events more spherical in shape, while the background events
are more collimated or have more jet activity. Variables in this class
include angular correlations, like the opening angle between the
tagged jets ($\Delta
R(\text{j}_m^{\text{tag}},\text{j}_n^{\text{tag}})$) or between the
lepton and closest jet ($\Delta R(\ell,\text{j}_{\text{closest}})$),
where in the dilepton analysis the angle is calculated with respect to
the lepton leading in $\pt$.  More complex event shape variables like
sphericity and aplanarity~\cite{sphericity}, as well as the
Fox--Wolfram moments $H_0,H_1,H_2,H_3$ ~\cite{FoxWolfram}, also exhibit
differences between signal and background.

The last class of variables used in the ANN involves the CSV
discriminant values of the tagged jets.  The signal events tend to
have more b jets than the dominant $\ttbar+\text{jets}$ background.  Beyond
the simple multiplicity of tagged jets we can, however, exploit the
overall b-jet content of the signal in several ways. For instance, the
average and squared-deviation from this average of the CSV
discriminant values for the tagged jets ($\mu^{\text{CSV}}$,
$(\sigma_n^{\text{CSV}})^2$ for the $n$-th tagged jet) are powerful
variables.  Events with genuine b jets will have higher average
CSV discriminant values and the b jets themselves will have CSV values
more tightly clustered around high values than those from
light-flavour or charm jets which are tagged.

Using the procedure discussed above, different variables are chosen
for use in each of the different event selection categories. This is
motivated by the fact that although the $\ttbar$+jets background is
dominant throughout, the kinematics of the events can be very distinct
in different jet multiplicity bins.  Similarly, the tagging
discriminant of the b jets clearly is different in events with 2, 3 or
$\geq$4~b-tags. Finally, the overall breakdown of the $\ttbar$+jets
background into $\ttbar+\bbbar$, $\ttbar+\ccbar$ and
$\ttbar$+light-flavor is different across the jet-tag categories,
implying different variables will be more effective in some categories
than others.

In nearly all event selection categories, the variables that
discriminate best between signal and background directly involve
b-tagging information, such as the average CSV output value for
b-tagged jets.  This is natural, since the largest fraction of the
backgrounds in all categories involve events with fewer b jets than
the $\ttbar \PH$ generally has.  However, when considering
specifically the $\ttbar + \bbbar$, a background very similar to the
signal, the b-tagging information alone is not as powerful, and
additional information from kinematic variables and angular
correlations, such as the minimum $\Delta R$ between all pairs of
b-tagged jets, become important.  Even so, the $\ttbar+\bbbar$
background remains difficult to separate from the $\ttbar \PH$ signal.

\begin{figure}[hbtp]
 \begin{center}
   \includegraphics[width=0.31\textwidth]{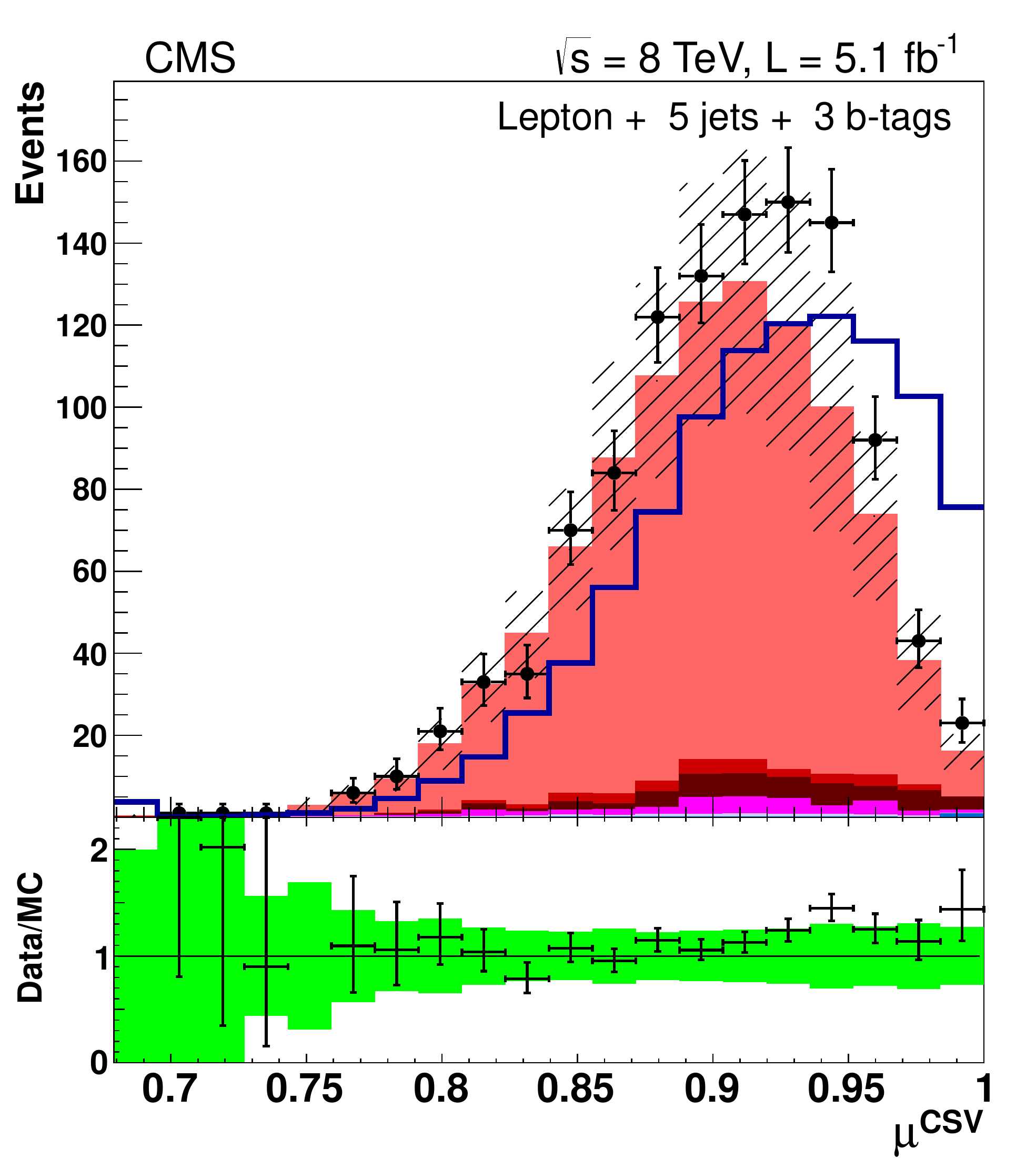}
   \includegraphics[width=0.31\textwidth]{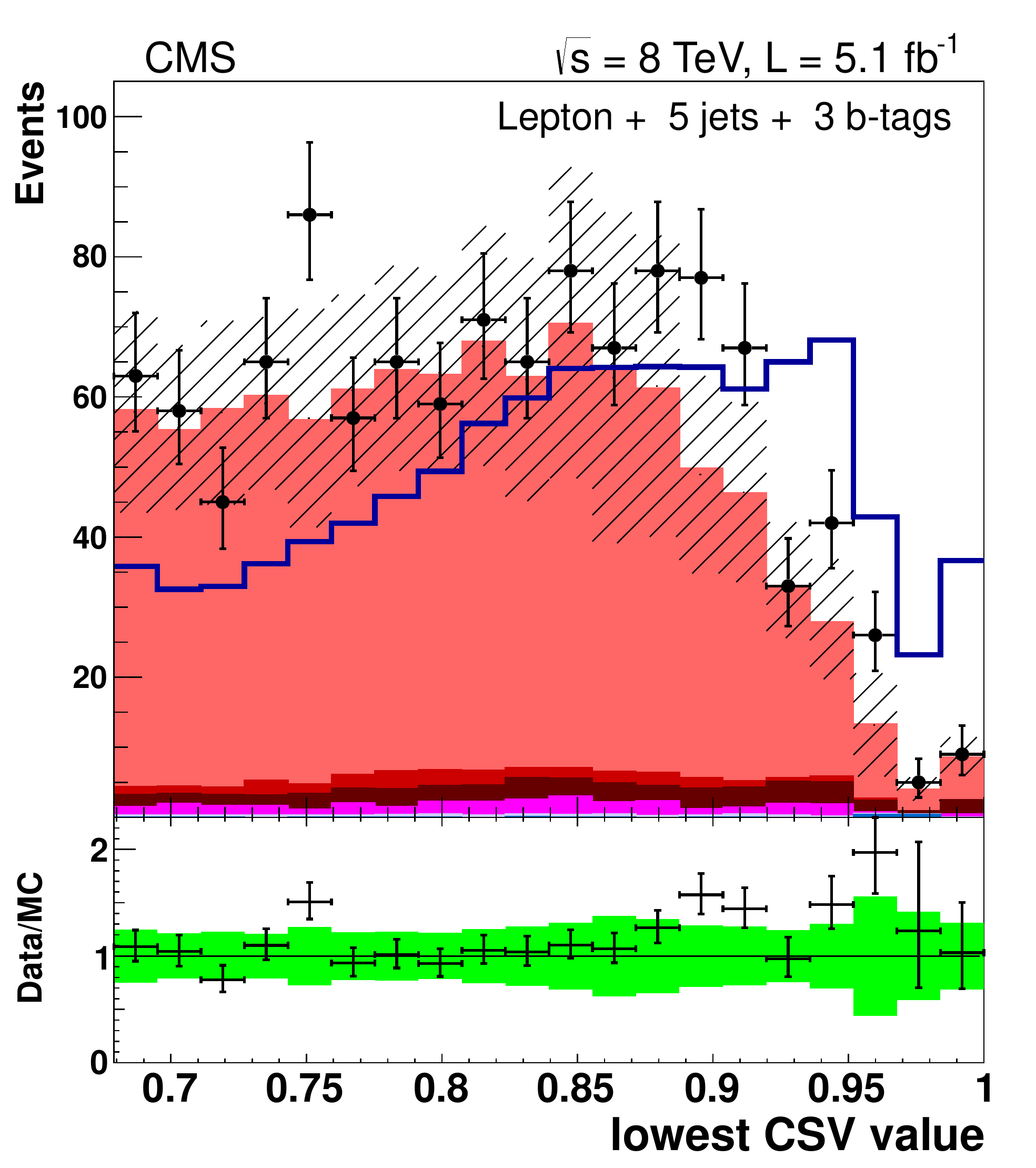}
   \includegraphics[width=0.31\textwidth]{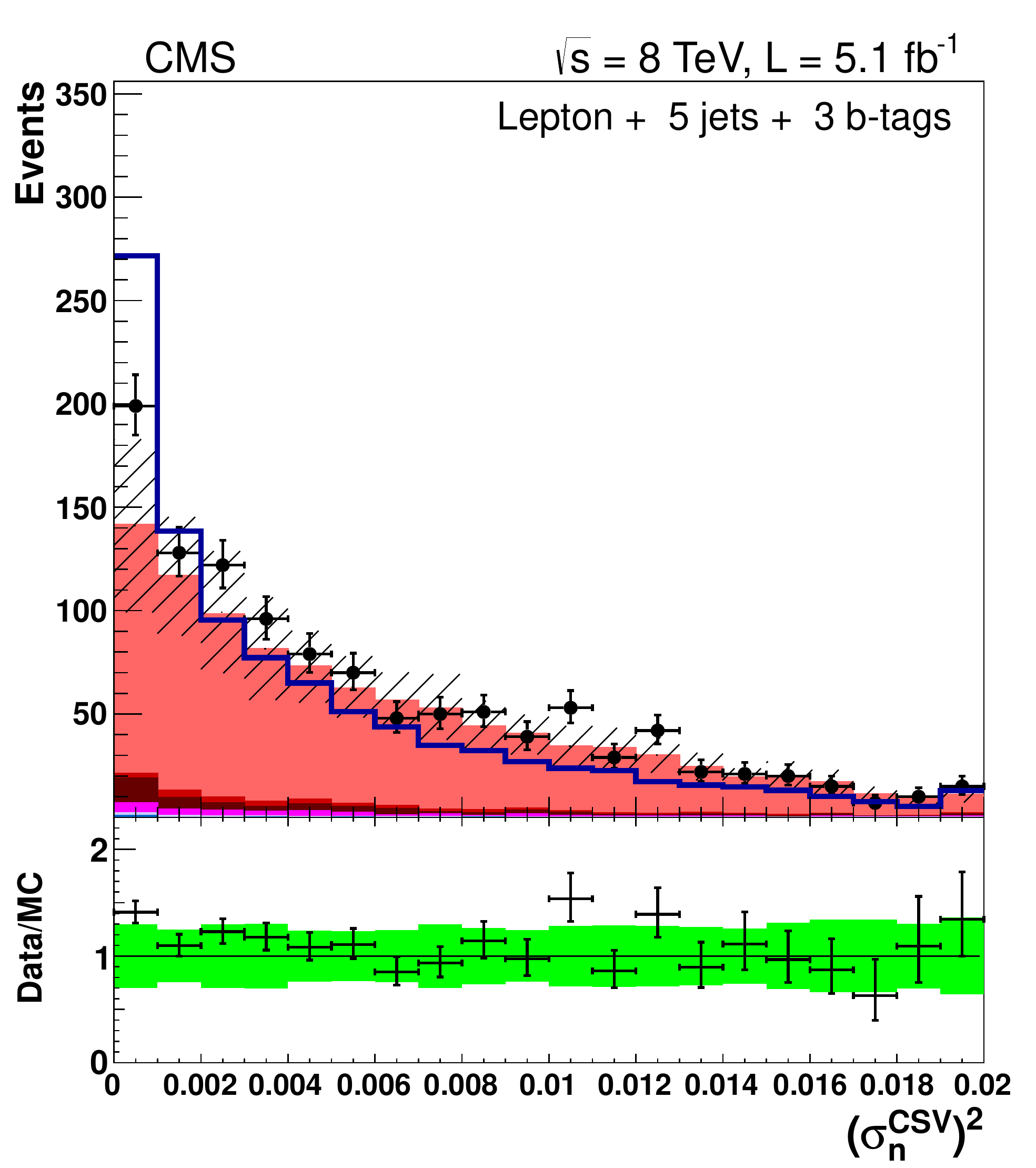}
   \includegraphics[width=0.31\textwidth]{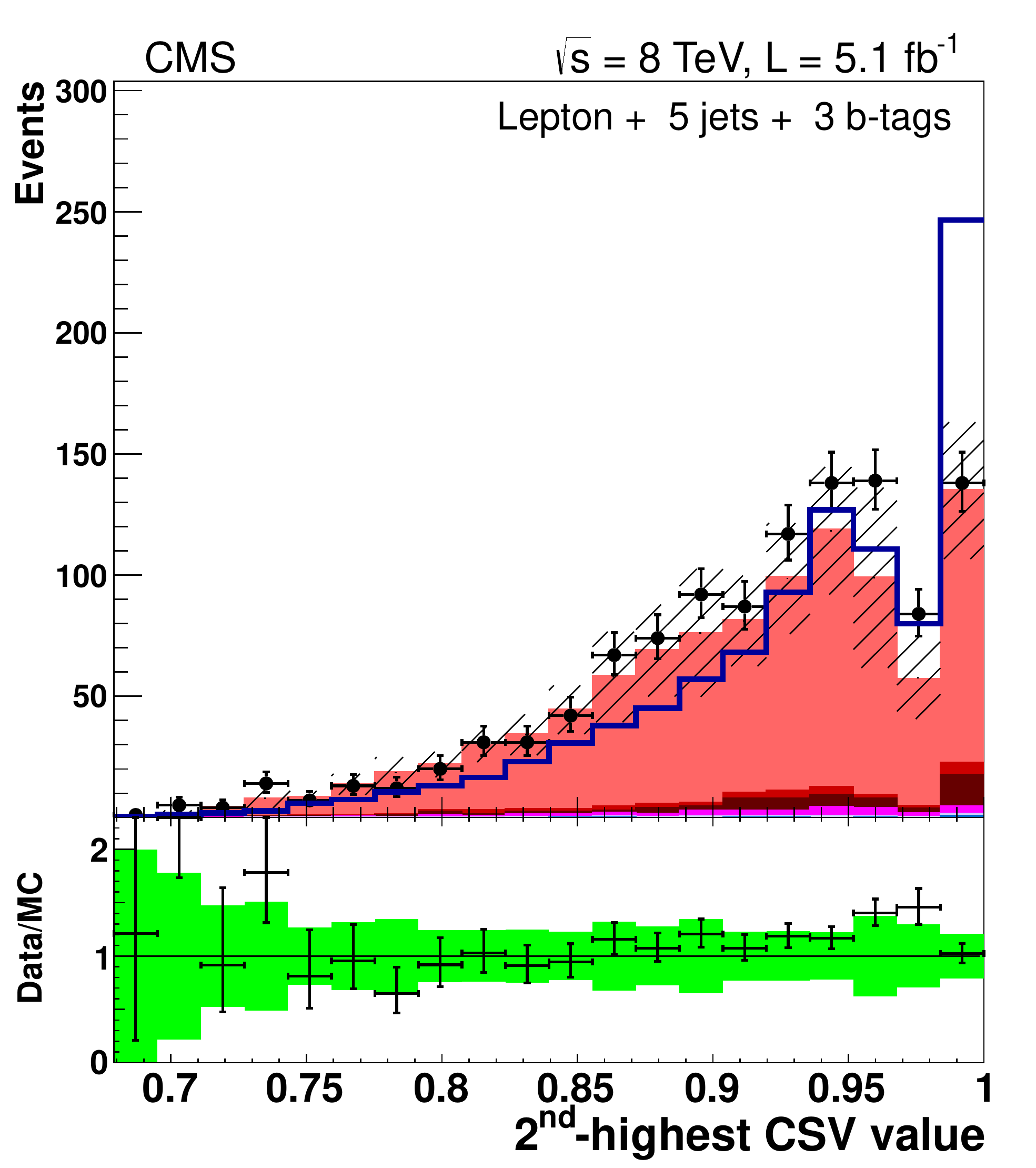}
   \includegraphics[width=0.31\textwidth]{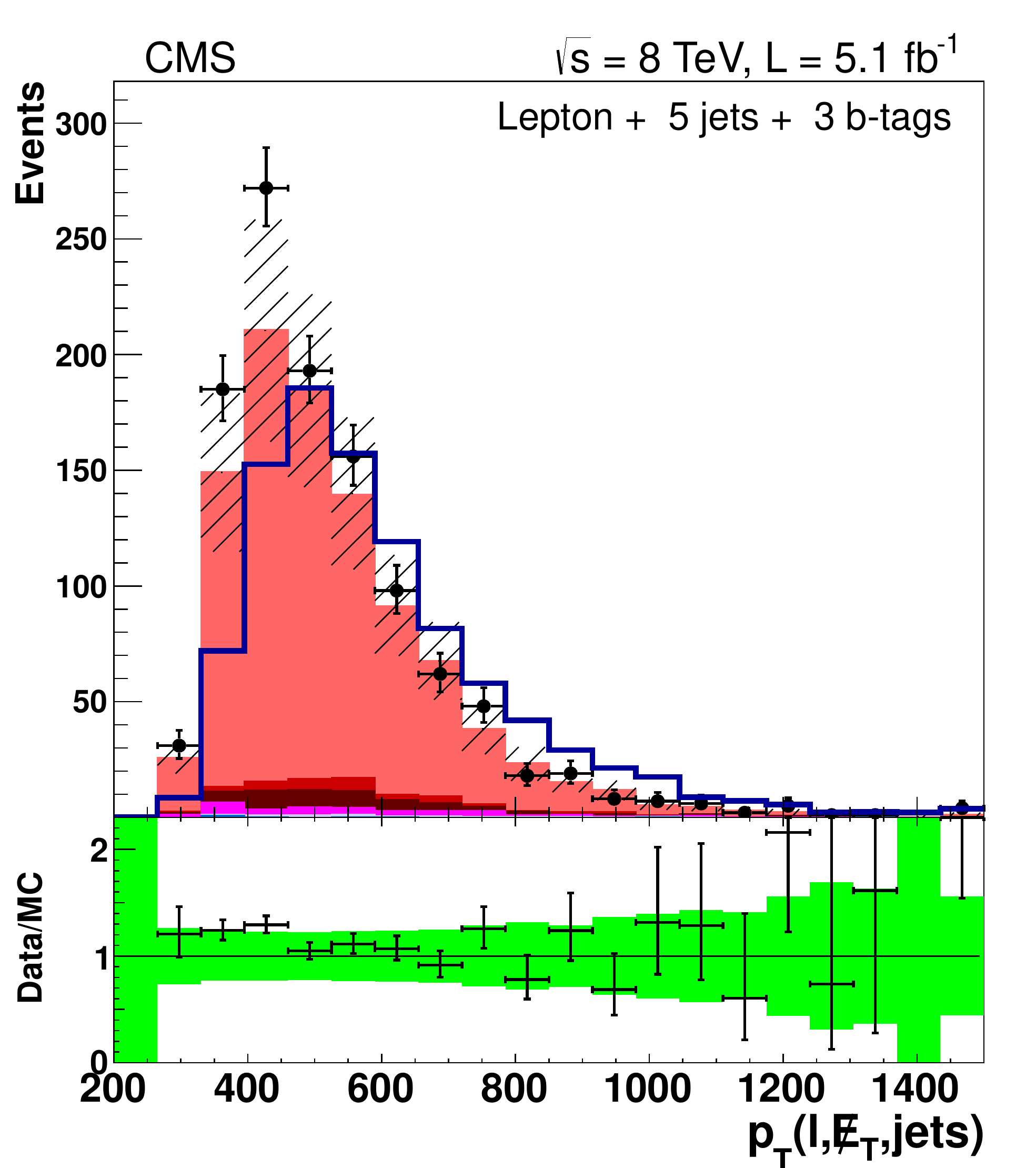}
   \raisebox{0.1\height}{\includegraphics[width=0.25\textwidth]{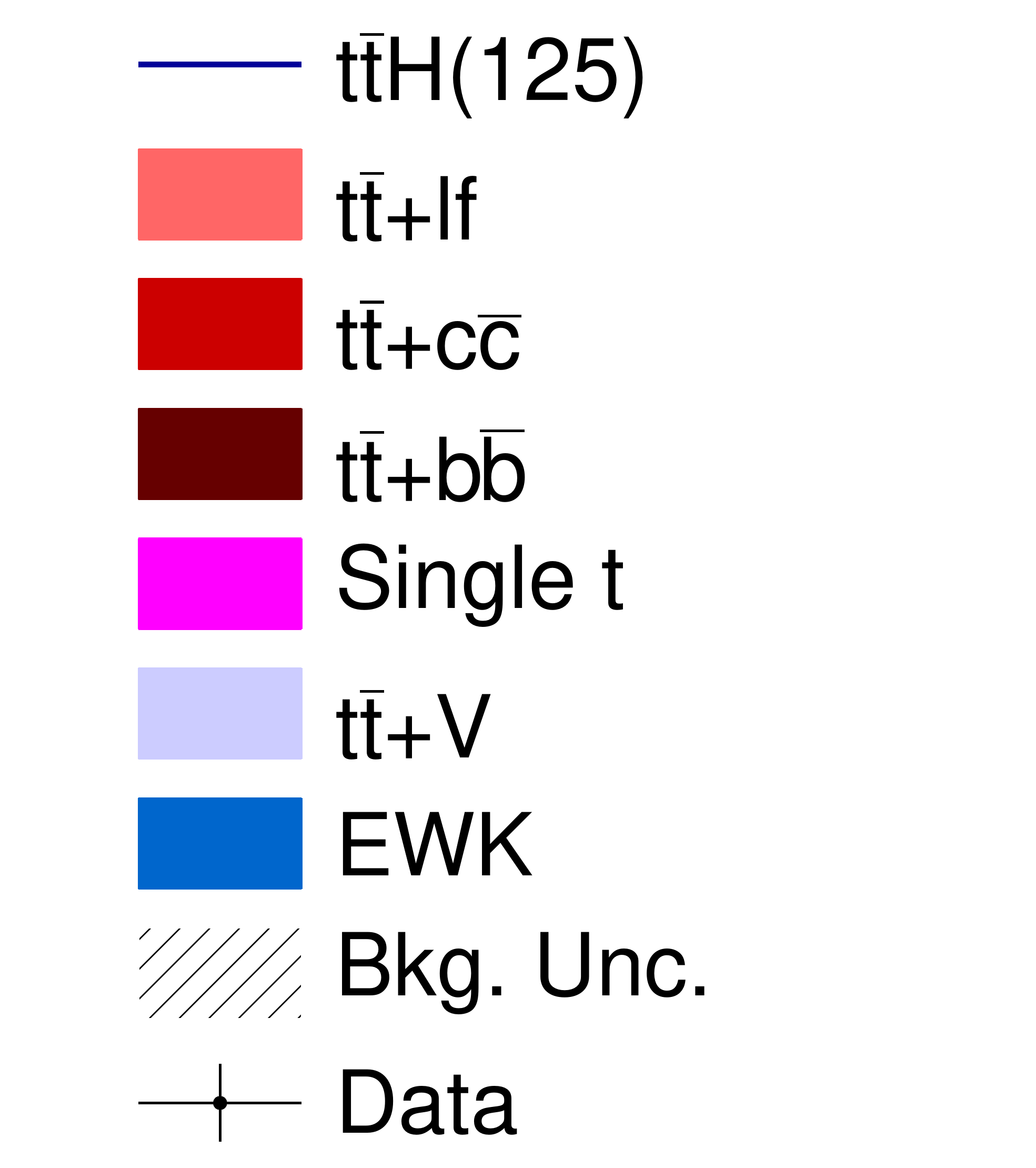}}
   \hspace{0.055\textwidth}
   \caption{Distributions of the five ANN input variables with
   rankings 1 through 5, in terms of separation, for the 5~jets +
   3~b-tags category of the lepton+jets channel at $8
  \TeV$. Definitions of the variables are given in the text. The
   background is normalized to the SM expectation; the
   uncertainty band (shown as a hatched band in the stack plot and a
   green band in the ratio plot) includes statistical and systematic
   uncertainties that affect both the rate and shape of the background
   distributions.  The $\ttbar \PH$ signal ($m_{\PH} = 125\GeVcc$) is
   normalized to $\sim$150 $\times$ SM expectation, equal to the total
   background yield, for easier comparison of the shapes.}
   \label{fig:lj_input_5j_3t_lep_8TeV_part1} \end{center}
\end{figure}

\begin{figure}[hbtp]
 \begin{center}
   \includegraphics[width=0.31\textwidth]{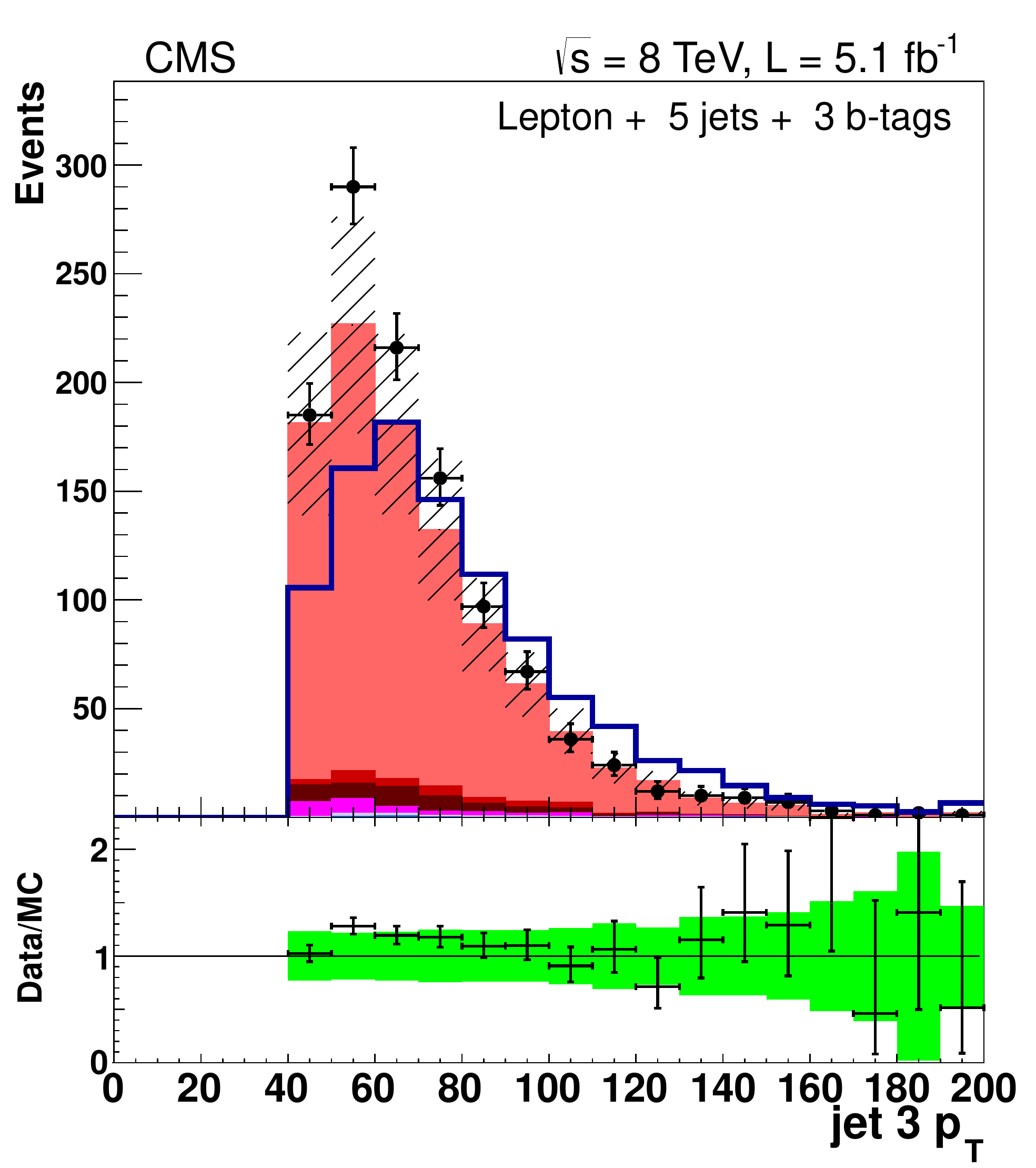}
   \includegraphics[width=0.31\textwidth]{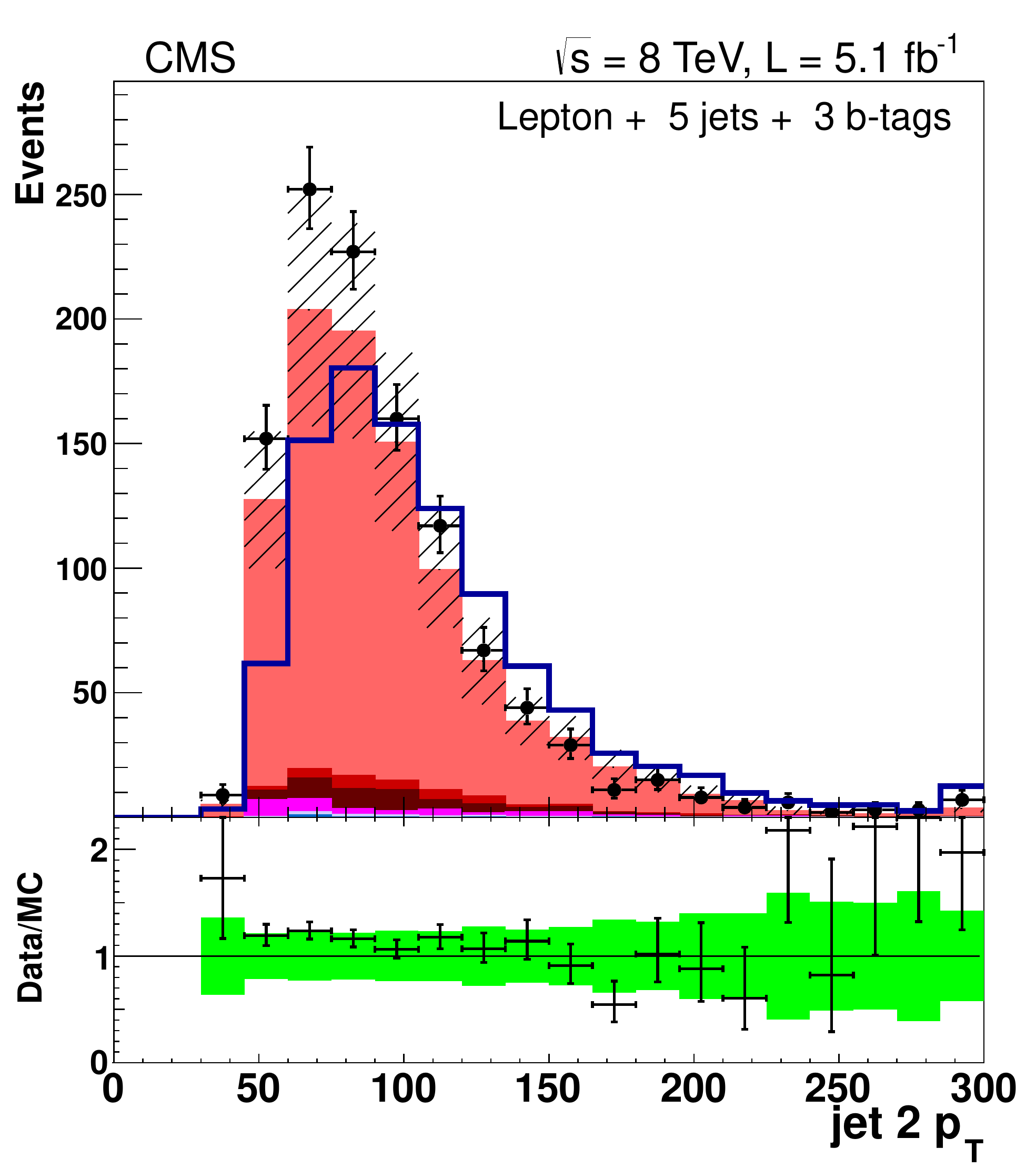}
   \includegraphics[width=0.31\textwidth]{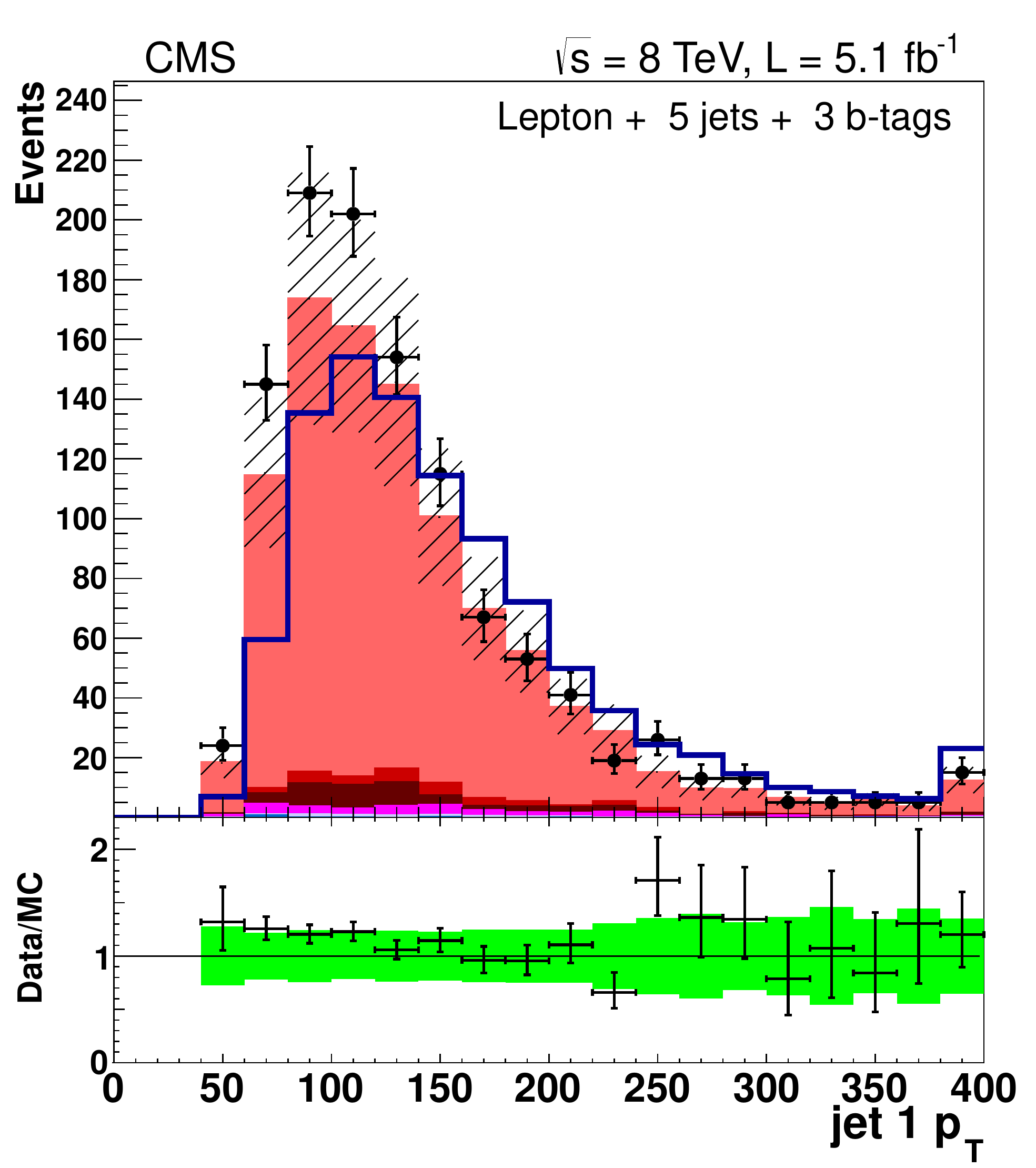}
   \includegraphics[width=0.31\textwidth]{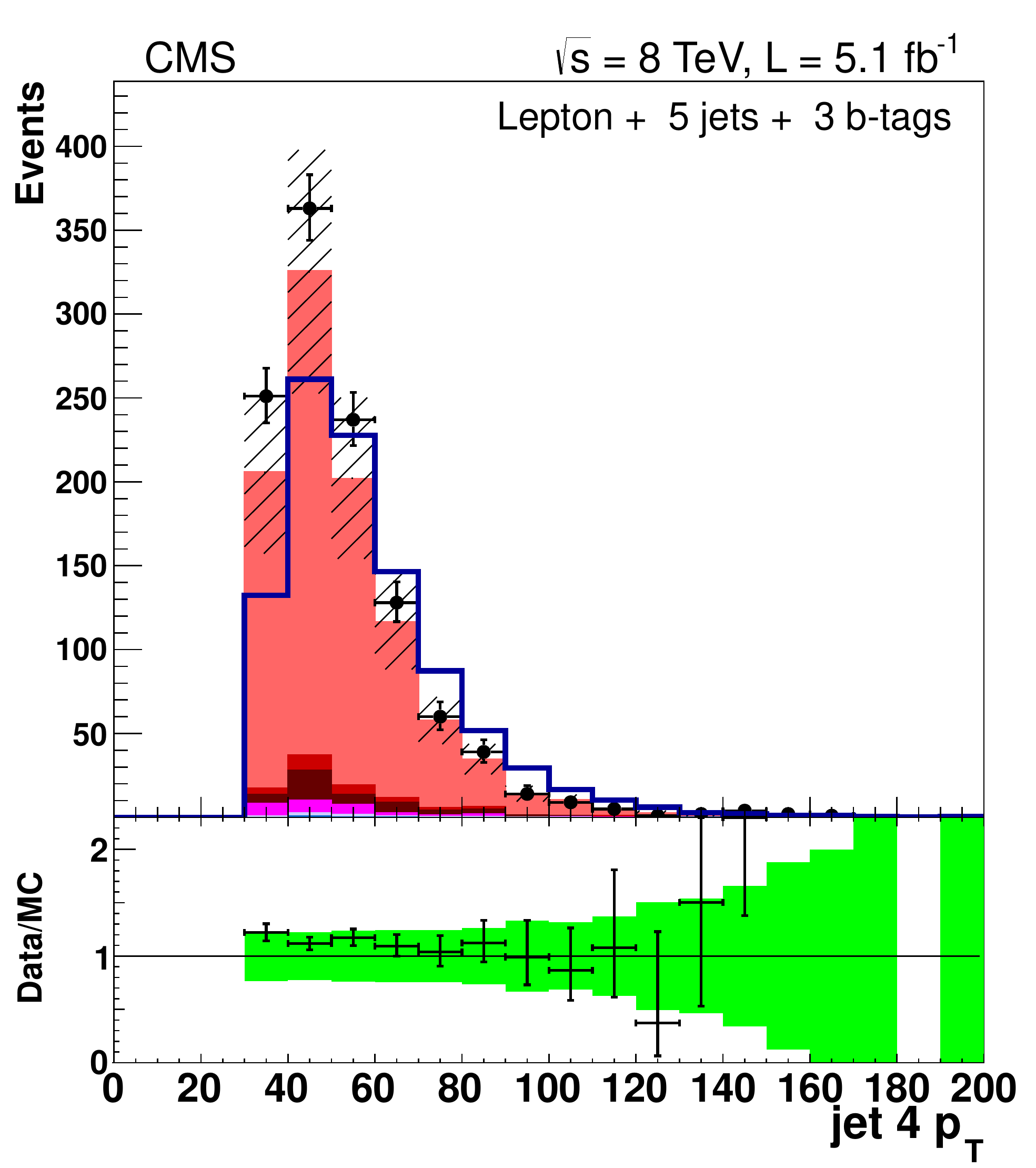}
   \includegraphics[width=0.31\textwidth]{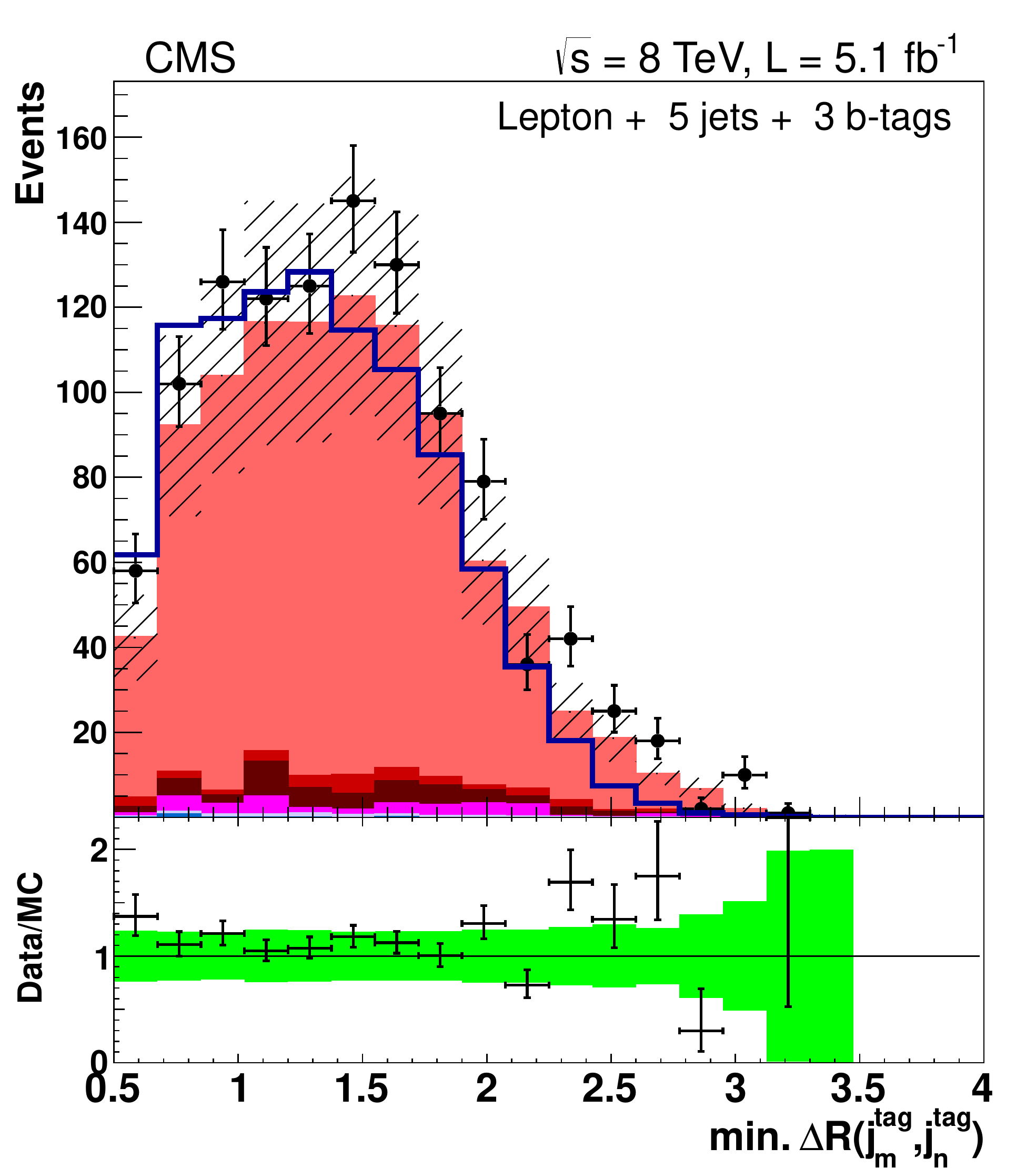}
   \raisebox{0.1\height}{\includegraphics[width=0.25\textwidth]{figures/samples_legend_tall_noTTHscale.pdf}}
   \hspace{0.055\textwidth}
   \caption{Distributions of the five ANN input variables with
     rankings 6 through 10, in terms of separation, for the 5~jets +
     3~b-tags category of the lepton+jets channel at $8
    \TeV$. Definitions of the variables are given in the text. The
     background is normalized to the SM expectation; the
     uncertainty band (shown as a hatched band in the stack plot and a
     green band in the ratio plot) includes statistical and systematic
     uncertainties that affect both the rate and shape of the
     background distributions.  The $\ttbar \PH$ signal ($m_{\PH} =
     125\GeVcc$) is normalized to $\sim$150 $\times$ SM expectation,
     equal to the total background yield, for easier comparison of the
     shapes.}  \label{fig:lj_input_5j_3t_lep_8TeV_part2} \end{center}
\end{figure}

\begin{figure}[hbtp]
 \begin{center}
 \includegraphics[width=0.31\textwidth]{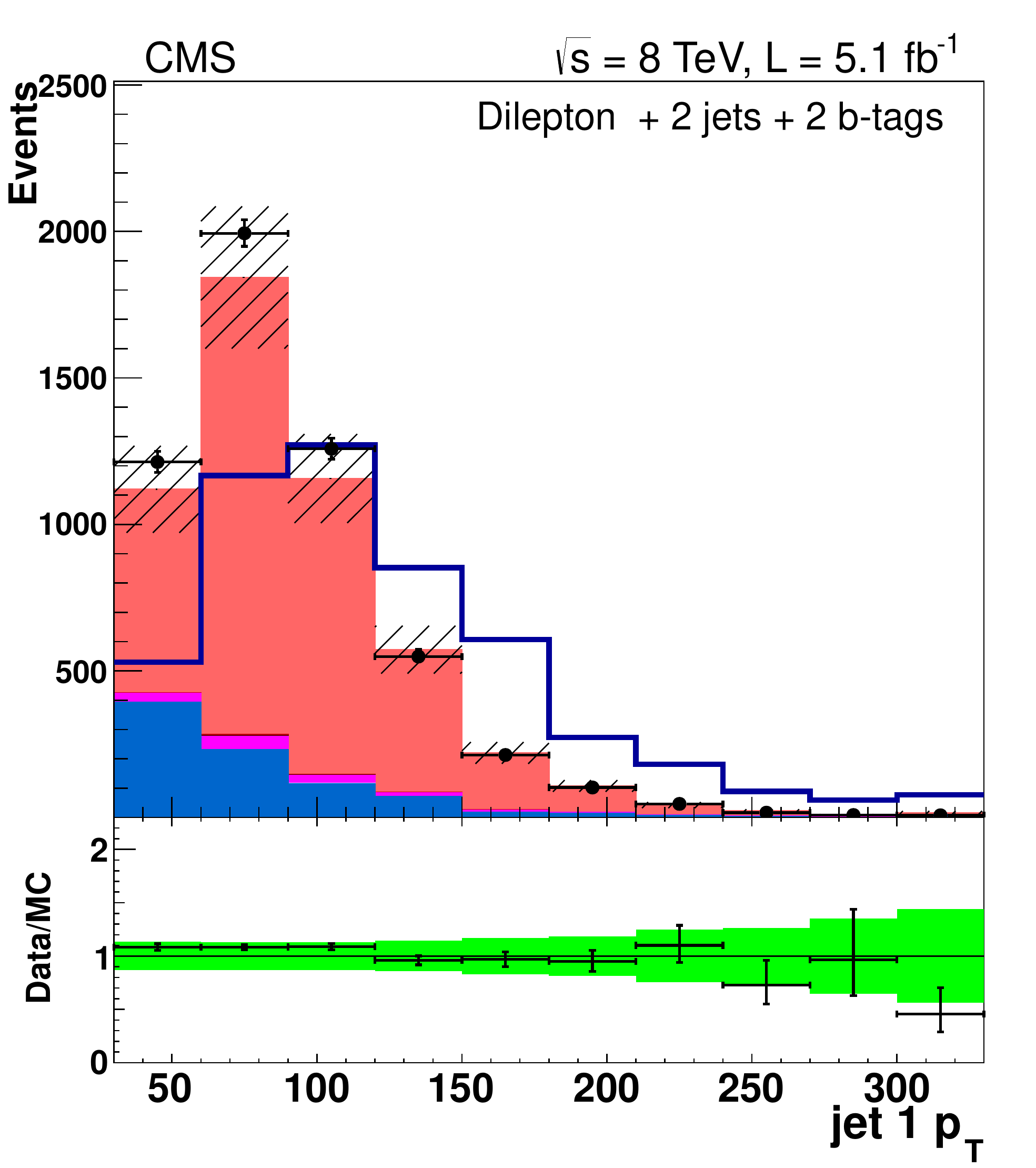}
 \includegraphics[width=0.31\textwidth]{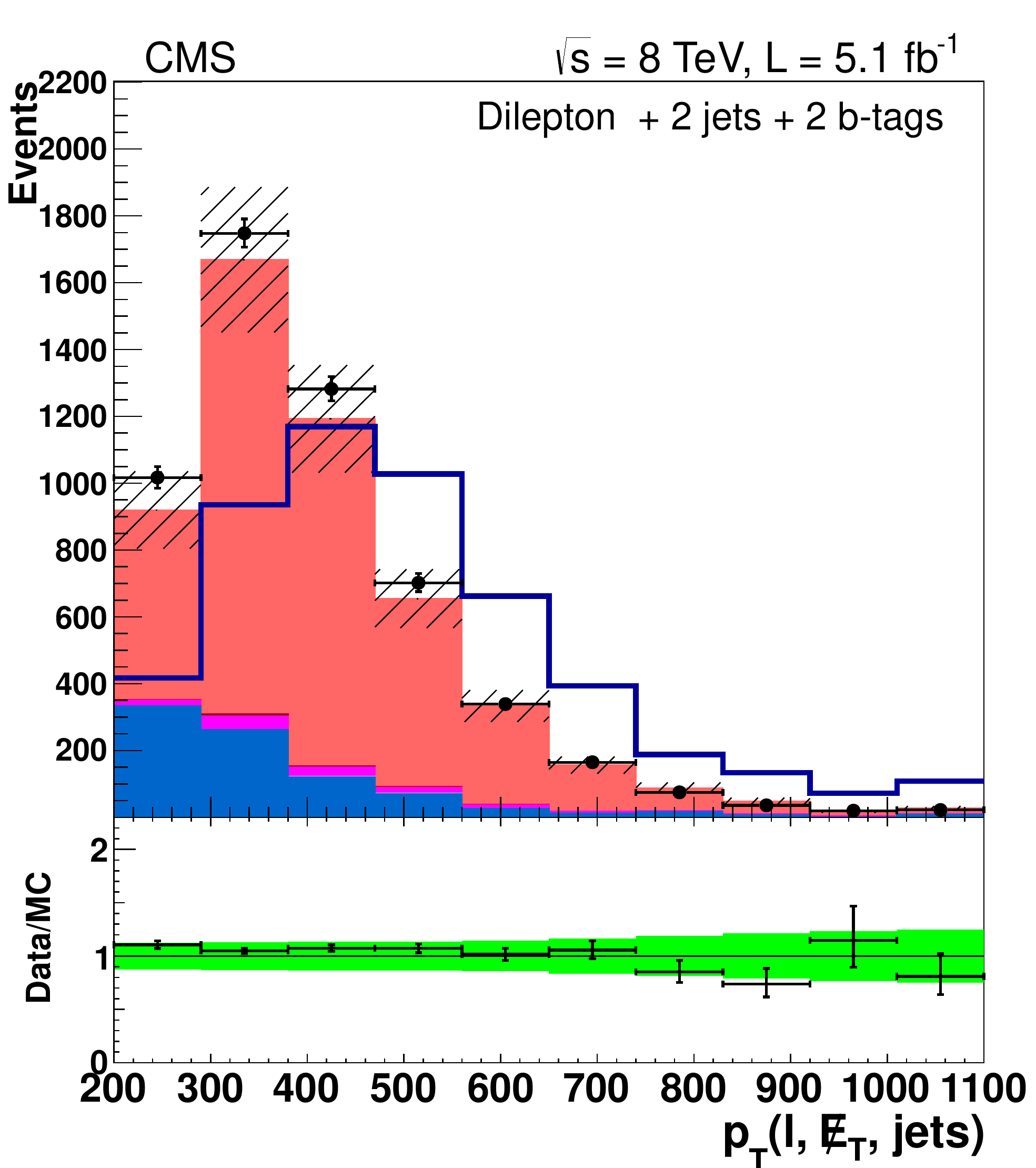}
 \includegraphics[width=0.31\textwidth]{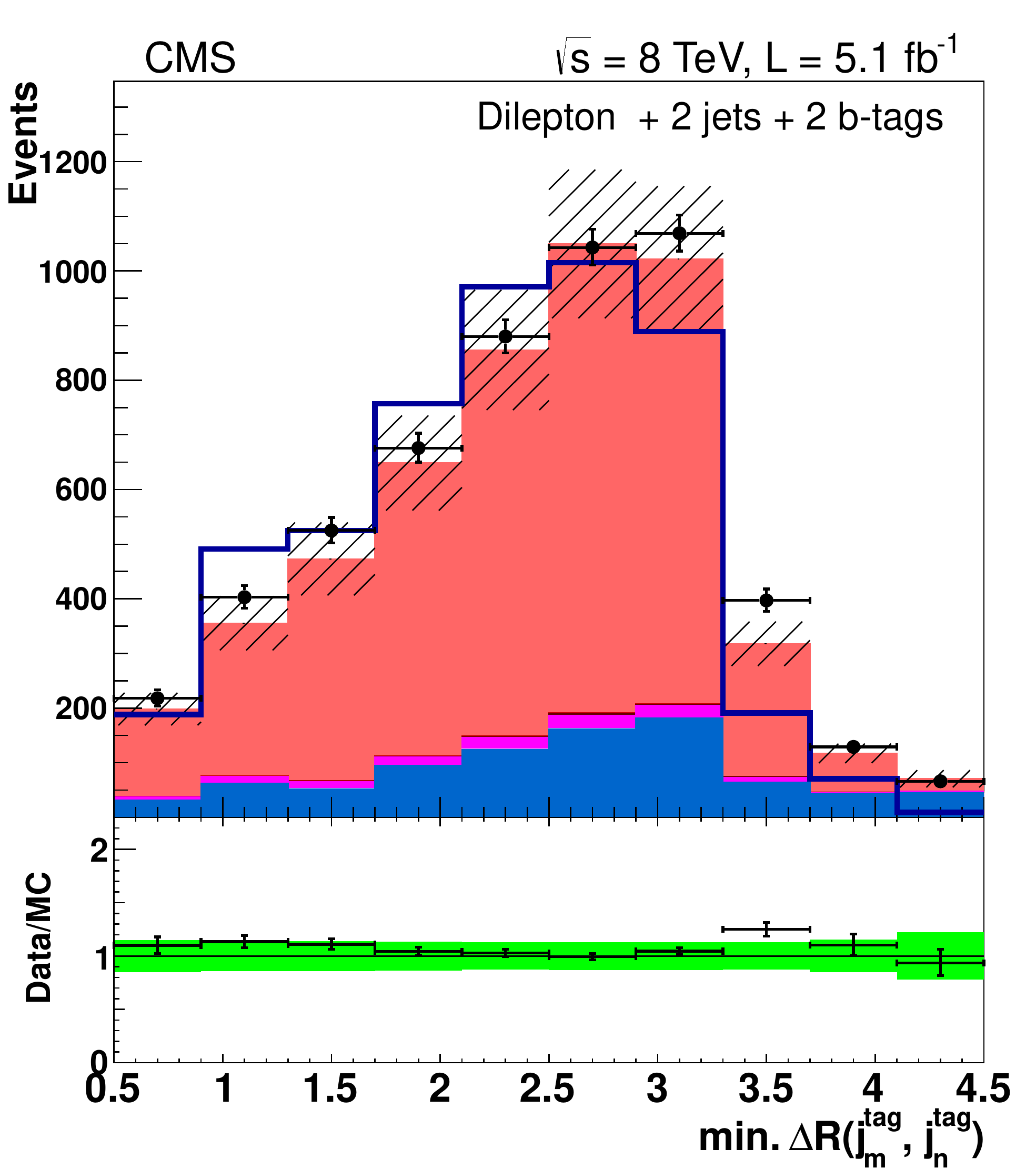}
 \vspace{0.2cm}

 \includegraphics[width=0.31\textwidth]{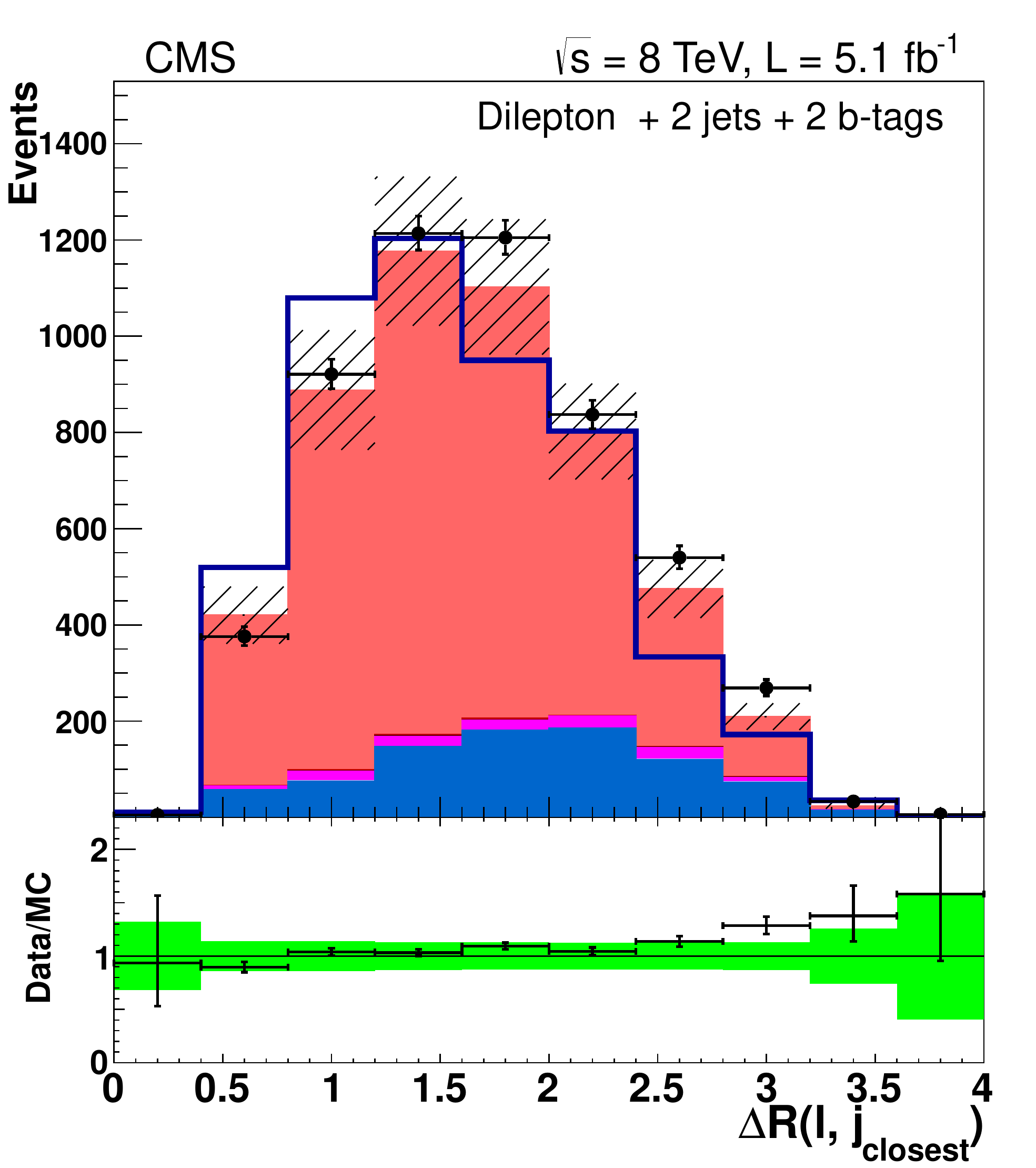}
 \includegraphics[width=0.31\textwidth]{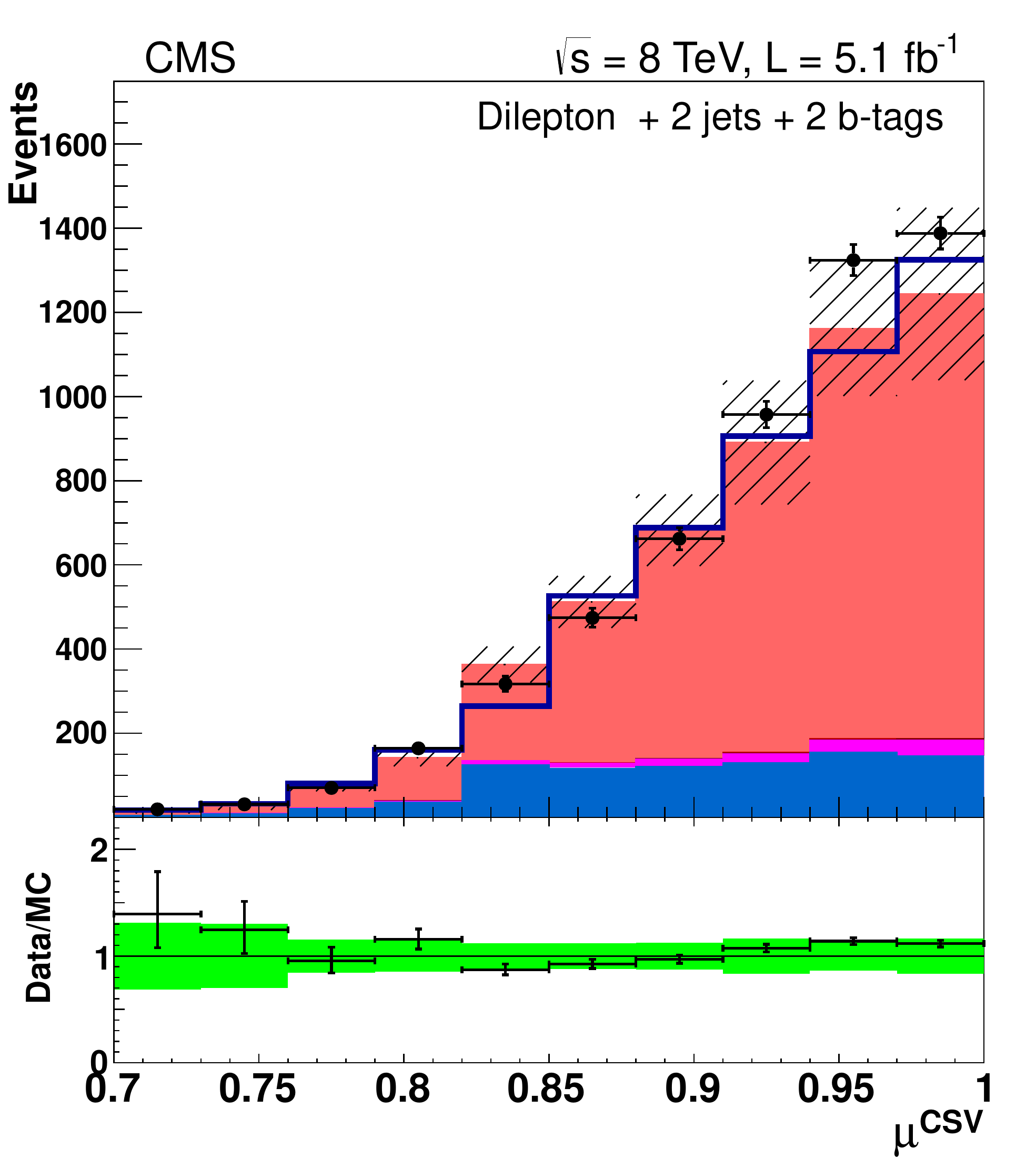}
    \raisebox{0.1\height}{\includegraphics[width=0.25\textwidth]{figures/samples_legend_tall_noTTHscale.pdf}}
 \hspace{0.055\textwidth}

 \caption{Distributions of ANN input variables for the 2~jets + 2~b-tags category of the dilepton channel at $8\TeV$. Definitions of
     the variables are given in the text. The
   background is normalized to the SM expectation; the uncertainty
   band (shown as a hatched band in the stack plot and a green band
   in the ratio plot) includes statistical and systematic
   uncertainties that affect both the rate and shape of the background
   distributions.  The $\ttbar \PH$ signal ($m_{\PH} =125\GeVcc$) is
   normalized to $\sim$7000 $\times$ SM expectation, equal to the
   total background yield, for easier comparison of the shapes.}
 \label{fig:dilep_input_5j_3t_lep_8TeV} \end{center}
\end{figure}

\begin{figure}[hbtp]
 \begin{center}
   \includegraphics[width=0.70\textwidth]{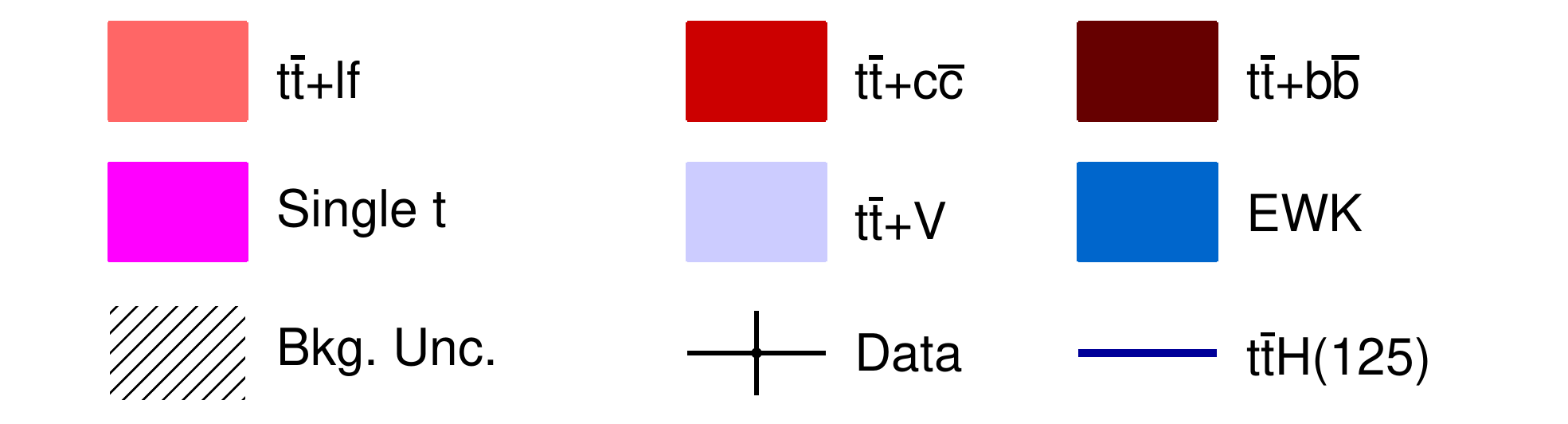}
   \includegraphics[width=0.31\textwidth]{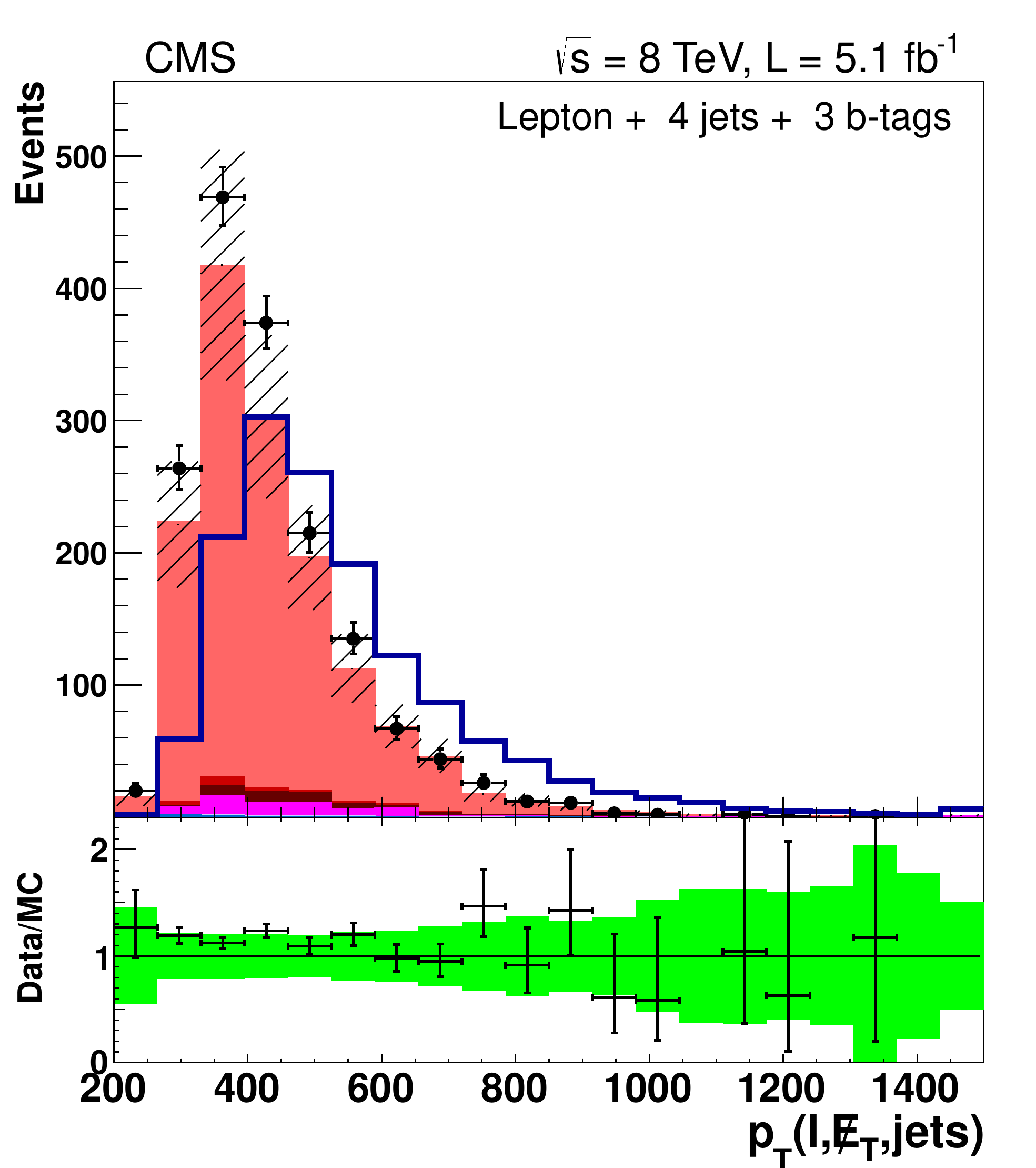}
   \includegraphics[width=0.31\textwidth]{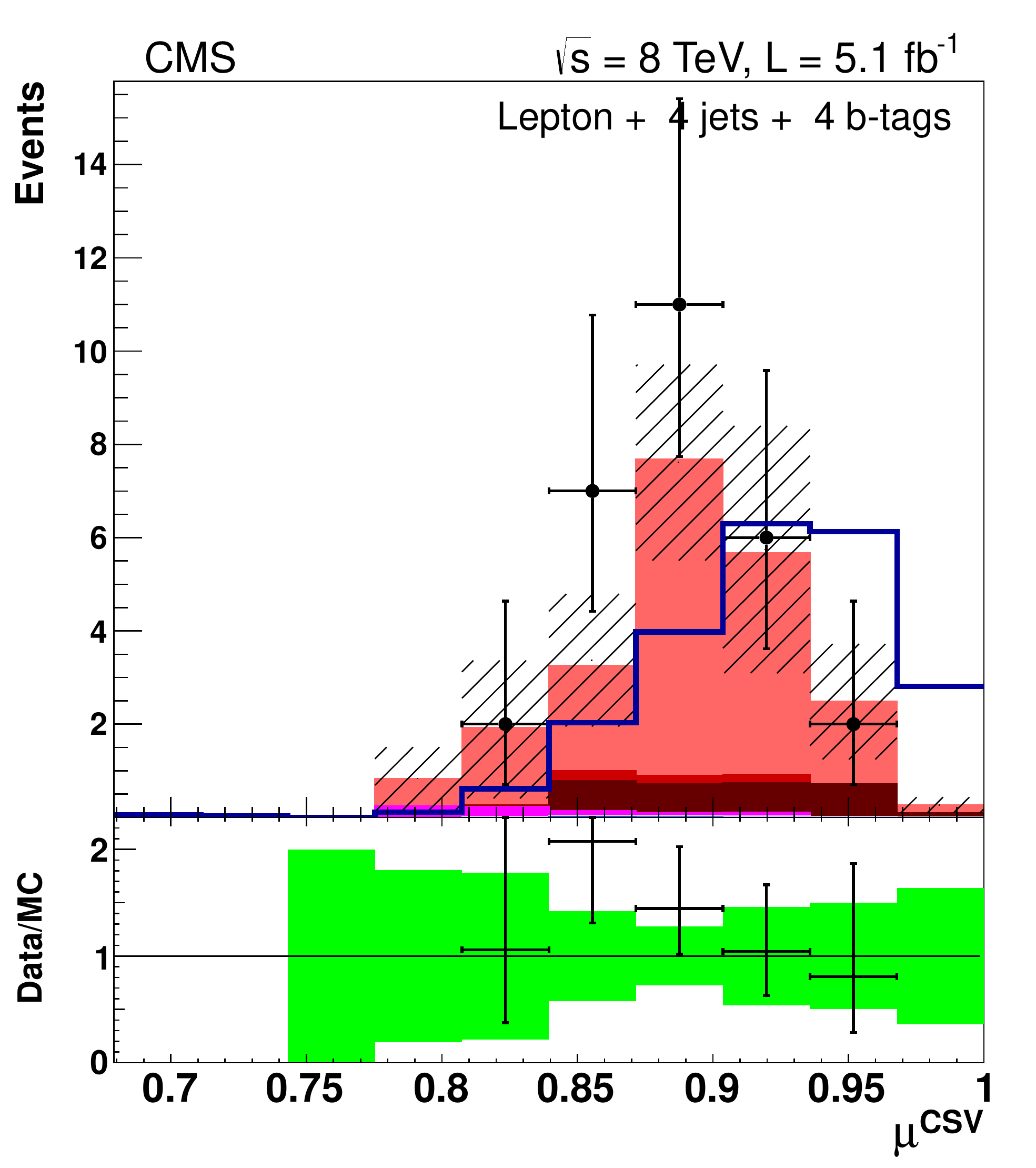}
   \includegraphics[width=0.31\textwidth]{figures/d2MCPlots_avg_btag_disc_btags_cut5_j5_t3_Combined_HtWgt.pdf}
   \includegraphics[width=0.31\textwidth]{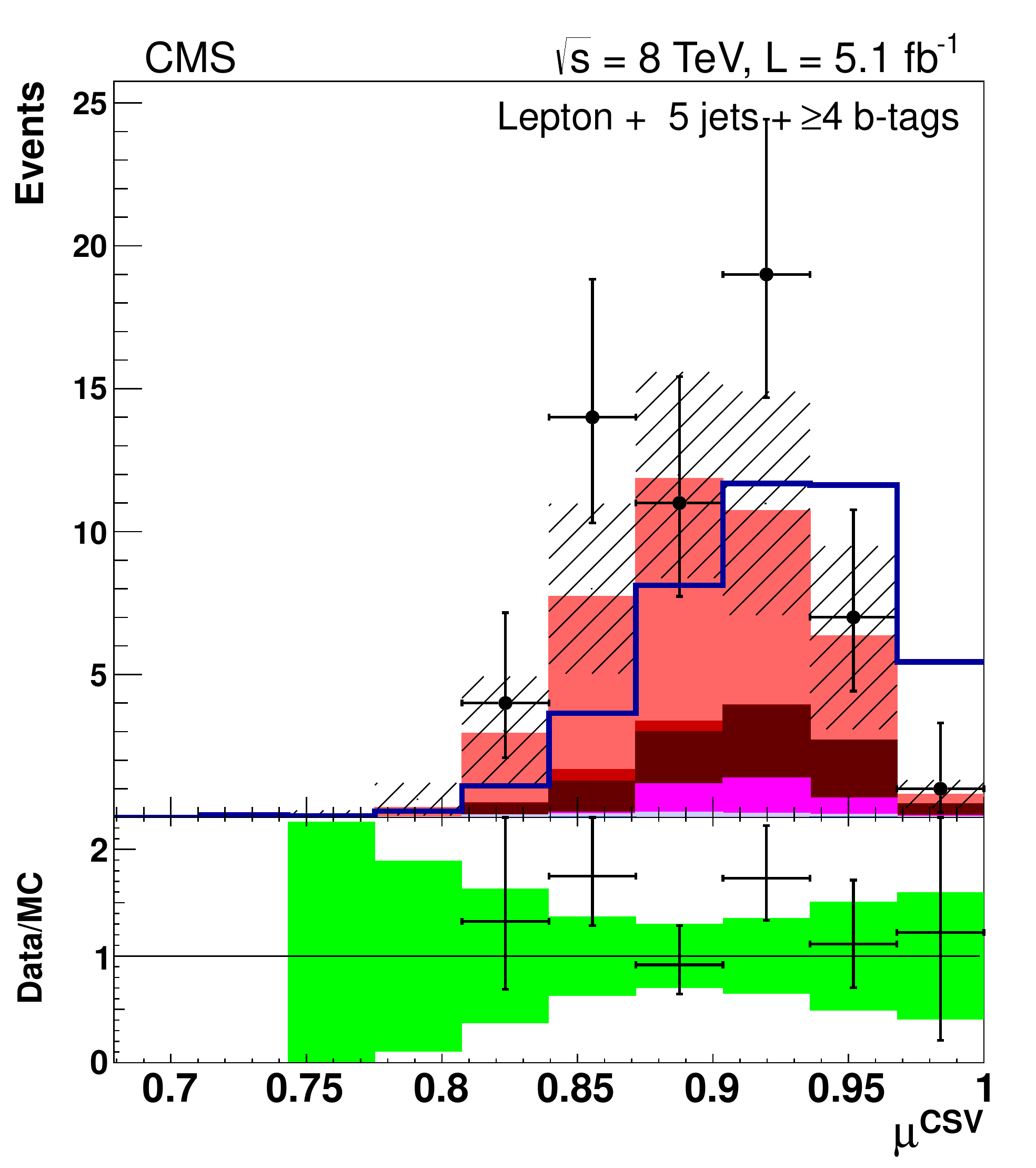}
   \includegraphics[width=0.31\textwidth]{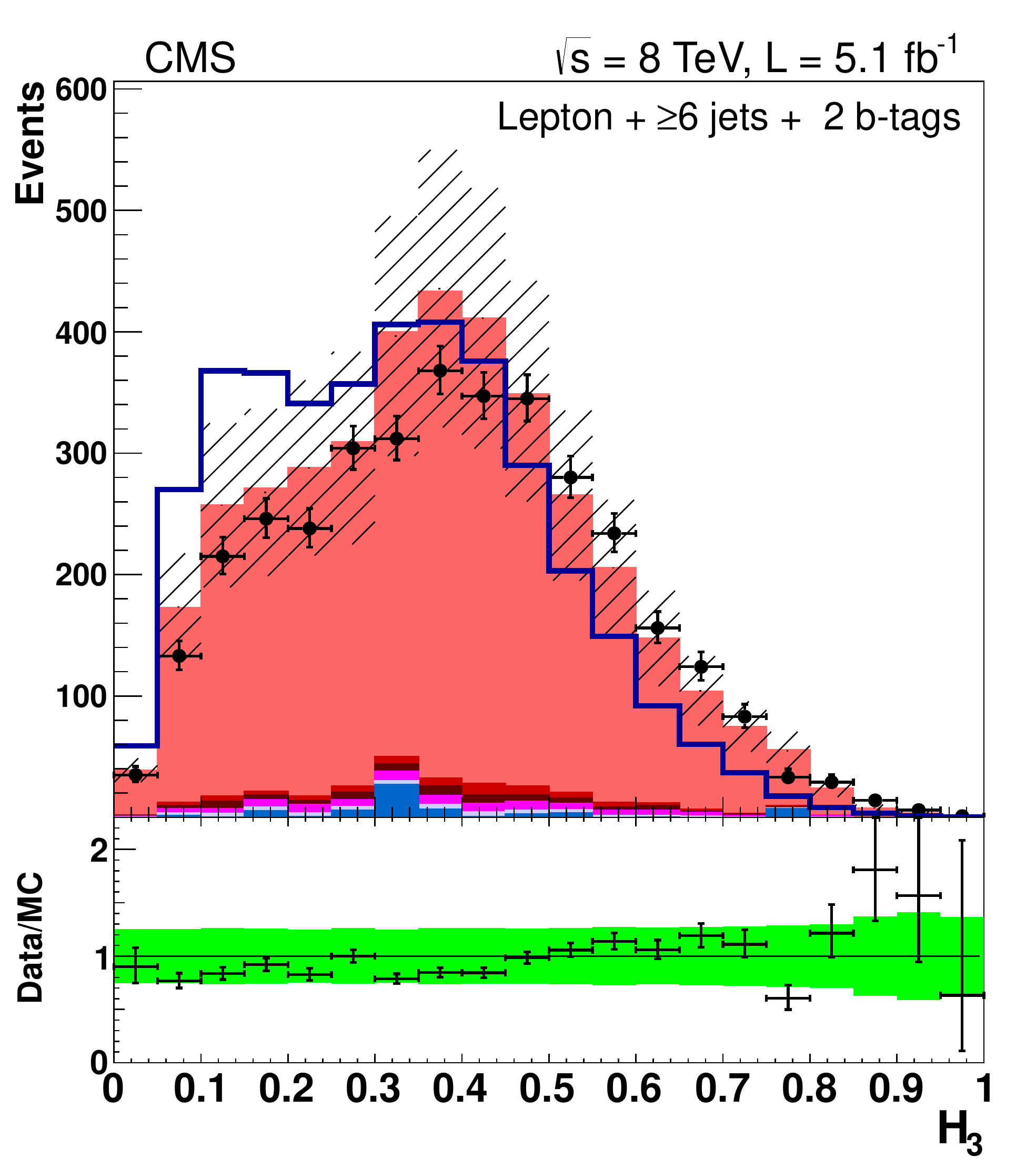}
   \includegraphics[width=0.31\textwidth]{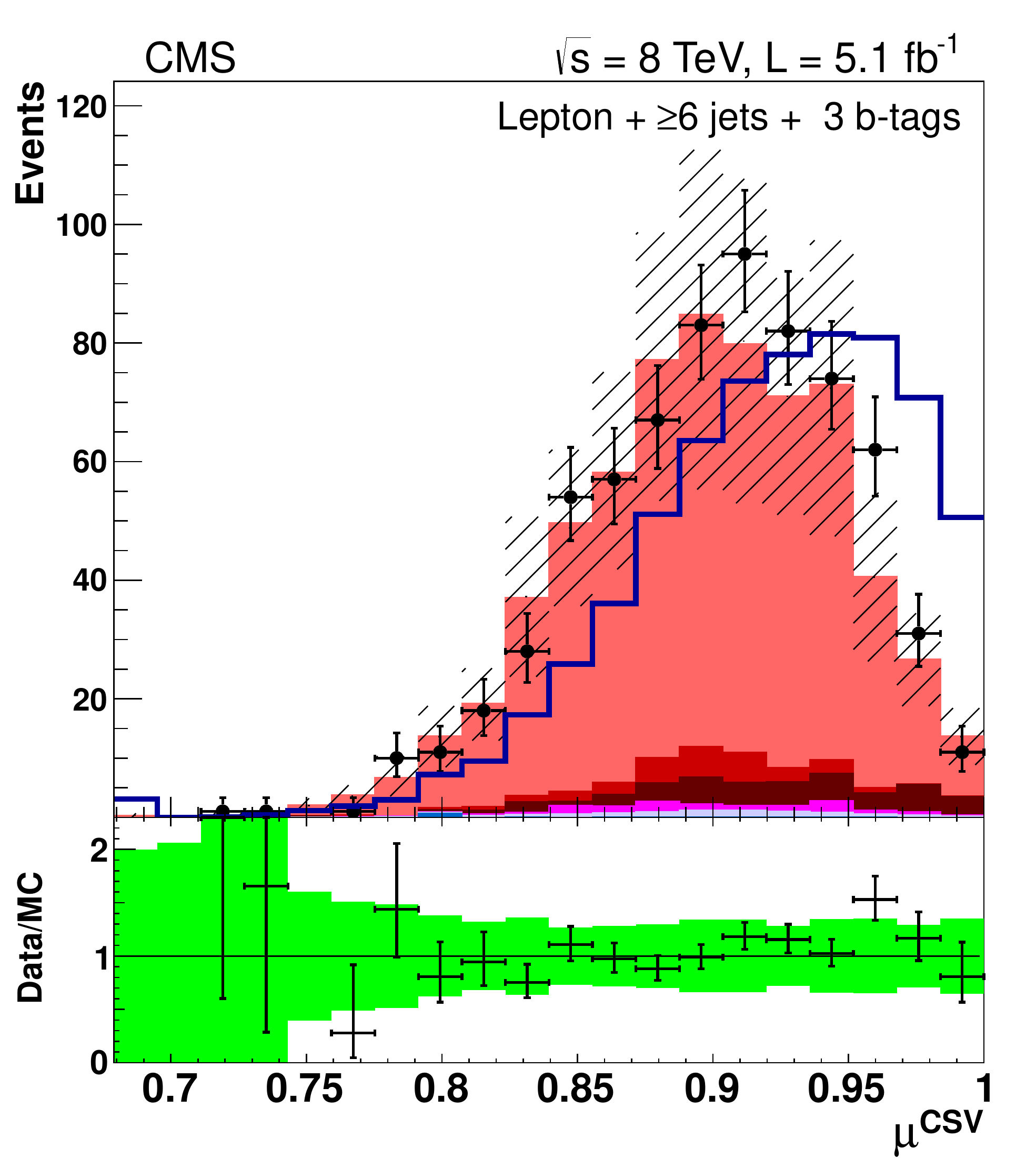}
   \includegraphics[width=0.31\textwidth]{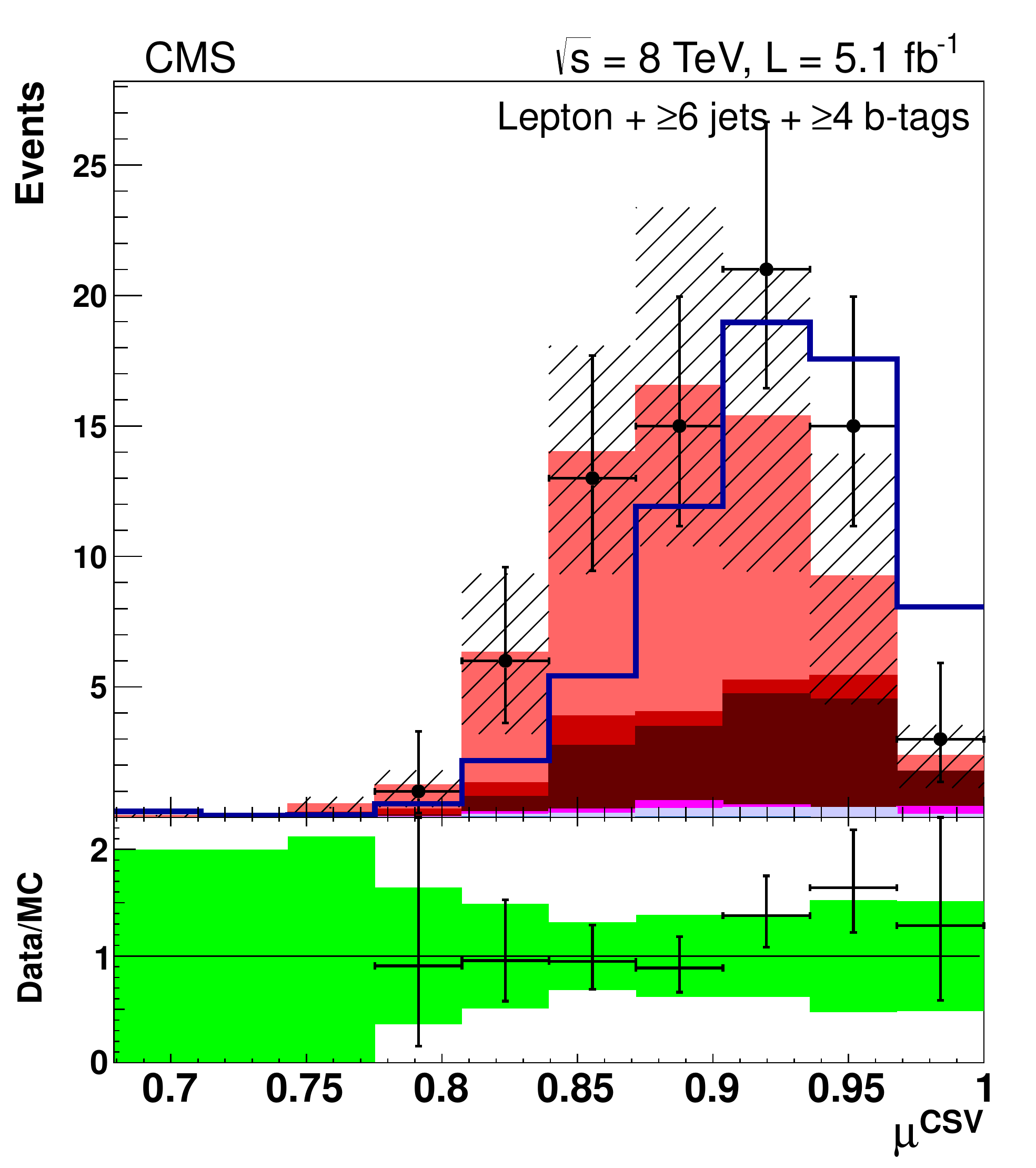}
   \includegraphics[width=0.31\textwidth]{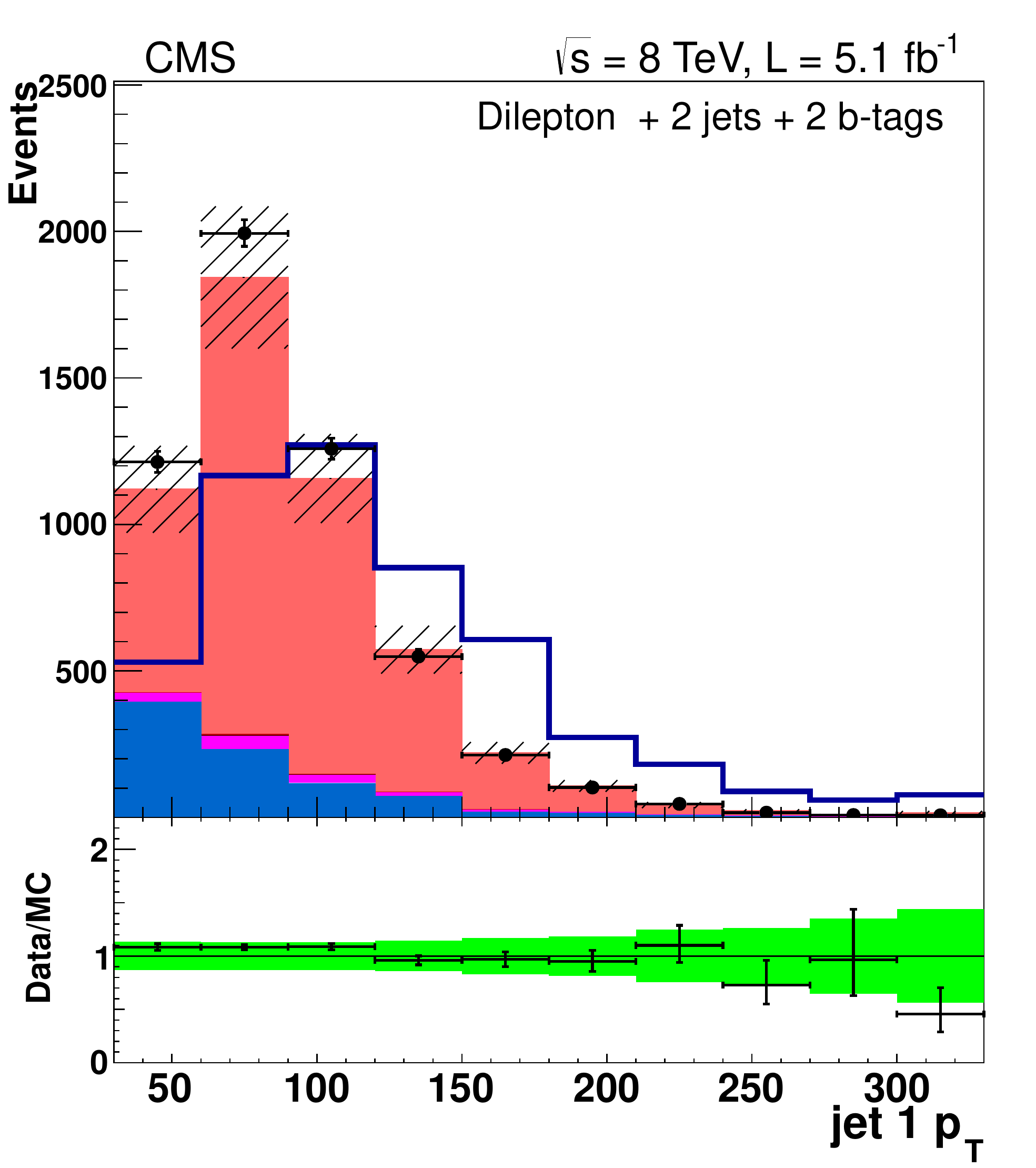}
   \includegraphics[width=0.31\textwidth]{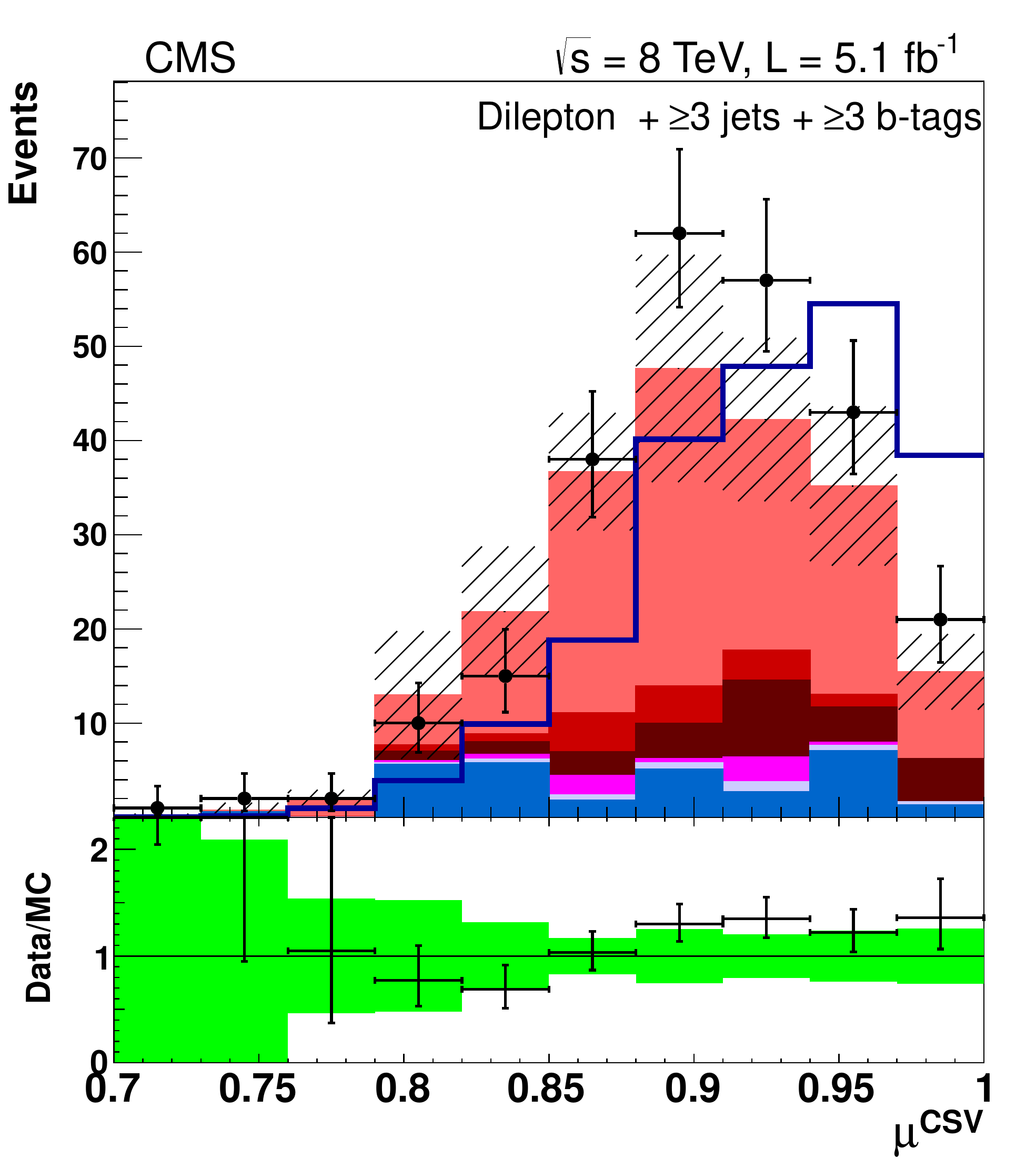}

   \caption{Input variables that give the best signal-background
     separation power for each of the lepton + jets and dilepton
     jet, b-tag categories used in the analysis at $8\TeV$. Definitions of
     the variables are given in the text. The
     background is normalized to the SM expectation; the
     uncertainty band (shown as a hatched band in the stack plot and
     a green band in the ratio plot) includes statistical and
     systematic uncertainties that affect both the rate and shape of
     the background distributions.  The $\ttbar \PH$ signal
     ($m_{\PH} =125\GeVcc$) is normalized to $\sim$25--7000 $\times$
     SM expectation, equal to the total background yield for that
     category, for easier comparison of the shapes.}
   \label{fig:bestVars_8TeV}
\end{center}
\end{figure}

Figures~\ref{fig:lj_input_5j_3t_lep_8TeV_part1} through
\ref{fig:dilep_input_5j_3t_lep_8TeV} show the variables used in the
ANN for the 5~jets + 3~b-tags category (lepton+jets channel) and the
2~jets + 2~b-tags (dilepton channel). The 5~jets + 3~b-tags category
is chosen for lepton+jets as a compromise between signal sensitivity
and adequate statistics for display purposes.  Also shown, in
Figure~\ref{fig:bestVars_8TeV}, are data-to-simulation comparisons of
the best input variables for each jet-tag category considered in the
$8\TeV$ analysis. The data-to-simulation ratio plots in
Figures~\ref{fig:lj_input_5j_3t_lep_8TeV_part1} through
\ref{fig:bestVars_8TeV} show that, within uncertainties, the
simulation reproduces well the shape and normalization of the
distributions of the variables used in the ANN before the final
maximum likelihood fit is performed (as discussed in
Section~\ref{sec:results}).  Correlations between input variables are
also well reproduced by simulation.

For ANN training, we use $\ttbar \PH$ ($m_{\PH} = 120\GeVcc$) as the
signal and $\ttbar$+jets as the background, such that there is an
equal amount of both for each category. The mass $m_{\PH} = 120
\GeVcc$ sample was chosen in the analysis of the $7\TeV$ data before
the observation of a Higgs-like particle at $m_{\PH} = 125\GeVcc$ was
announced. This mass point was preserved in the $8\TeV$ ANN training
for consistency. The signal and background events used to train an ANN
are split in half: one half is used to do the training itself, while
the other is used as an independent test sample to monitor performance
during training. The ANN method used is the ``multilayer perceptron'',
available as part of the \textsc{tmva}~\cite{Hocker:2007ht} package in
\textsc{root}~\cite{Brun:1997pa}.  A multilayer perceptron is a specific
kind of neural network in which the neurons in each layer only have
connections to neurons in the following layer.  The network
architecture used here consists of two hidden layers, with $N$ neurons
in the first layer and $N-1$ neurons in the second layer, where $N$ is
the number of input variables.  Standard tests were completed during
ANN training to look for evidence of overtraining; no such evidence
was found in any jet-tag category, providing confidence that our
training statistics were satisfactory given the number of input
variables used in each.

\begin{figure}[hbtp]
 \begin{center}
   \hspace{0.32\textwidth}
   \includegraphics[width=0.31\textwidth]{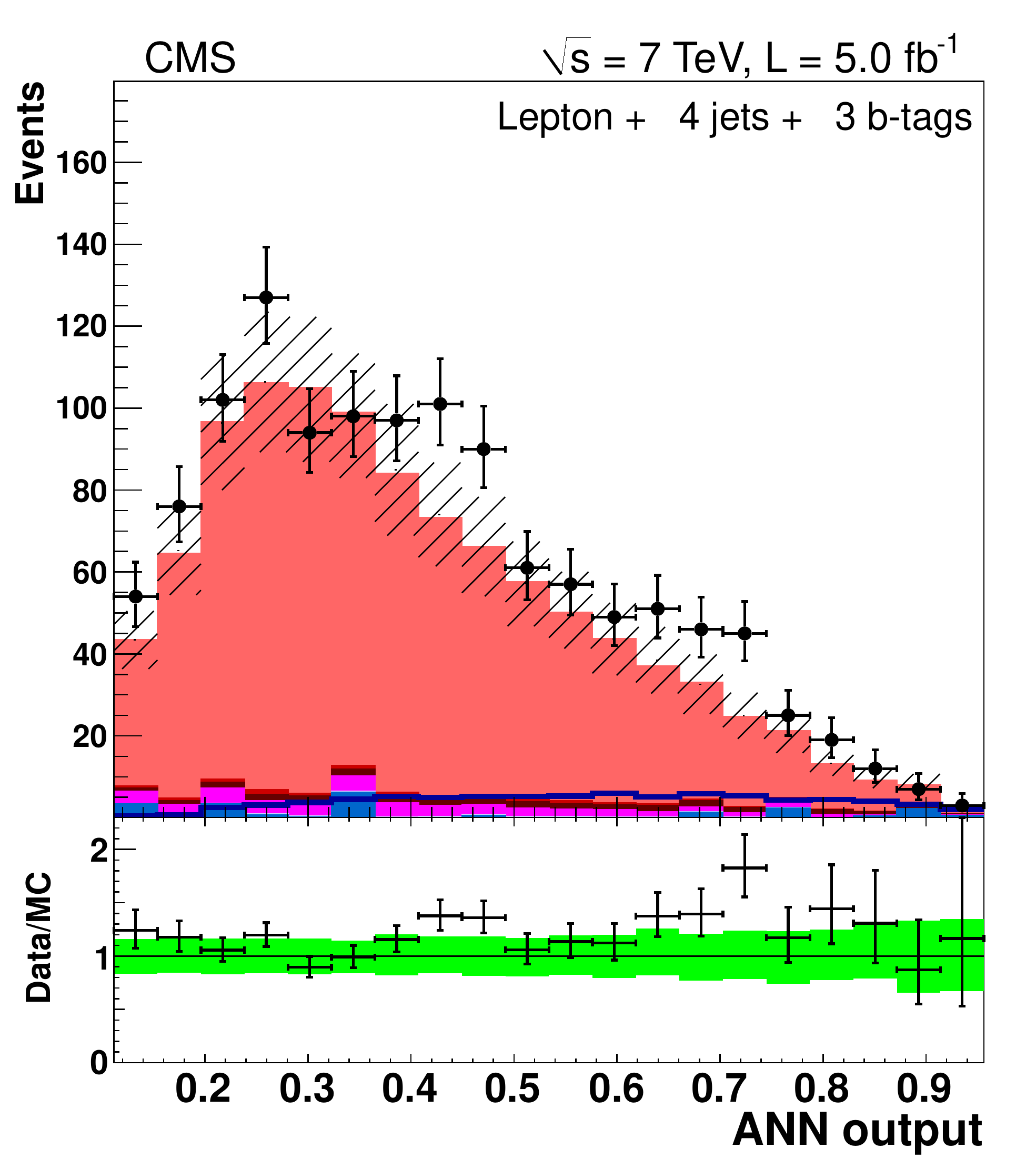}
   \includegraphics[width=0.31\textwidth]{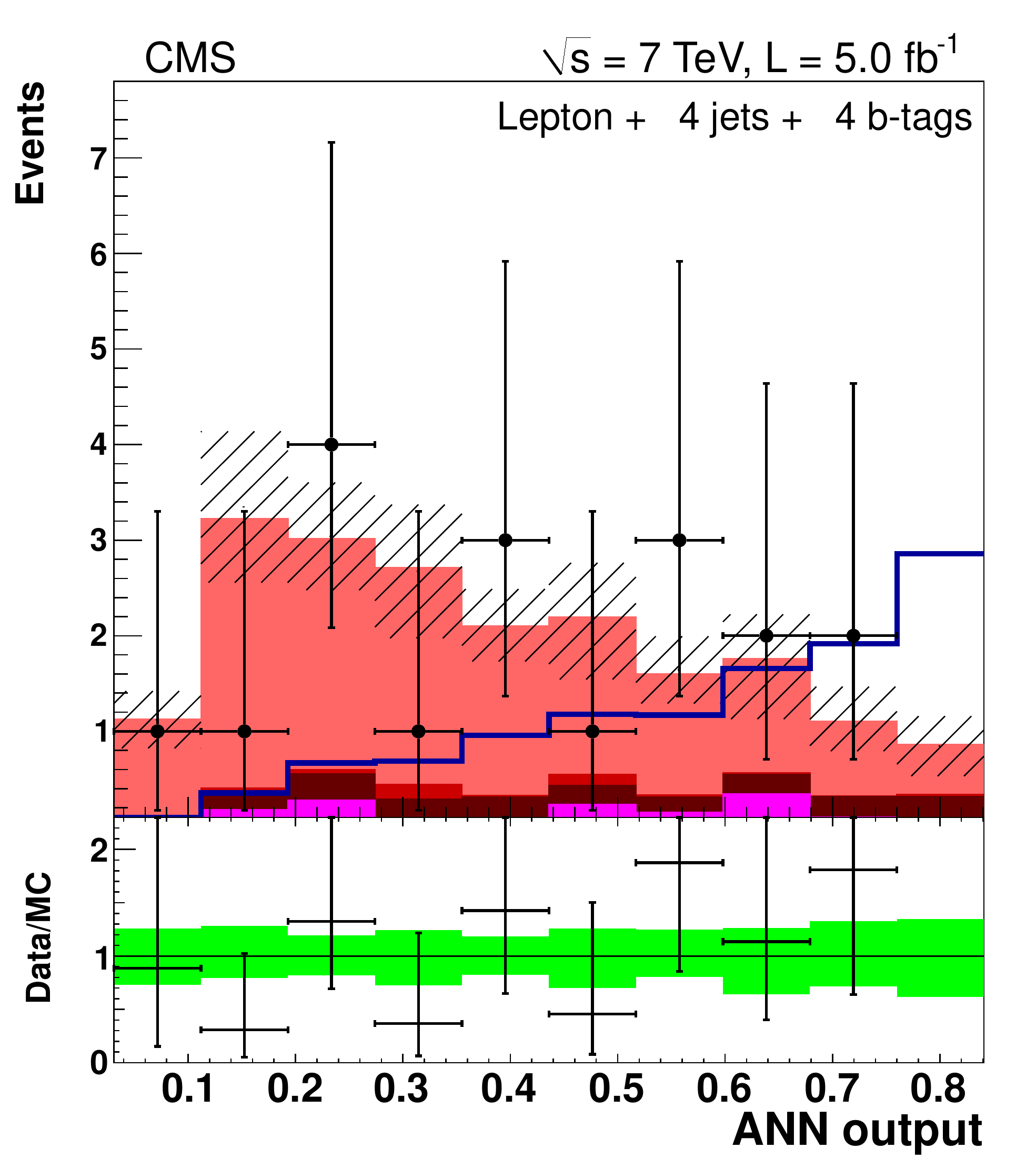}
   \vspace{0.2cm}

   \raisebox{0.1\height}{\includegraphics[width=0.25\textwidth]{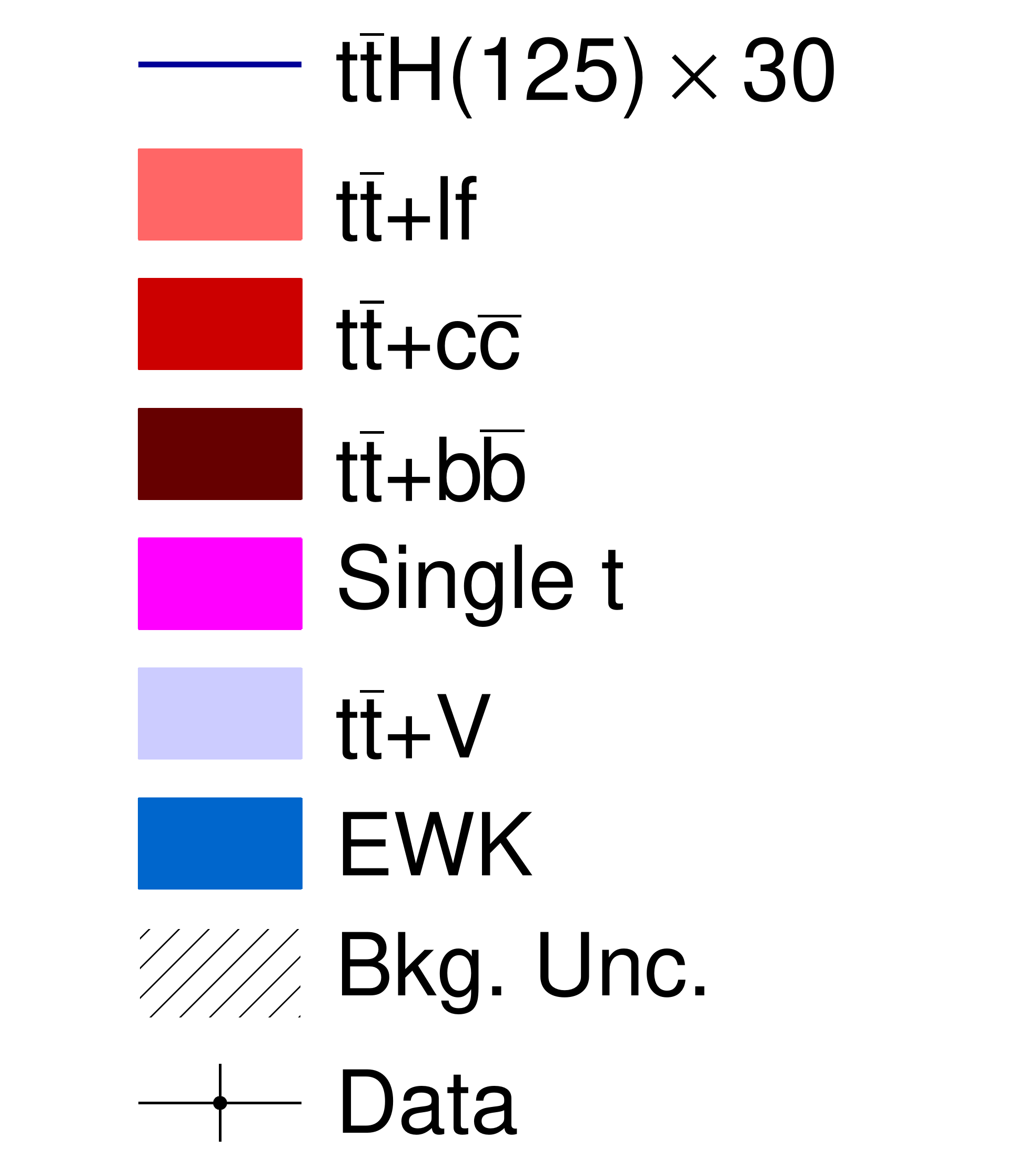}}
   \hspace{0.055\textwidth}
   \includegraphics[width=0.31\textwidth]{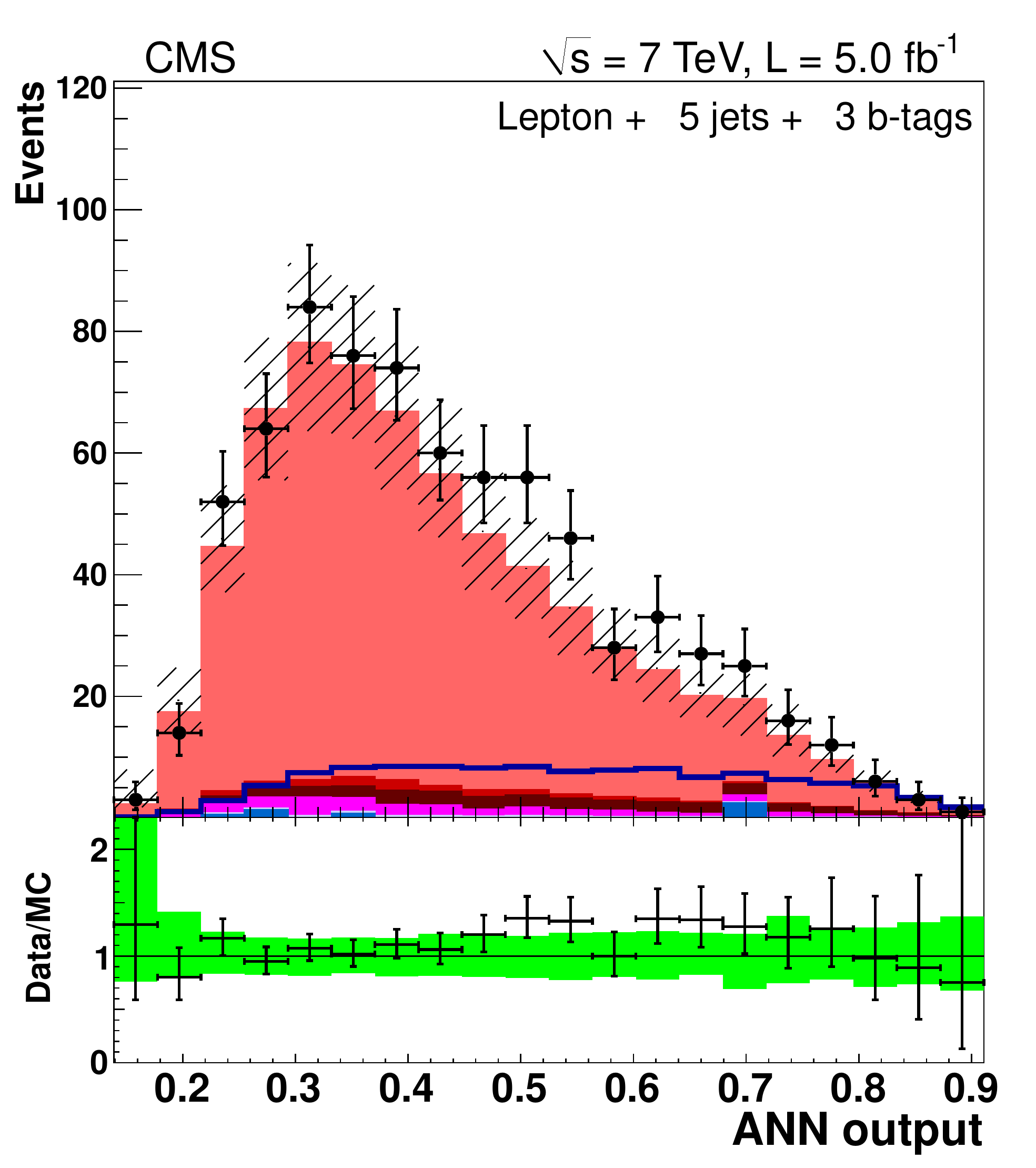}
   \includegraphics[width=0.31\textwidth]{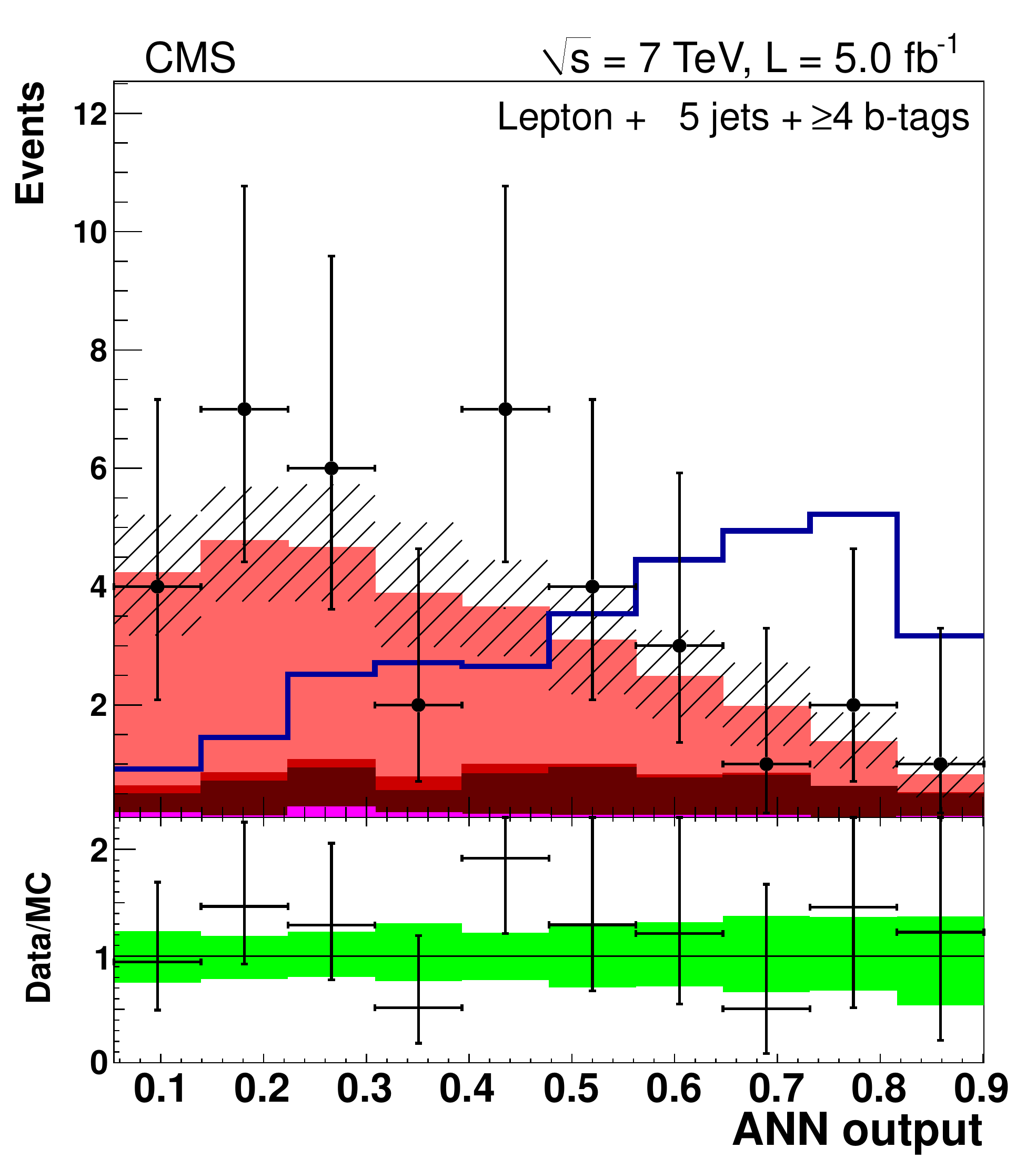}
   \vspace{0.2cm}

   \includegraphics[width=0.31\textwidth]{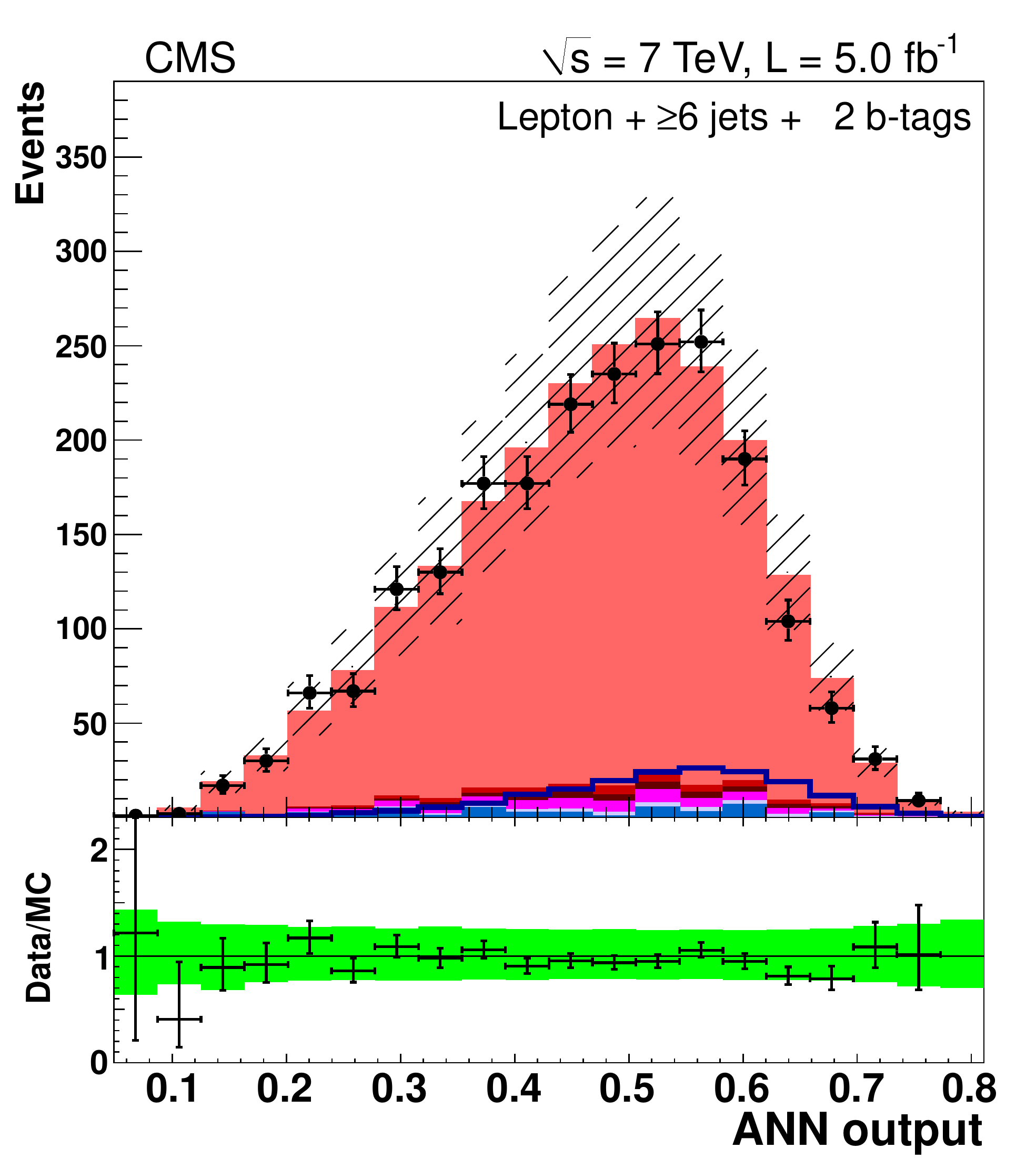}
   \includegraphics[width=0.31\textwidth]{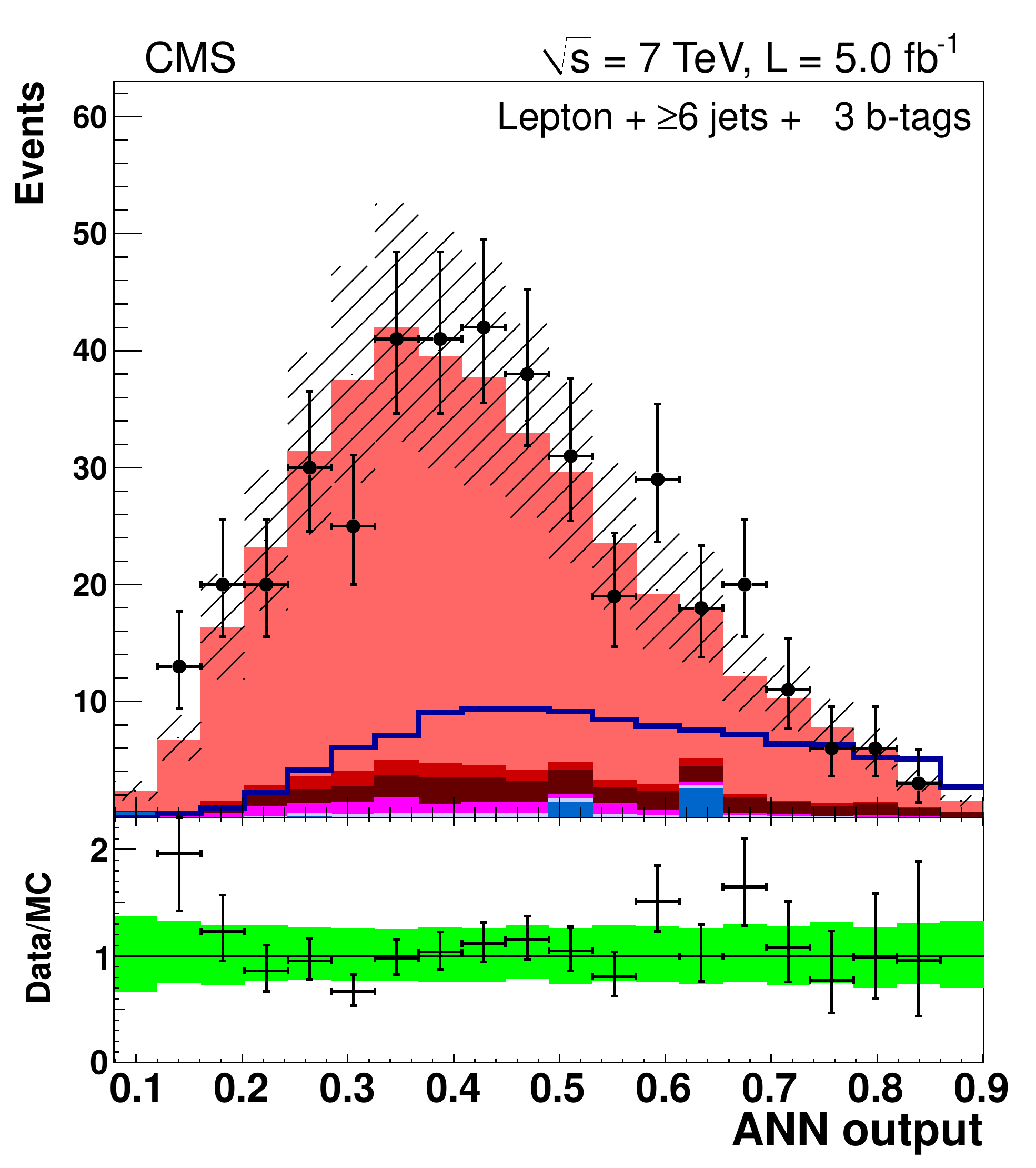}
   \includegraphics[width=0.31\textwidth]{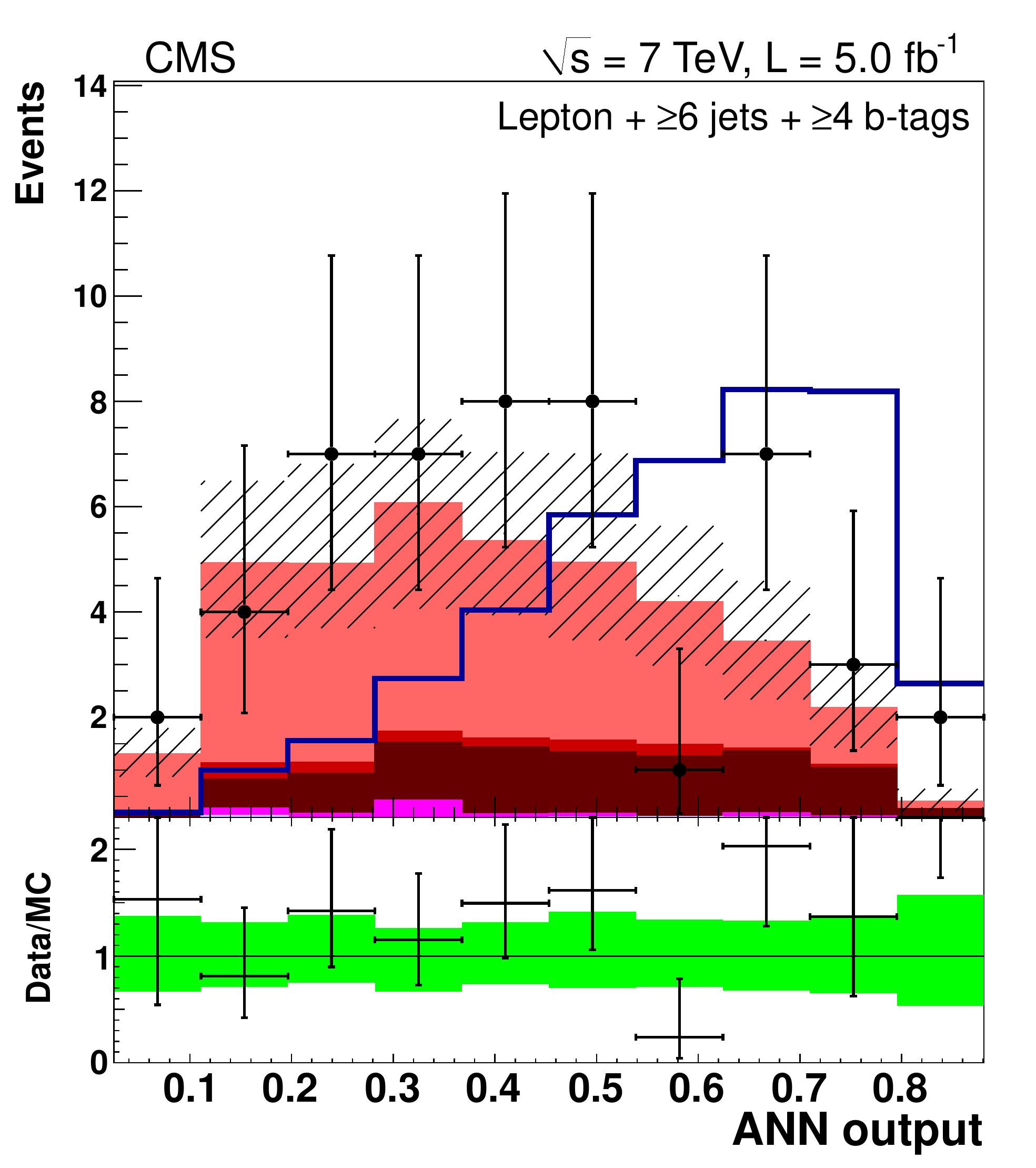}
   \caption{The distributions of the ANN output for  lepton+jets
     events at $7\TeV$ in the various analysis categories.
     The top, middle, and
     bottom rows are events with 4, 5, and $\ge$6 jets, respectively,
     while the left, middle, and right-hand columns are events with 2,
     3, and $\ge$4 b-tags, respectively. Background-like events have a
     low ANN output value. Signal-like events have a high
     ANN output value.  The background is normalized to the SM expectation;
     the uncertainty band (shown as a
     hatched band in the stack plot and a green band in the ratio
     plot) includes statistical and systematic uncertainties that
     affect both the rate and shape of the background distributions.
     The $\ttbar \PH$ signal ($m_{\PH} = 125\GeVcc$) is normalized to
     30 $\times$ SM expectation.}  \label{fig:lj_ANNoutput_7TeV}
\end{center}
\end{figure}

\begin{figure}[hbtp]
 \begin{center}
   \hspace{0.32\textwidth}
   \includegraphics[width=0.31\textwidth]{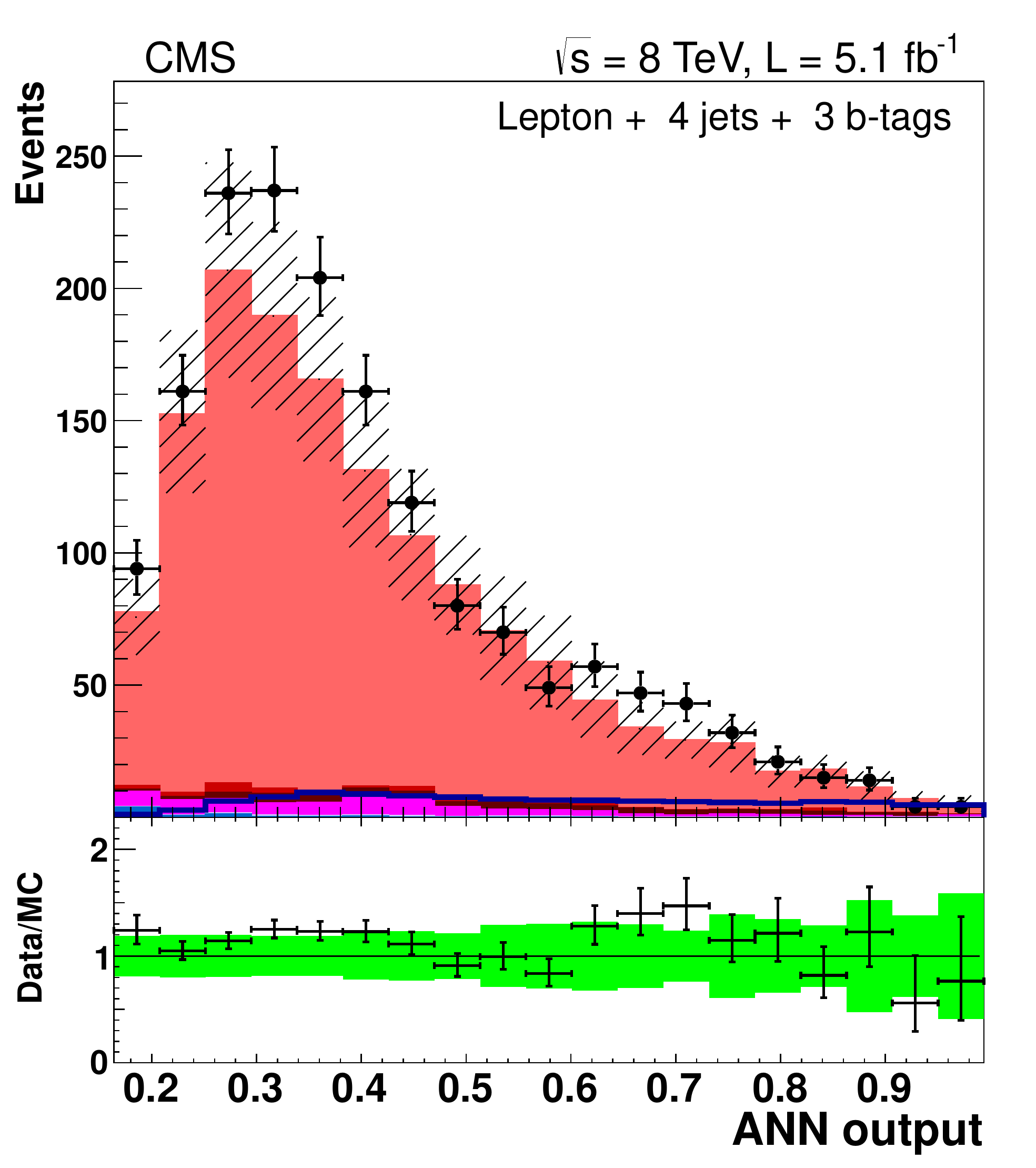}
   \includegraphics[width=0.31\textwidth]{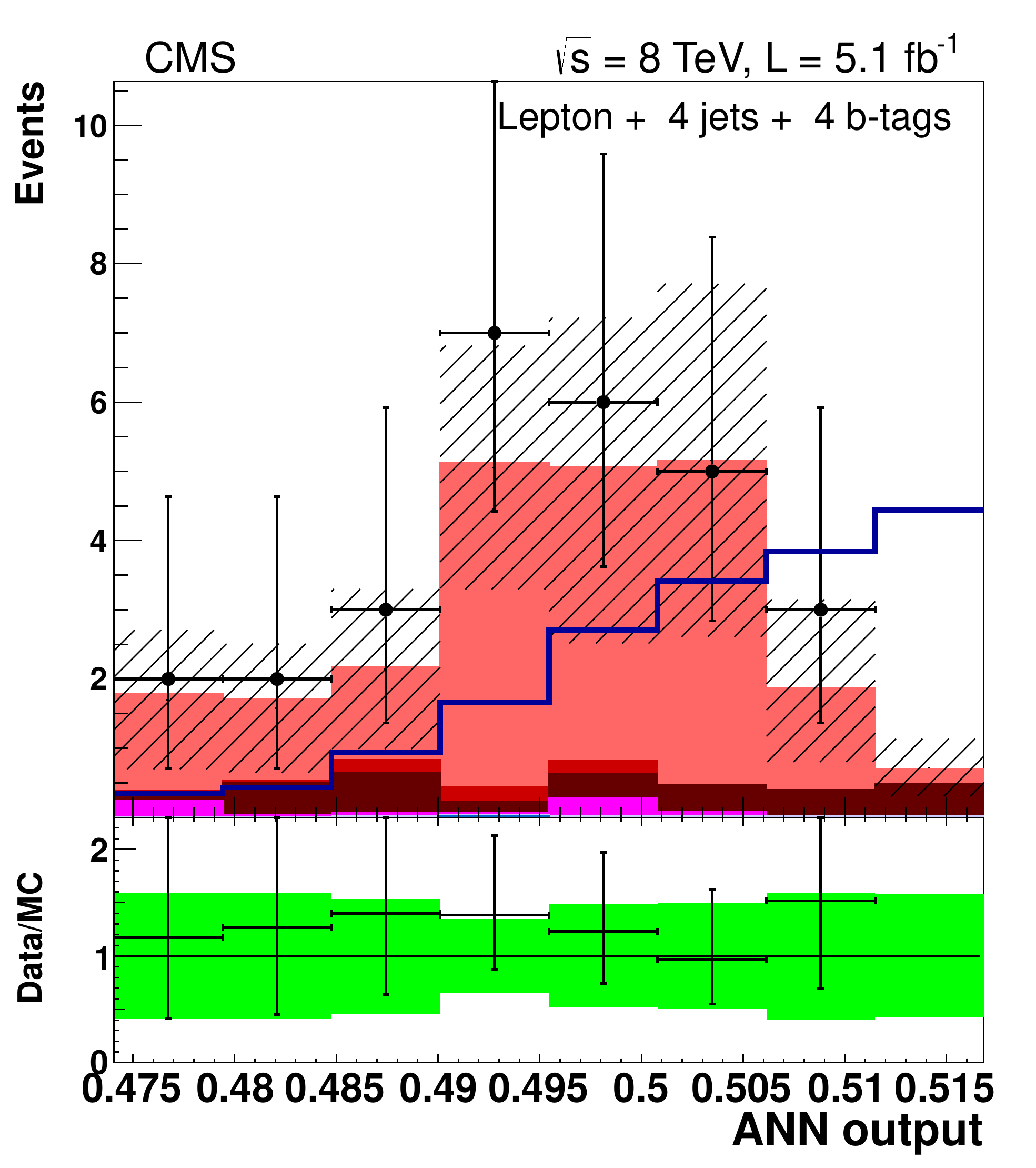} \hfill
   \vspace{0.2cm}

   \raisebox{0.1\height}{\includegraphics[width=0.25\textwidth]{figures/samples_legend_tall.pdf}}
   \hspace{0.055\textwidth}
   \includegraphics[width=0.31\textwidth]{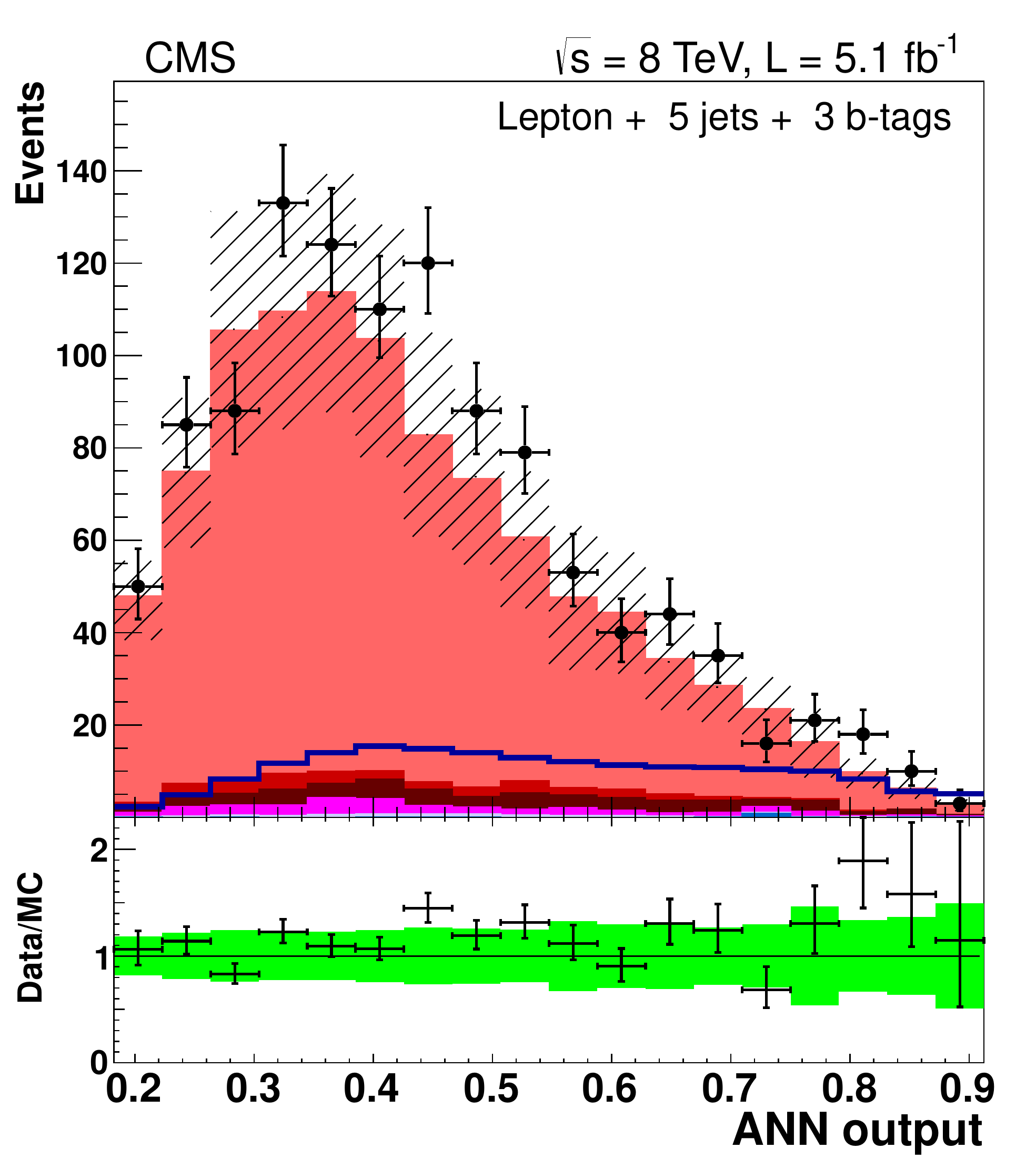}
   \includegraphics[width=0.31\textwidth]{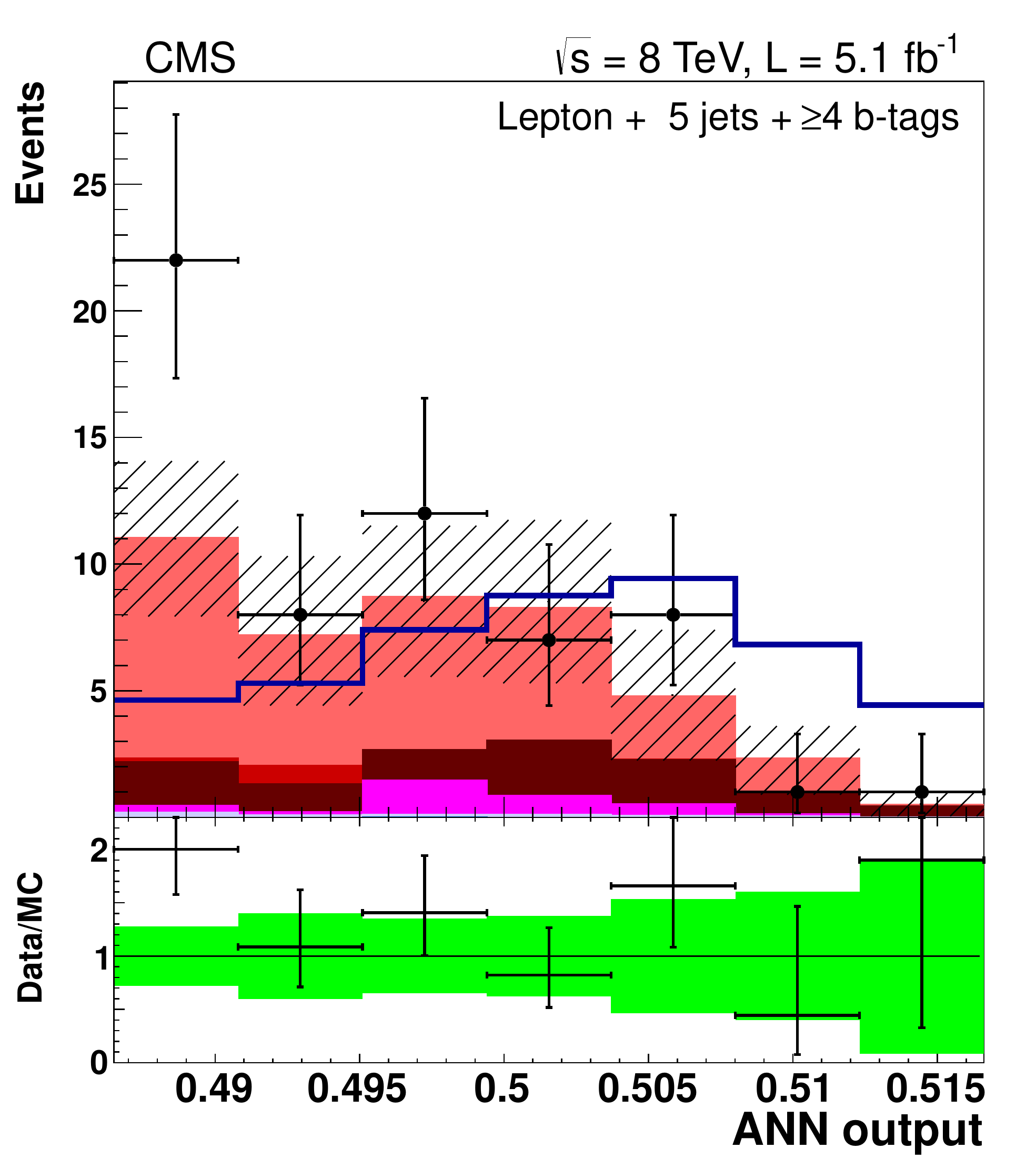}
   \vspace{0.2cm}

   \includegraphics[width=0.31\textwidth]{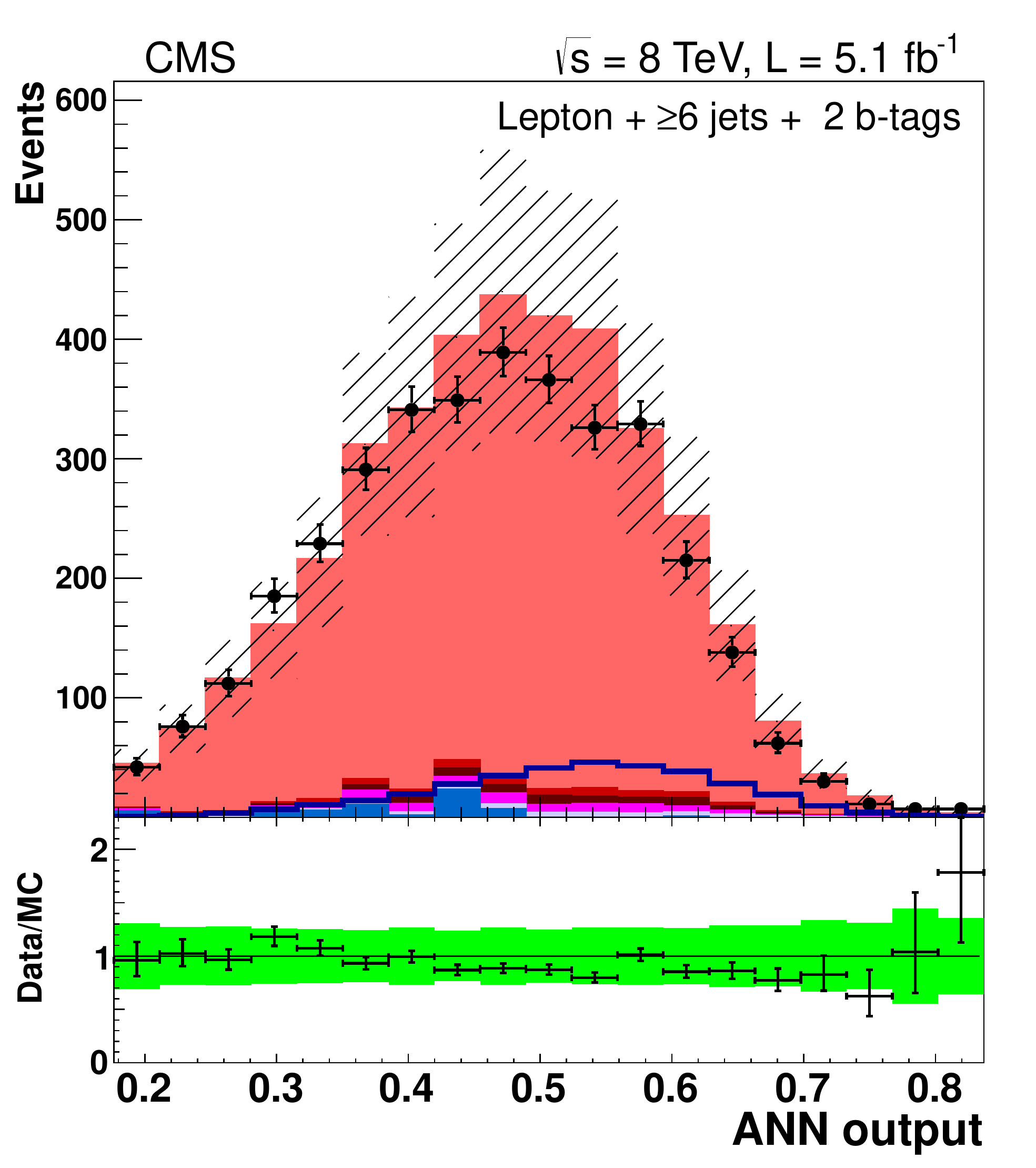}
   \includegraphics[width=0.31\textwidth]{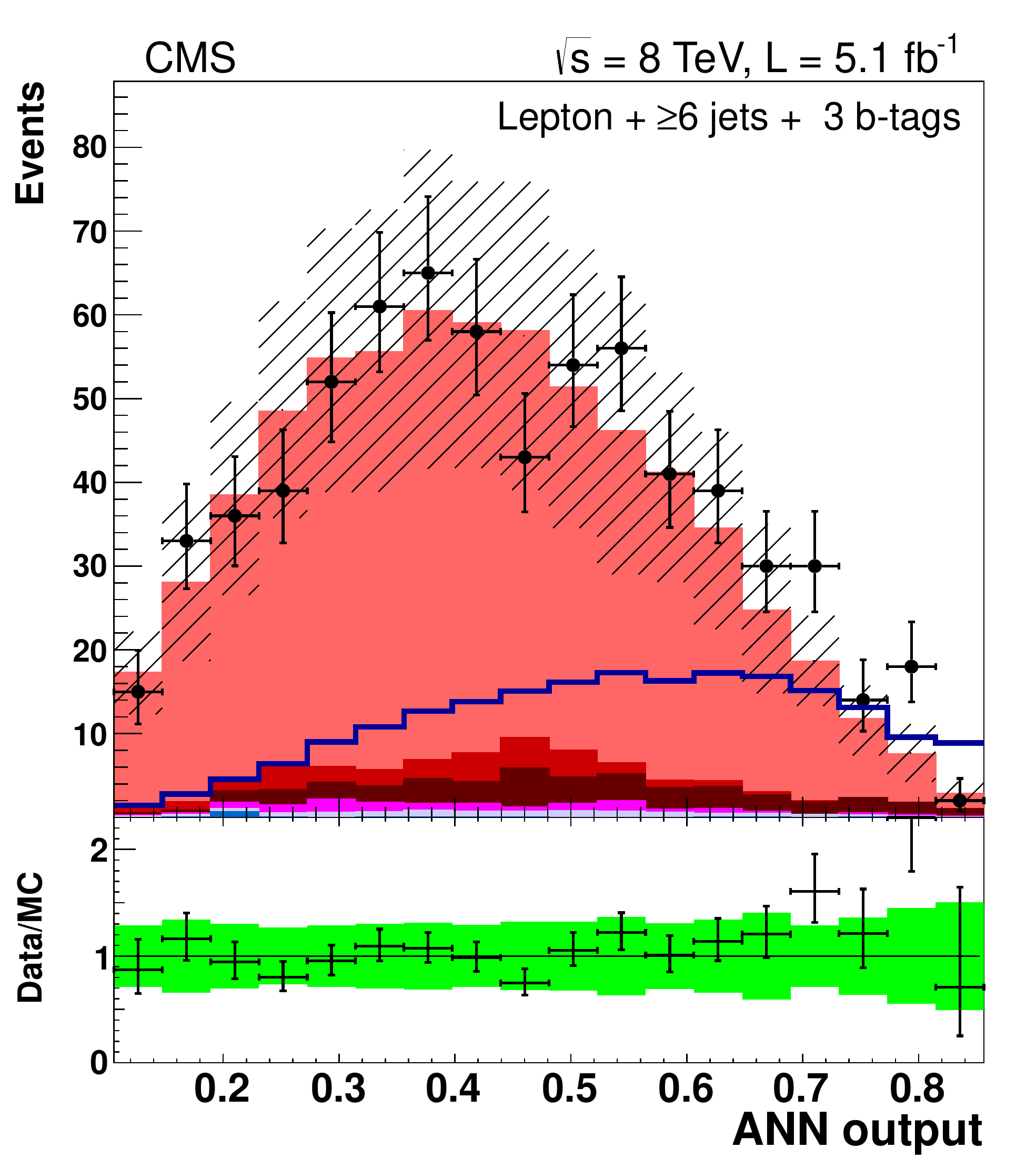}
   \includegraphics[width=0.31\textwidth]{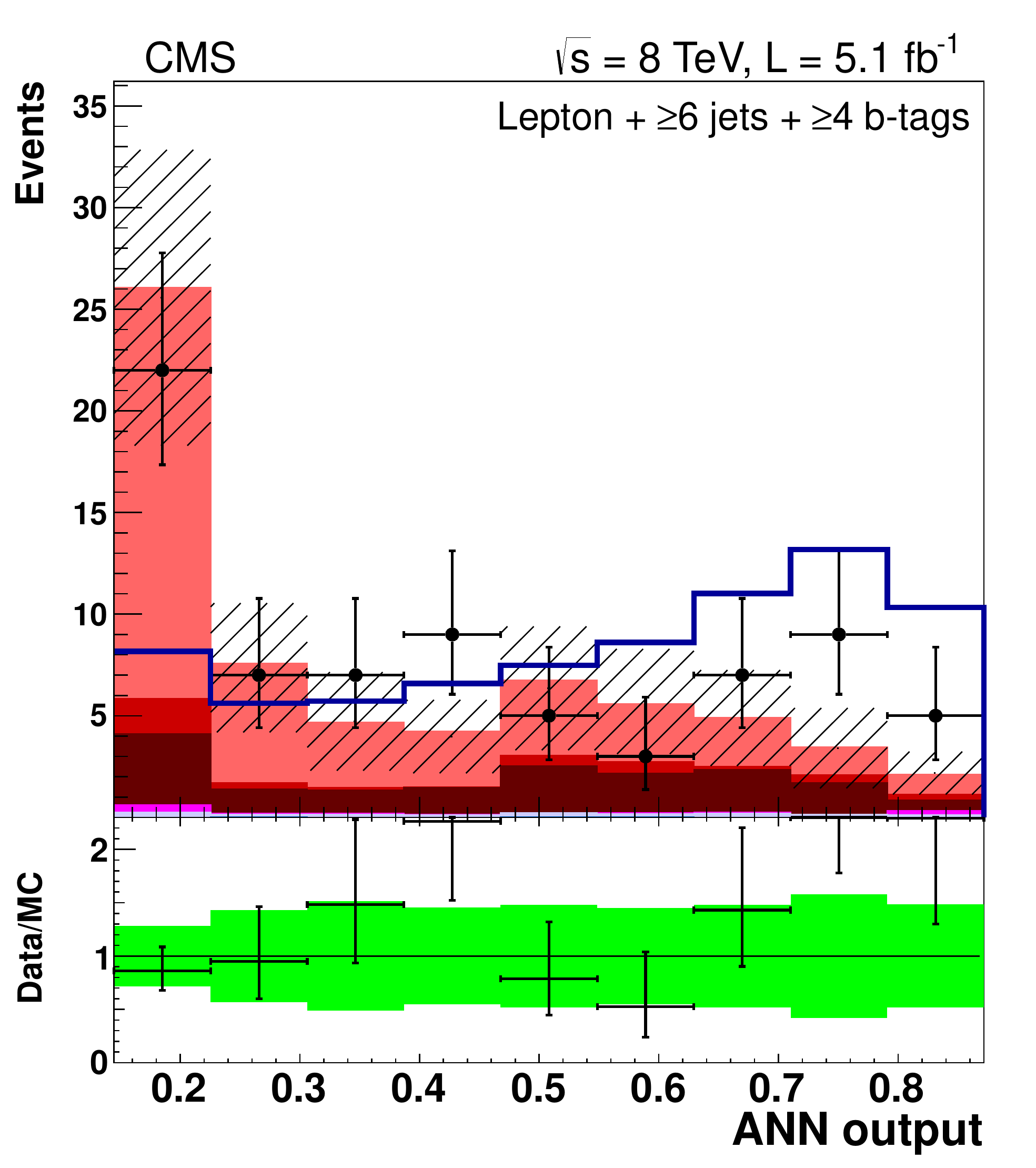}
   \caption{The distributions of the ANN output for lepton+jets
     events at $8\TeV$ in the various analysis categories.
     The top, middle and,
     bottom rows are events with 4, 5, and $\ge$6 jets, respectively,
     while the left, middle, and right-hand columns are events with 2,
     3, and $\ge$4 b-tags, respectively. Background-like events have a
     low ANN output value. Signal-like events have a high
     ANN output value.  The background is normalized to the SM expectation;
     the uncertainty (shown as a
     hatched band in the stack plot and a green band in the ratio
     plot) includes statistical and systematic uncertainties that
     affect both the rate and shape of the background distributions.
     The $\ttbar \PH$ signal ($m_{\PH} = 125\GeVcc$) is normalized to
     30 $\times$ SM expectation.}  \label{fig:lj_ANNoutput_8TeV}
\end{center}
\end{figure}

\begin{figure}[hbtp]
\begin{center}
 \includegraphics[width=0.31\textwidth]{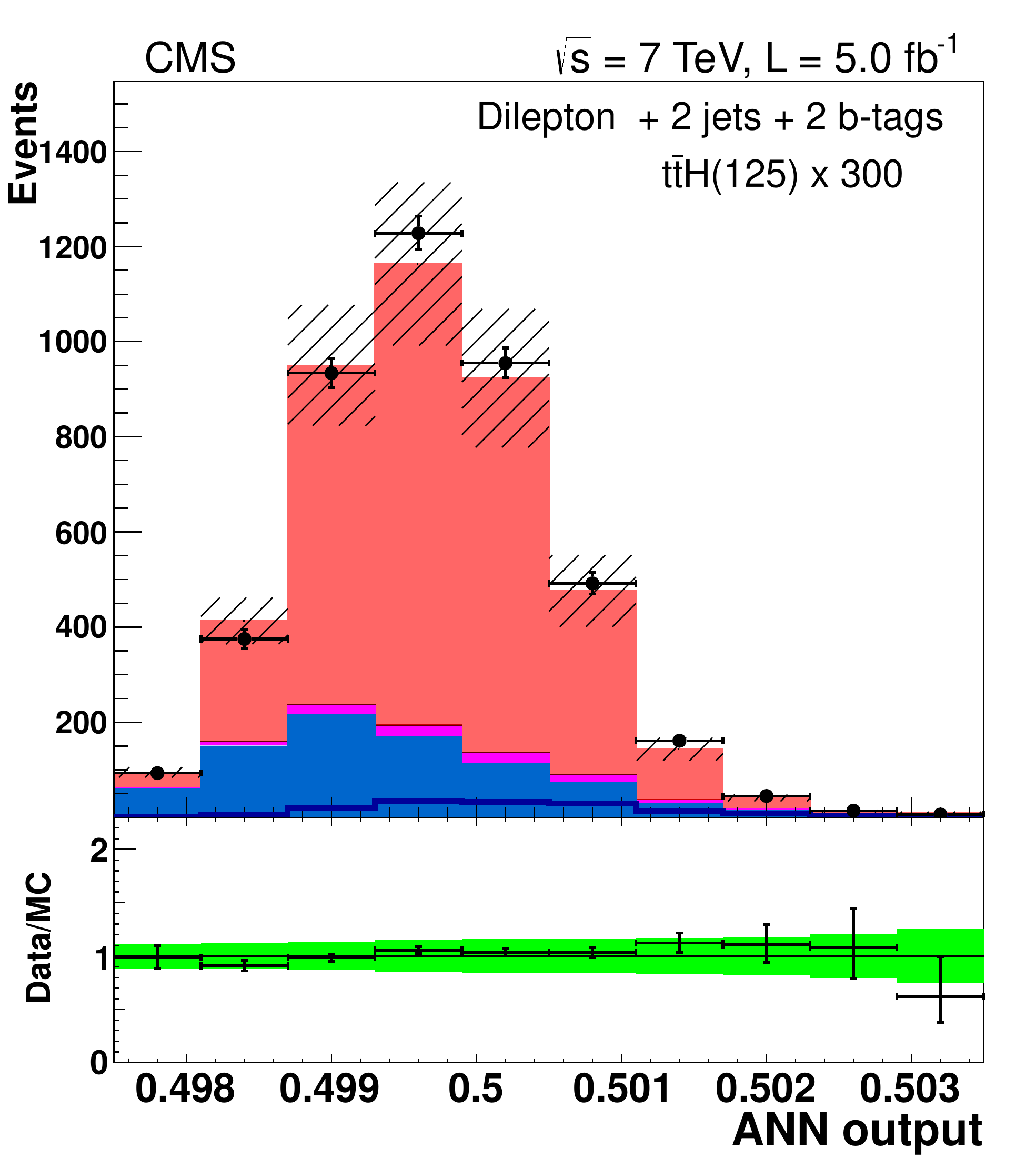}
 \includegraphics[width=0.31\textwidth]{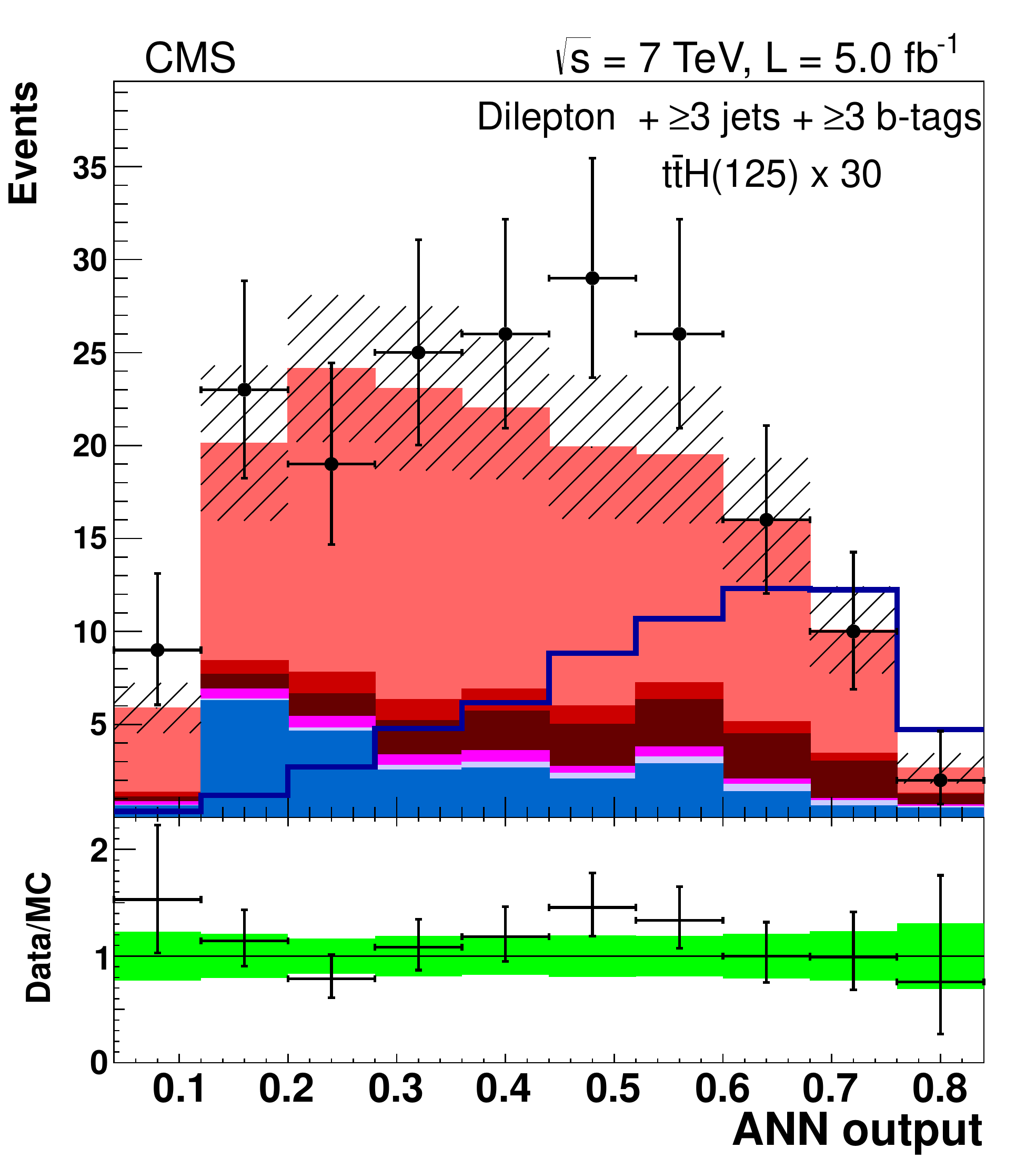}
 \raisebox{0.1\height}{\includegraphics[width=0.25\textwidth]{figures/samples_legend_tall_noTTHscale.pdf}}
 \caption{The distributions of the ANN output for  dilepton
     events at $7\TeV$ in the various analysis categories.
   The left plot shows events with 2~jets + 2~b-tags and right plot shows events with $\geq$3~jets + $\geq$3~b-tags.
   The background is normalized to the SM expectation; the uncertainty (shown as
   a hatched band in the stack plot and a green band in the ratio
   plot) band includes statistical and systematic uncertainties that
   affect both the rate and shape of the background distributions.
   The $\ttbar \PH$ signal ($m_{\PH} = 125\GeVcc$) is normalized to
   300 or 30 $\times$ SM expectation for the 2~jets + 2~b-tags and the $\geq$3~jets + $\geq$3~b-tags categories, respectively.}
 \label{fig:dilep_ANNoutput_7TeV}
\end{center}
\end{figure}

\begin{figure}[hbtp]
\begin{center}
 \includegraphics[width=0.31\textwidth]{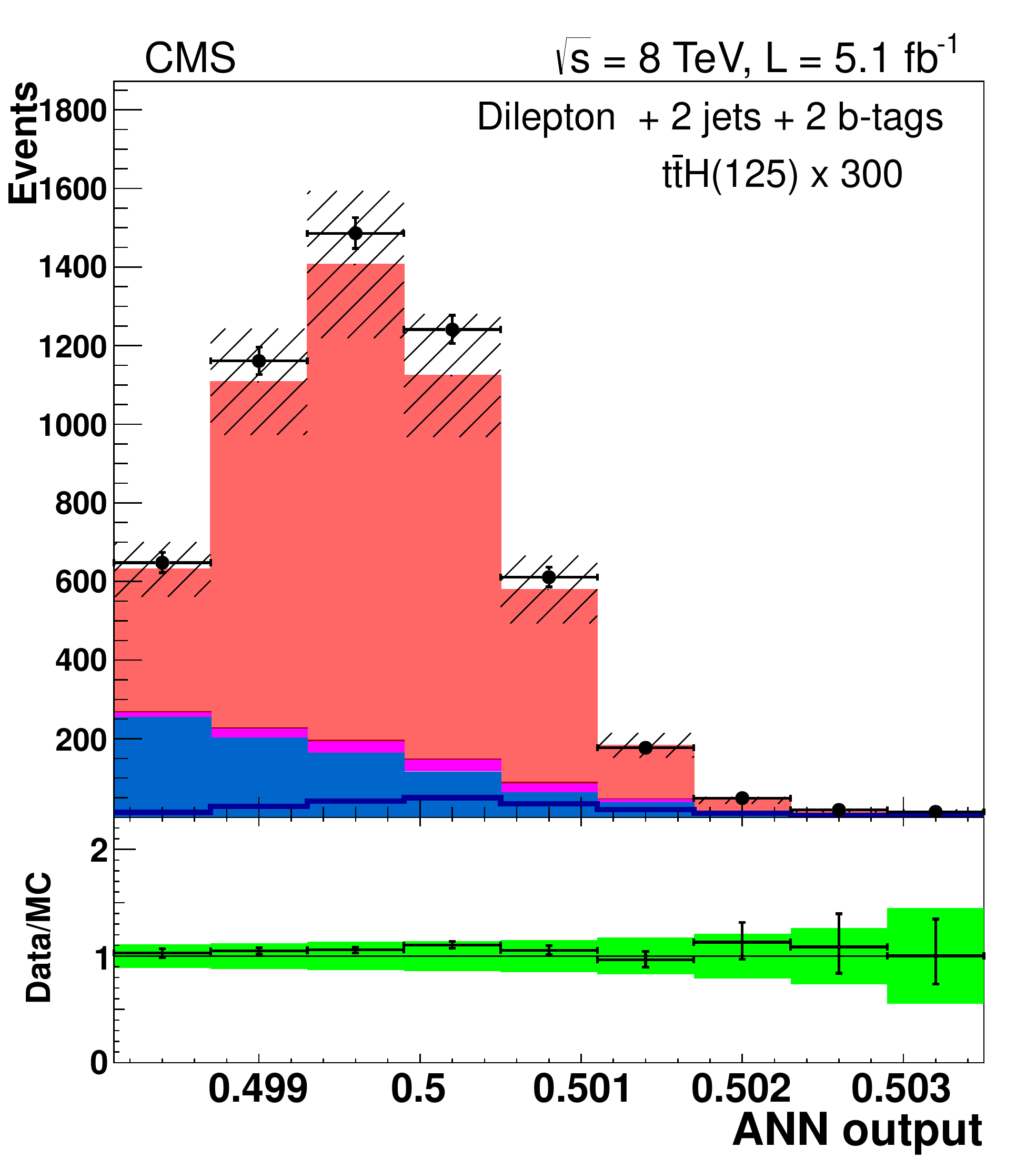}
 \includegraphics[width=0.31\textwidth]{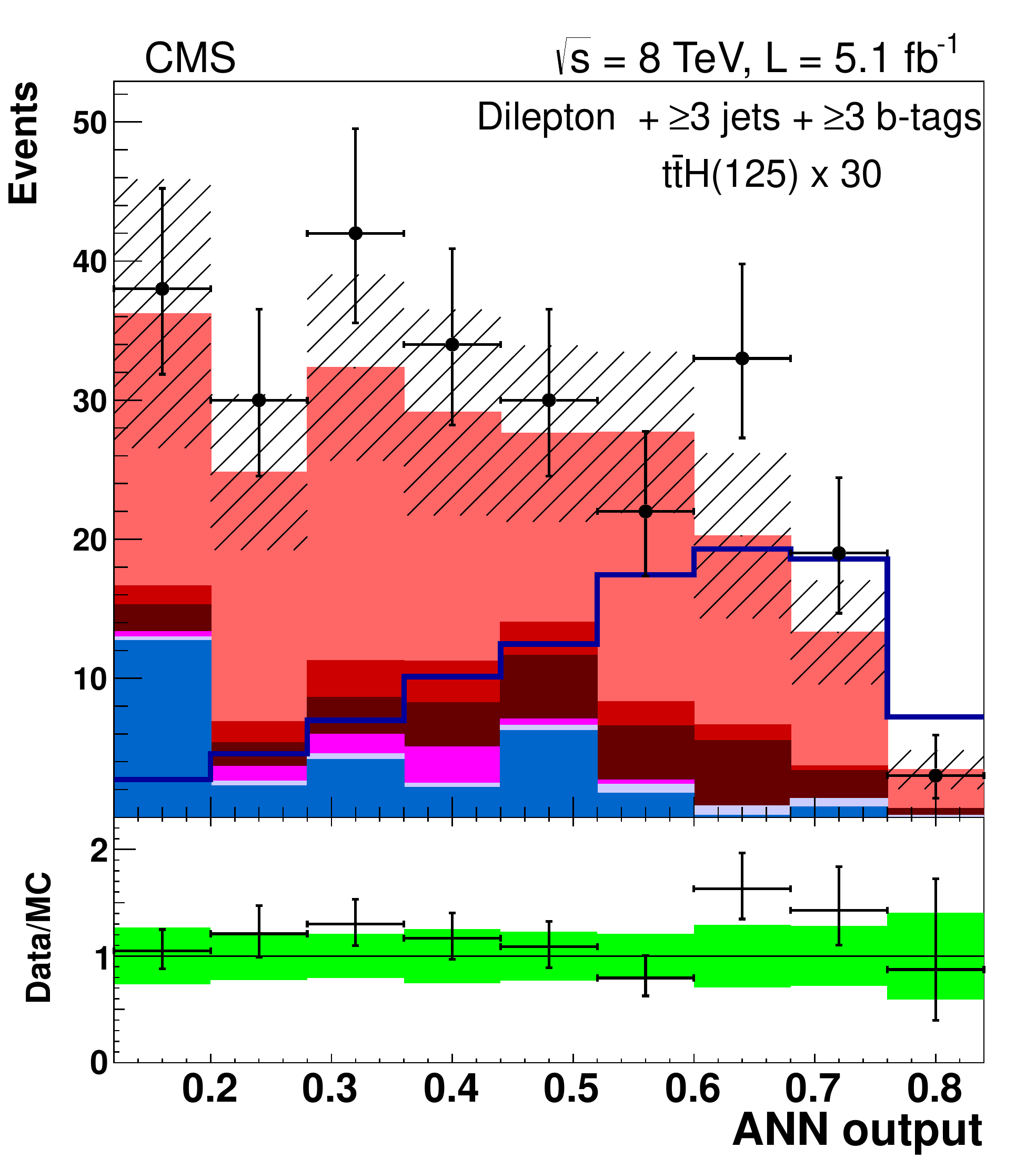}
 \raisebox{0.1\height}{\includegraphics[width=0.25\textwidth]{figures/samples_legend_tall_noTTHscale.pdf}}
 \caption{The distributions of the ANN output for  dilepton
   events at $8\TeV$ in the various analysis categories.
   The left plot shows events with 2~jets + 2~b-tags and right plot shows events with $\geq$3~jets + $\geq$3~b-tags.
   The background is normalized to the SM expectation; the uncertainty
   (shown as a hatched band in the stack plot and a green band in
   the ratio plot) includes statistical and systematic uncertainties
   that affect both the rate and shape of the background
   distributions.  The $\ttbar \PH$ signal ($m_{\PH} = 125\GeVcc$)
    is normalized to 300 or 30 $\times$ SM expectation for the 2~jets + 2~b-tags and the $\geq$3~jets + $\geq$3~b-tags categories, respectively.}
 \label{fig:dilep_ANNoutput_8TeV}
\end{center}
\end{figure}

The ANN output provides better discrimination between signal and
background than any one of the input variables individually.
Figures~\ref{fig:lj_ANNoutput_7TeV} and \ref{fig:lj_ANNoutput_8TeV}
show the ANN output for all the categories of the lepton+jets channel
in $7\TeV$ and $8\TeV$ data, respectively, and
Figs.~\ref{fig:dilep_ANNoutput_7TeV} and
\ref{fig:dilep_ANNoutput_8TeV} show output distributions for dilepton
events. We use these ANN output distributions for the signal
extraction as described in Section~\ref{sec:results}.

\section{Systematic uncertainties}
\label{sec:systematics}
Table~\ref{tab:syst} lists the systematic uncertainties that affect
signal and background yields, the shape of the ANN output, or both.
The effects of these uncertainties are evaluated specifically for each
event selection category, and the effects from the same source are
treated as completely correlated across the categories. The impact on
the rate is the relative change in expected yield due to each
uncertainty. Some sources of uncertainty affect predicted yields for
all processes in each category uniformly, while in some cases the
uncertainty affects the predicted yield of some processes in certain
categories more than others; in the latter cases the range of the
effect on the predicted yield is given across all processes in all
categories. Hence large relative rate changes listed in Table 5 can
typically be attributed to processes with small expected yields in a
single category that change significantly when considering a source of
uncertainty.

\begin{table}[hbtp]
\small
\centering
\topcaption{Summary of the systematic uncertainties considered on the inputs to the limit calculation.  Except where noted, each row in this table will be treated as a single, independent nuisance parameter.}
\label{tab:syst}
\begin{tabular}{|l|c|c|l|}
\hline
Source & Rate Uncertainty & Shape & Remarks \\
\hline
Luminosity ($7\TeV$) & 2.2\% & No & All signal and backgrounds \\
Luminosity ($8\TeV$) & 4.4\% & No & All signal and backgrounds \\
Lepton ID/Trig & 4\% & No & All signal and backgrounds \\
Pileup & 1\% & No & All signal and backgrounds \\
Additional Pileup Corr. & -- & Yes & All signal and backgrounds \\
Jet Energy Resolution & 1.5\% & No & All signal and backgrounds \\
Jet Energy Scale & 0--60\% & Yes & All signal and backgrounds \\
b-Tag SF ($\cPqb / \cPqc$) & 0--33.6\% & Yes & All signal and backgrounds \\
b-Tag SF (mistag) & 0--23.5\% & Yes & All signal and backgrounds \\
MC Statistics & -- & Yes & All backgrounds \\
\hline
PDF ($\cPg \cPg$) & 9\% & No & For $\cPg \cPg$ initiated processes ($\ttbar$, $\ttbar \cPZ$, $\ttbar \PH$) \\
PDF ($\cPq \cPaq$) & 4.2--7\% & No & For $\cPq\cPaq$ initiated processes ($\ttbar \PW$, $\PW$, $\cPZ$). \\
PDF ($\cPq \cPg$) & 4.6\% & No & For $\cPq \cPg$ initiated processes (single top) \\
\hline
QCD Scale ($\ttbar \PH$) & 15\% & No & For NLO $\ttbar\PH$ prediction \\
QCD Scale ($\ttbar$) & 2--12\% & No & For NLO $\ttbar$ and single top predictions \\
QCD Scale (V) & 1.2--1.3\% & No & For NNLO $\PW$ and $\cPZ$ prediction \\
QCD Scale (VV) & 3.5\% & No & For NLO diboson prediction \\
\hline
Madgraph Scale ($\ttbar$) & 0--20\% & Yes & $\ttbar+\text{jets}$$/\bbbar/\ccbar$ uncorrelated. Varies by jet bin.\\
Madgraph Scale (V) & 20--60\% & No & Varies by jet bin. \\
\hline
$\ttbar + \bbbar$ & 50\% & No & Only $\ttbar + \bbbar$. \\
\hline
\end{tabular}
\end{table}

Lepton identification and trigger efficiency uncertainties were found
to have a small impact on the analysis. The uncertainties were
estimated by comparing variations in the difference in performance
between data and MC simulation using a high-purity sample
of $\cPZ$-boson decays. The largest
variations were at most 4\% for a small fraction of
events, such as electrons at low $\pt$. The analysis conservatively uses 4\%
uncertainty on the lepton scale overall.  To ascertain the effects of the
uncertainty on the pileup distribution, the cross section used to
predict the distribution of pileup interactions in MC is varied by 8\%
from its nominal value, and the resulting change in the number of
pileup interactions is propagated through the analysis. The systematic
uncertainty due to the additional pileup correction, based on the
scalar sum of the $\pt$ of the jets, is evaluated by doubling or
removing the correction applied. The uncertainty on the luminosity
estimate corresponding to the $7\TeV$ dataset is
2.2\%~\cite{CMS-PAS-SMP-12-008} and, for the $8\TeV$ dataset,
4.4\%~\cite{CMS-PAS-LUM-12-001}.

The uncertainty from the jet energy scale~\cite{Chatrchyan:2011ds} is
evaluated by varying the energy scale for all jets in the signal and
background predictions up and down by one standard deviation as a
function of jet $\pt$ and $\eta$ and re-evaluating the yields and ANN
shapes of all processes.  Similarly, the uncertainty on the jet energy
resolution is obtained by varying the jet energy resolution correction
up and down by one standard deviation, although in this case the
effect on shape is negligible and therefore not included.

The b-tagging scale factor corrects the b-tagging efficiency in
simulation to match that measured in data~\cite{CMS:2012hd}.  The
uncertainty on this scale factor is evaluated by varying it up and
down by one standard deviation and the new CSV output value
corresponding to that uncertainty is recalculated. This new CSV value
is used to determine both the number of tags associated with that
systematic and the new shape of variables that use the CSV output,
such as the average CSV value for b-tagged jets.  This uncertainty
affects both rate and shape estimates. Since the b-tagging scale
factor uncertainty affects the ANN shape differently for events with
different number of jets or number of b-tagged jets, we conservatively
assume no correlations among all the categories.

We account for the effect of background MC statistics in our analysis
using the approach described
in~\cite{BarlowBeeston,BarlowBeeston2}. To make the limit computation
more efficient and stable, we do not evaluate this uncertainty for any
bin in the ANN shapes for which the MC statistical uncertainty is
negligible compared to the data statistics or where there is no
appreciable contribution from signal.  In total, there are 64 nuisance
parameters used to describe the MC statistics for the $8\TeV$
results, but only five are needed for $7\TeV$, due to the larger MC
statistics available for those samples.  Tests show that the effect of
neglecting bins as described above is smaller than 5\%.

Theoretical uncertainties on the cross sections used to predict the
rates of various processes are propagated to the yield estimates.  All
rates are estimated using cross sections of at least NLO accuracy,
which have uncertainties arising primarily from PDFs and the choice of
factorization and renormalization scales.  The cross section
uncertainties are each separated into their PDF and scale components
and correlated where appropriate between processes.  For example, the
PDF uncertainty for processes originating primarily from gluon-gluon
initial states, e.g., $\ttbar$ and $\ttbar \PH$ production, are treated
as 100\% correlated.

In addition, for the $\ttbar+\text{jets}$ (including $\ttbar+\bbbar$ and
$\ttbar+\ccbar$) and the V+jets processes, the inclusive NLO or better
cross section prediction are extrapolated to exclusive rates for
particular jet or tag categories using the \MADGRAPH
tree-level matrix element generator matched to the \PYTHIA
parton shower MC program.  Although
\textsc{Madgraph} incorporates contributions from higher-order
diagrams, because it does so only at tree-level, it is subject to
fairly large uncertainties arising from the choice of scale.  These
uncertainties are evaluated using samples for which the factorization
and renormalization scales have been varied up and down by a factor of
two.  The rate uncertainty arising from this source varies with the
number of additional jets in the production diagram, and is larger for
events with more jets.  The effect of scale variations on the ANN
output shape is also included for the $\ttbar+\text{jets}$ sample.  Scale
variations are treated as uncorrelated for the $\ttbar+$light flavour,
$\ttbar+\bbbar$, and $\ttbar+\ccbar$ components to cover the
uncertainty in the relative yields of those processes; the impact on
the ANN output shape from scale variation in the V+jets processes is
neglected, since this contribution is small in most categories. The
scale variations for $\PW+\text{jets}$ and $Z+\text{jets}$ are treated as correlated
with each other, but uncorrelated with $\ttbar+\text{jets}$.

As the background due to the $\ttbar+\bbbar$ contribution is very
similar to the signal, the uncertainty on its rate and shape will have
a substantial impact on our search. Due to the lack of more accurate higher
order theoretical predictions for this process, we estimated this
background and assessed its uncertainty based on the inclusive
$\ttbar$ sample and the most important contribution to the uncertainty
comes from the factorization and renormalization scale
systematics. Neither control region studies nor higher-order
theoretical calculations~\cite{Bredenstein:2010rs} can currently
constrain the normalization of the $\ttbar+\bbbar$ contribution to
better than 50\% accuracy.  Therefore, to be conservative, an extra
50\% rate uncertainty is assigned to $\ttbar+\bbbar$ for both $7\TeV$
and $8\TeV$.

\section{Results}
\label{sec:results}

A maximum likelihood fit is performed on the ANN output distributions
from the nine jet-tag categories considered in the analysis. We
consider the model including the SM backgrounds and a Higgs boson
signal, as well as a model with only SM backgrounds but no Higgs boson
signal.  As we currently lack sensitivity to detect a SM Higgs boson
signal, and observe no significant excess in the data, we focus here
on setting 95\% confidence level (CL) upper limits on the possible
presence of a SM-like signal.

The statistical methodology employed by this analysis is identical to
that used for other CMS
searches~\cite{CMS:2012gu,Chatrchyan:2012tx,CMS_ATLAS_Combo}. In
brief, we use a modified frequentist
CL$_s$~\cite{Junk:1999kv,Read:2002hq} approach in which the test
statistic involves the ratio of the likelihood functions constructed
from the background expectations plus the SM Higgs boson signal scaled
by an arbitrary parameter $\mu$, where $\mu \geq 0$.  The parameter
$\mu = \sigma/\sigma_{\text{SM}}$ is the ratio of the cross section of
our signal process ($\sigma$) to the expected SM Higgs boson cross
section ($\sigma_{\text{SM}}$).  The likelihood function describes the
expected yield of signal and background in bins of the ANN output for
each event selection category.  The systematic uncertainties described
in Section~\ref{sec:systematics} are incorporated into the likelihood
by means of nuisance parameters that affect each background's rate,
shape or both.  Shape variations are handled by means of template
morphing.  A vertical template morphing approach is used where the
shapes are smoothly interpolated between the ${\pm}1\sigma$ varied
shapes and linearly extrapolated outside that region.  This is the
standard template morphing approach used by all CMS Higgs analyses.
As appropriate for the frequentist approach taken here, the nuisance
parameters are profiled during the limit extraction. The nuisance
parameter correlations are implemented in a way that accounts for
event migrations between the selection categories.  Furthermore, in
cases involving shape systematics, where high-statistics,
background-rich categories might overconstrain certain systematic
effects in the lower-statistics, higher-sensitivity categories, we
take the approach of decorrelating the nuisance parameters to avoid
overly aggressive constraints.

When combining the results from the $7\TeV$ and $8\TeV$ datasets, the
proper correlation in systematic effects must be represented in the
nuisance parameter choices.  Given that for all theoretical predictions and
many experimental uncertainties, exactly the same calculation or
calibration is applied to the $7\TeV$ and $8\TeV$ datasets, the associated
systematic uncertainties are treated as completely correlated and a
single nuisance parameter is used to implement the effect.  There are
two exceptions to this approach. The luminosity is evaluated
separately for the two analyses and the dominant uncertainties are
largely independent, so the luminosity uncertainty is treated as
uncorrelated between $7\TeV$ and $8\TeV$.  Furthermore, as separate MC
samples are used for the two datasets, the MC statistical
uncertainties are treated as uncorrelated between the two datasets.

\begin{figure}[hbtp]
  {\centering
    \includegraphics[width=0.75\textwidth]{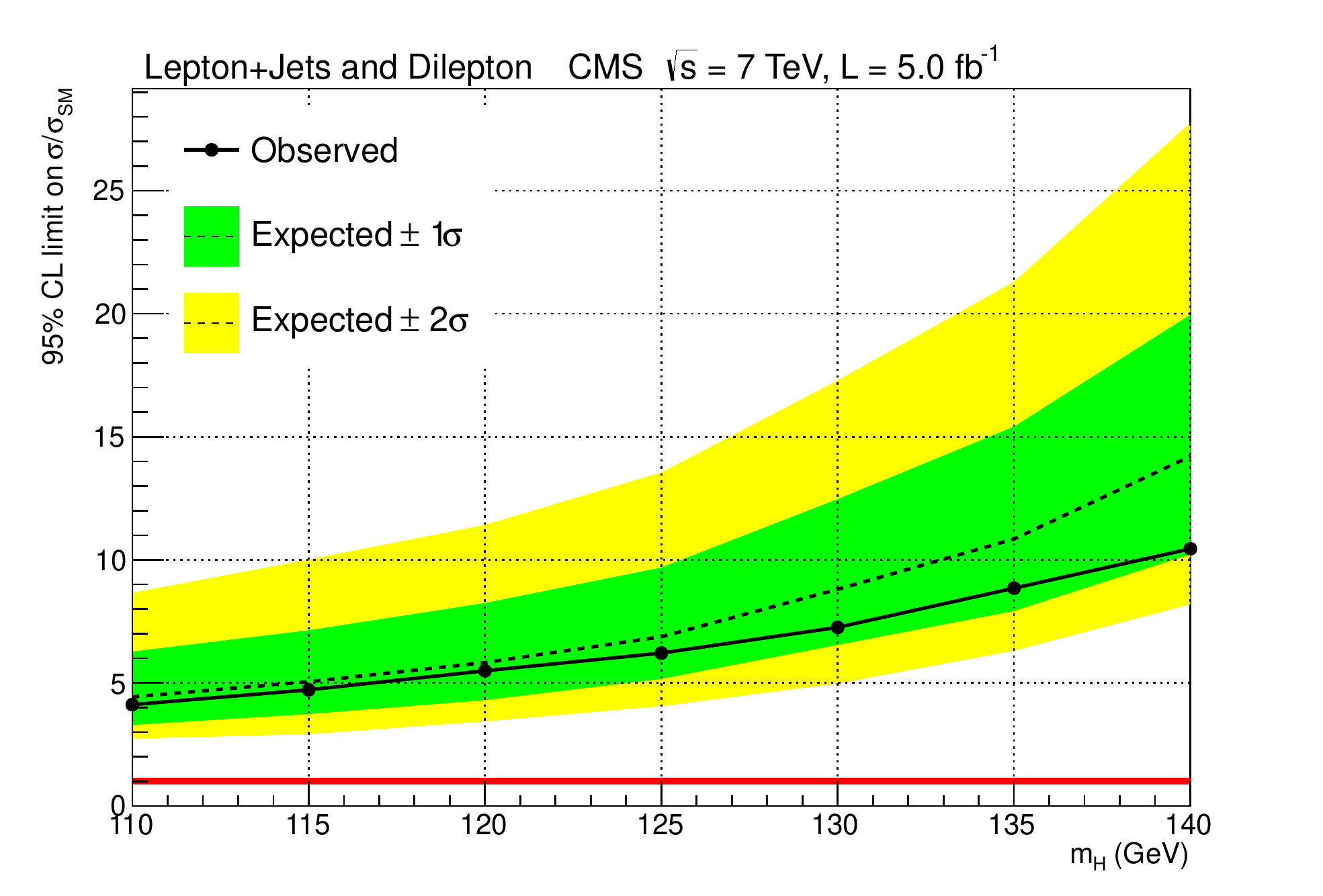}
    \includegraphics[width=0.75\textwidth]{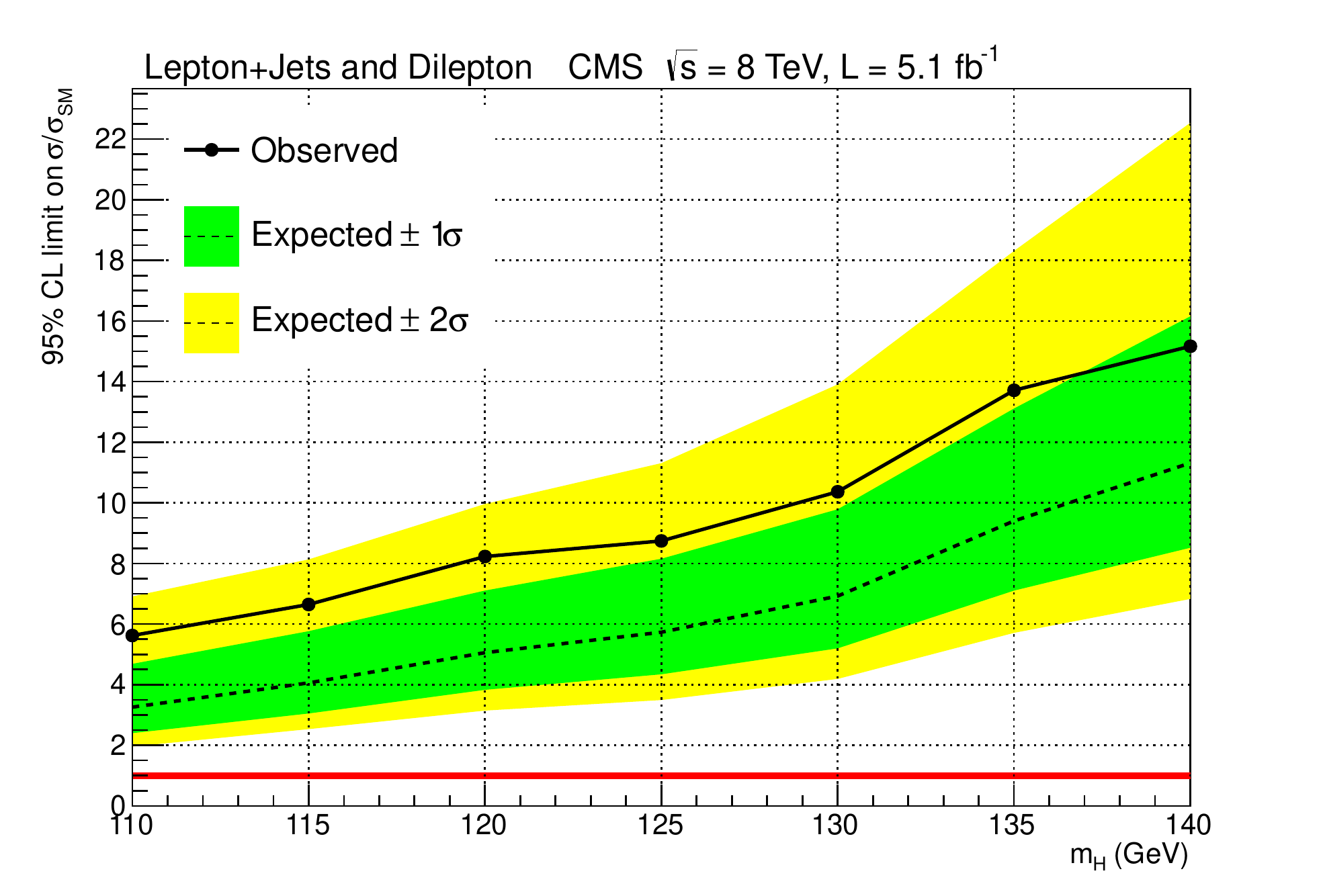}
   \caption{The observed and expected 95\% CL upper limits on the signal strength parameter $\mu = \sigma / \sigma_{\text{SM}}$ for lepton+jets and dilepton channels combined using
    the 2011 dataset at $7\TeV$ (above) and the 2012 dataset at $8\TeV$ (below).}
    \label{fig:comb_limit_7TeV}}
\end{figure}

\begin{figure}[hbtp]
   {\centering
     \includegraphics[width=0.75\textwidth]{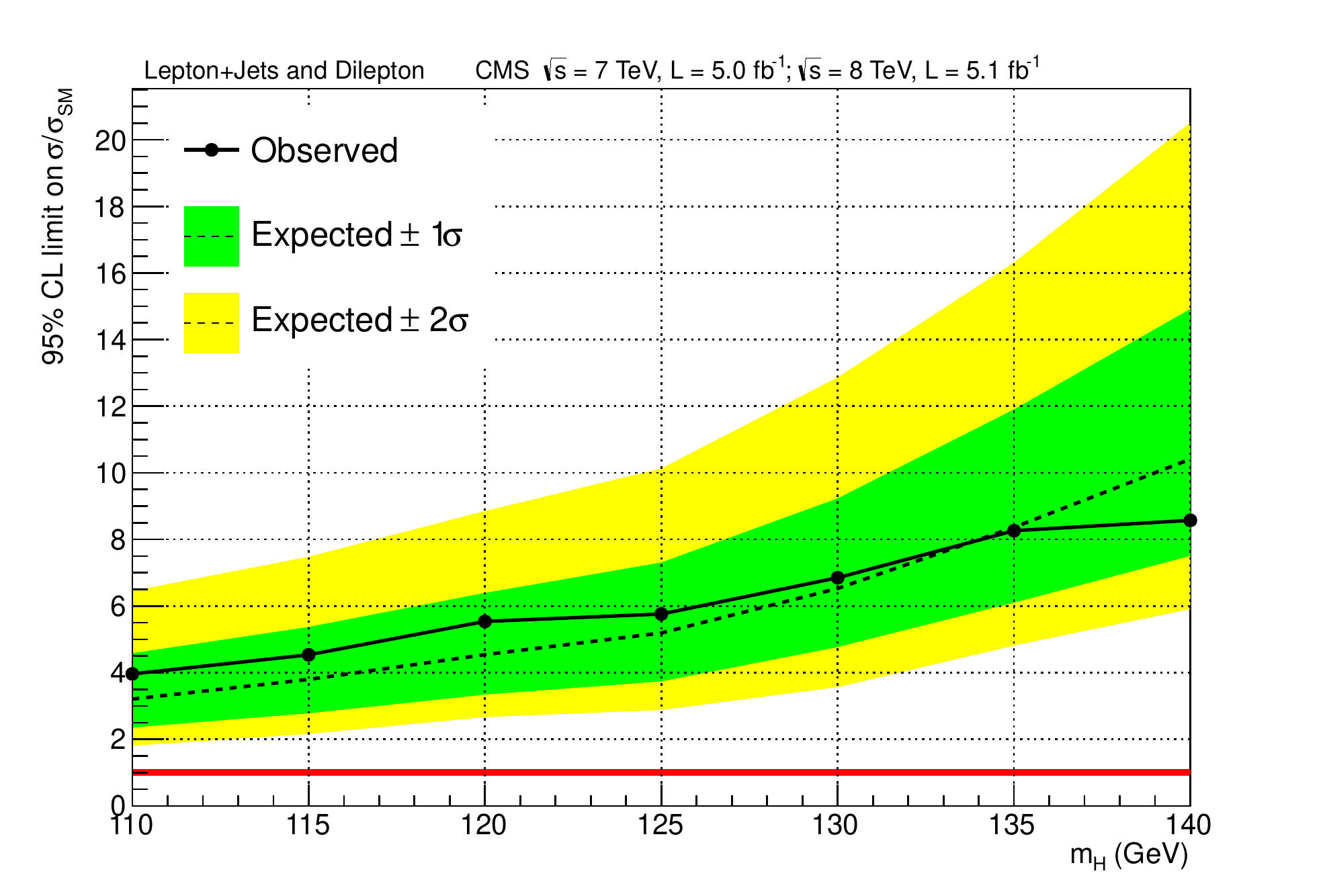}
     \caption{Using the 2011+2012 datasets, the observed and expected
     95\% CL upper limits on the signal strength parameter $\mu =
     \sigma / \sigma_{\text{SM}}$ for lepton+jets and dilepton channels
     combined.}  \label{fig:comb_bothYears_limit}}
\end{figure}

Background-dominated categories are used to constrain the fitted background
contributions in the signal-enhanced categories. The prediction from
the fit for the composition of the selected sample in each category
more accurately describes the data than the prediction directly from
simulation, and the uncertainties on the final composition are
reduced. The resulting distributions are driven by the shape from
$\ttbar$+light flavor, the dominant background in each category. No significant
excesses of data above the background-only predictions are observed,
and we use our statistical treatment to extract upper limits on the
amount of $\ttbar \PH$ production consistent with our data.
Figure~\ref{fig:comb_limit_7TeV} shows the 95\% CL upper limit on the
ratio $\mu$ of the $\ttbar \PH$ cross section with respect to that
predicted by the SM as a function of $m_{\PH}$ for the $7\TeV$ and $8
\TeV$ samples, separately, combining both lepton+jets and dilepton
channels in each dataset. Figure~\ref{fig:comb_bothYears_limit} shows
the upper limit obtained by combining both data
samples. Table~\ref{tab:comb_bothYears_limitTable} shows the expected
and observed limits for $7\TeV$, $8\TeV$, and combined analysis,
using both the lepton+jets and the dilepton channels. The expected
limit is extracted from the background-only hypothesis with no Higgs
signal present.  In addition to the median expected limit, the bands
that contain 68\% (1 standard deviation) and 95\% (2 standard
deviations) around the median are also quoted.  The median expected
limit for a Higgs boson mass of $125\GeVcc$ is $5.2 \times \sigma_{\text{SM}}$ while the observed limit is
$5.8 \times \sigma_{\text{SM}}$.

As a cross check, we extracted the limit using the best single
variable according to Table~\ref{tab:inputs} and plotted in
Fig.~\ref{fig:bestVars_8TeV} instead of the ANN output.  Otherwise,
the analysis was performed in exactly the same way as the version
based on the ANN, including the event selection categories, systematic
uncertainties, and treatment of the nuisance parameters.  The resulting median
expected limit, for a Higgs boson mass of $125\GeVcc$ is $6.6 \times
\sigma_{\text{SM}}$, approximately 27\% higher than the
limit obtained with the ANN.  The primary reason for this decrease in
sensitivity is the loss of separating power and the increased
susceptibility to individual systematic effects coming from using
fewer variables.  The observed limit obtained using the best single
variable analysis is $10.4 \times \sigma_{\text{SM}}$, which is beyond
the 68\% CL range (on $\mu$) of the expected ($[5.0,9.2]$) but within the 95\%
CL range ($[4.0,12.7]$).

 \begin{table}[hbtp]
   \centering
{\small
   \topcaption{Observed and expected 95\% CL upper limits on $\mu=\sigma/\sigma_{\text{SM}}$ for the SM Higgs boson in the lepton+jets and dilepton channels combined, using
    the 7~TeV dataset, $8\TeV$ dataset, and both datasets combined.}
   \label{tab:comb_bothYears_limitTable}
\begin{tabular}{|c|c|ccc|} \hline
\multicolumn{5}{|c|}{7\TeV Lepton+Jets and Dilepton} \\ \hline
          &          & \multicolumn{3}{c|}{Expected} \\
$m_{H}$ & Observed & Median & 68\% CL Range &  95\% CL Range  \\ \hline
$110\GeVcc$ & 4.1 & 4.4 & [3.3, 6.3] & [2.7, 8.7] \\
$115\GeVcc$ & 4.7 & 5.0 & [3.7, 7.1] & [2.9, 10.0] \\
$120\GeVcc$ & 5.5 & 5.8 & [4.3, 8.2] & [3.4, 11.4] \\
$125\GeVcc$ & 6.2 & 6.9 & [5.2, 9.7] & [4.1, 13.6] \\
$130\GeVcc$ & 7.3 & 8.8 & [6.5, 12.5] & [5.0, 17.3] \\
$135\GeVcc$ & 8.8 & 10.8 & [7.9, 15.4] & [6.3, 21.3] \\
$140\GeVcc$ & 10.4 & 14.2 & [10.2, 20.0] & [8.2, 27.8] \\
\hline
\end{tabular}}

\vspace{0.2cm}
{\small
\begin{tabular}{|c|c|ccc|} \hline
\multicolumn{5}{|c|}{8\TeV Lepton+Jets and Dilepton} \\ \hline
           &          & \multicolumn{3}{c|}{Expected} \\
$m_{H}$ & Observed & Median & 68\% CL Range &  95\% CL Range  \\ \hline
$110\GeVcc$ & 5.6 & 3.3 & [2.4, 4.7] & [2.0, 6.9] \\
$115\GeVcc$ & 6.6 & 4.1 & [3.0, 5.8] & [2.5, 8.1] \\
$120\GeVcc$ & 8.2 & 5.1 & [3.8, 7.1] & [3.1, 10.0] \\
$125\GeVcc$ & 8.7 & 5.7 & [4.3, 8.2] & [3.5, 11.3] \\
$130\GeVcc$ & 10.4 & 6.9 & [5.2, 9.8] & [4.2, 13.9] \\
$135\GeVcc$ & 13.7 & 9.4 & [7.1, 13.1] & [5.7, 18.3] \\
$140\GeVcc$ & 15.2 & 11.3 & [8.5, 16.2] & [6.8, 22.5] \\
\hline
\end{tabular}}

\vspace{0.2cm}
{\small
\begin{tabular}{|c|c|ccc|} \hline
\multicolumn{5}{|c|}{7\TeV + 8\TeV Lepton+Jets and Dilepton combined} \\ \hline
           &          & \multicolumn{3}{c|}{Expected} \\
$m_{H}$ & Observed & Median & 68\% CL Range &  95\% CL Range  \\ \hline
$110\GeVcc$ & 4.0 & 3.2 & [2.4, 4.6] & [1.8, 6.5] \\
$115\GeVcc$ & 4.5 & 3.8 & [2.8, 5.4] & [2.2, 7.5] \\
$120\GeVcc$ & 5.5 & 4.5 & [3.3, 6.4] & [2.7, 8.9] \\
$125\GeVcc$ & 5.8 & 5.2 & [3.7, 7.3] & [2.9, 10.1] \\
$130\GeVcc$ & 6.8 & 6.5 & [4.8, 9.2] & [3.6, 12.9] \\
$135\GeVcc$ & 8.3 & 8.4 & [6.1, 11.9] & [4.8, 16.3] \\
$140\GeVcc$ & 8.6 & 10.4 & [7.5, 14.9] & [5.9, 20.5] \\
\hline
\end{tabular}}
 \end{table}

\section{Summary}
\label{sec:conclusions}

A search for the standard model Higgs boson produced in association
with a top-quark pair has been performed at the CMS experiment using
data samples corresponding to an integrated luminosity of $5.0\fbinv$
($5.1\fbinv$) collected in $\Pp\Pp$ collisions at the center-of-mass
energy of $7\TeV$ ($8\TeV$). Events are considered where the
top-quark pair decays to either one lepton+jets ($\ttbar \to \ell\cPgn
\cPq \cPaq^{\prime} \bbbar $) or dileptons ($\ttbar \to \ell^{+} \cPgn \ell^{-} \cPagn
\bbbar$), $\ell$ being an electron or a muon. The search has been optimized
for the decay mode $\PH \to \bbbar$, however sensitivity to other
decay modes has been preserved. Artificial neural networks are used to
discriminate between signal and background events. Combining the
results from the $7\TeV$ and $8\TeV$ samples, the observed
(expected) limit on the cross section for Higgs boson production in
association with top-quark pairs for a Higgs boson mass of $125\GeVcc$ is 5.8 (5.2) times the standard model expectation. This is the
first such search at the LHC.

\section*{Acknowledgements}
\hyphenation{Bundes-ministerium Forschungs-gemeinschaft Forschungs-zentren} We congratulate our colleagues in the CERN accelerator departments for the excellent performance of the LHC and thank the technical and administrative staffs at CERN and at other CMS institutes for their contributions to the success of the CMS effort. In addition, we gratefully acknowledge the computing centres and personnel of the Worldwide LHC Computing Grid for delivering so effectively the computing infrastructure essential to our analyses. Finally, we acknowledge the enduring support for the construction and operation of the LHC and the CMS detector provided by the following funding agencies: the Austrian Federal Ministry of Science and Research; the Belgian Fonds de la Recherche Scientifique, and Fonds voor Wetenschappelijk Onderzoek; the Brazilian Funding Agencies (CNPq, CAPES, FAPERJ, and FAPESP); the Bulgarian Ministry of Education, Youth and Science; CERN; the Chinese Academy of Sciences, Ministry of Science and Technology, and National Natural Science Foundation of China; the Colombian Funding Agency (COLCIENCIAS); the Croatian Ministry of Science, Education and Sport; the Research Promotion Foundation, Cyprus; the Ministry of Education and Research, Recurrent financing contract SF0690030s09 and European Regional Development Fund, Estonia; the Academy of Finland, Finnish Ministry of Education and Culture, and Helsinki Institute of Physics; the Institut National de Physique Nucl\'eaire et de Physique des Particules~/~CNRS, and Commissariat \`a l'\'Energie Atomique et aux \'Energies Alternatives~/~CEA, France; the Bundesministerium f\"ur Bildung und Forschung, Deutsche Forschungsgemeinschaft, and Helmholtz-Gemeinschaft Deutscher Forschungszentren, Germany; the General Secretariat for Research and Technology, Greece; the National Scientific Research Foundation, and National Office for Research and Technology, Hungary; the Department of Atomic Energy and the Department of Science and Technology, India; the Institute for Studies in Theoretical Physics and Mathematics, Iran; the Science Foundation, Ireland; the Istituto Nazionale di Fisica Nucleare, Italy; the Korean Ministry of Education, Science and Technology and the World Class University program of NRF, Republic of Korea; the Lithuanian Academy of Sciences; the Mexican Funding Agencies (CINVESTAV, CONACYT, SEP, and UASLP-FAI); the Ministry of Science and Innovation, New Zealand; the Pakistan Atomic Energy Commission; the Ministry of Science and Higher Education and the National Science Centre, Poland; the Funda\c{c}\~ao para a Ci\^encia e a Tecnologia, Portugal; JINR (Armenia, Belarus, Georgia, Ukraine, Uzbekistan); the Ministry of Education and Science of the Russian Federation, the Federal Agency of Atomic Energy of the Russian Federation, Russian Academy of Sciences, and the Russian Foundation for Basic Research; the Ministry of Science and Technological Development of Serbia; the Secretar\'{\i}a de Estado de Investigaci\'on, Desarrollo e Innovaci\'on and Programa Consolider-Ingenio 2010, Spain; the Swiss Funding Agencies (ETH Board, ETH Zurich, PSI, SNF, UniZH, Canton Zurich, and SER); the National Science Council, Taipei; the Thailand Center of Excellence in Physics, the Institute for the Promotion of Teaching Science and Technology of Thailand and the National Science and Technology Development Agency of Thailand; the Scientific and Technical Research Council of Turkey, and Turkish Atomic Energy Authority; the Science and Technology Facilities Council, UK; the US Department of Energy, and the US National Science Foundation.

Individuals have received support from the Marie-Curie programme and the European Research Council (European Union); the Leventis Foundation; the A. P. Sloan Foundation; the Alexander von Humboldt Foundation; the Belgian Federal Science Policy Office; the Fonds pour la Formation \`a la Recherche dans l'Industrie et dans l'Agriculture (FRIA-Belgium); the Agentschap voor Innovatie door Wetenschap en Technologie (IWT-Belgium); the Ministry of Education, Youth and Sports (MEYS) of Czech Republic; the Council of Science and Industrial Research, India; the Compagnia di San Paolo (Torino); and the HOMING PLUS programme of Foundation for Polish Science, cofinanced from European Union, Regional Development Fund.

\bibliography{auto_generated}   

\clearpage

\cleardoublepage \appendix\section{The CMS Collaboration \label{app:collab}}\begin{sloppypar}\hyphenpenalty=5000\widowpenalty=500\clubpenalty=5000\textbf{Yerevan Physics Institute,  Yerevan,  Armenia}\\*[0pt]
S.~Chatrchyan, V.~Khachatryan, A.M.~Sirunyan, A.~Tumasyan
\vskip\cmsinstskip
\textbf{Institut f\"{u}r Hochenergiephysik der OeAW,  Wien,  Austria}\\*[0pt]
W.~Adam, T.~Bergauer, M.~Dragicevic, J.~Er\"{o}, C.~Fabjan\cmsAuthorMark{1}, M.~Friedl, R.~Fr\"{u}hwirth\cmsAuthorMark{1}, V.M.~Ghete, N.~H\"{o}rmann, J.~Hrubec, M.~Jeitler\cmsAuthorMark{1}, W.~Kiesenhofer, V.~Kn\"{u}nz, M.~Krammer\cmsAuthorMark{1}, I.~Kr\"{a}tschmer, D.~Liko, I.~Mikulec, D.~Rabady\cmsAuthorMark{2}, B.~Rahbaran, C.~Rohringer, H.~Rohringer, R.~Sch\"{o}fbeck, J.~Strauss, A.~Taurok, W.~Treberer-treberspurg, W.~Waltenberger, C.-E.~Wulz\cmsAuthorMark{1}
% ...
\vskip\cmsinstskip
% ...
1:~~Also at Vienna University of Technology, Vienna, Austria\\
2:~~Also at CERN, European Organization for Nuclear Research, Geneva, Switzerland\\
% ...

\end{sloppypar}
\end{document}